\documentclass[11pt]{article}
\pdfoutput=1

\usepackage{graphics, color,soul}
\usepackage{graphicx}
\usepackage{amssymb}

\usepackage{lmodern,mathtools}

\usepackage{booktabs}
\usepackage[english]{babel}
\usepackage{amsmath,amssymb,amsbsy,amstext, amsthm, simplewick}
\usepackage{hyperref}
\usepackage{tikz}

\usetikzlibrary{decorations.pathmorphing,shapes.misc}
\tikzset{snake it/.style={decorate, decoration=snake}}
\tikzset{cross/.style={cross out, draw=black, minimum size=2*(#1-\pgflinewidth), inner sep=0pt, outer sep=0pt},
cross/.default={1pt}}

\usepackage{amsfonts}
\usepackage{amssymb}
\usepackage{upgreek}
\usepackage{simplewick}
 \usepackage{exscale,relsize}
\usepackage{mathtools}
\usepackage{comment}

\usepackage[margin=1cm,labelfont={sf,bf,scriptsize},textfont={sf,scriptsize}]{caption}

\usepackage{colortbl}
\definecolor{lightgreen}{cmyk}{0.2, 0, 0.2, 0.2}
\definecolor{lightgray}{cmyk}{0.1,0.2,0,0.1}
\definecolor{lightgray2}{cmyk}{0.1,0.1,0,0.1}

\makeatletter
\newlength{\apb@width}
\newcommand{\autoparbox}[2][c]{\settowidth{\apb@width}{#2}\parbox[#1]{\apb@width}{#2}}

\makeatother

\numberwithin{equation}{section}

\def\beq{\begin{equation}}
\def\eeq{\end{equation}}

\def\bea{\begin{eqnarray}}
\def\eea{\end{eqnarray}}

\def\d{{\rm d}}

\def\beq{\begin{equation}}
\def\eeq{\end{equation}}
\def\be{\begin{equation}}
\def\ee{\end{equation}}
\def\bea{\begin{eqnarray}}
\def\eea{\end{eqnarray}}

\def\d{{\rm d}}

\def\d{{\rm d}}

\def\0{{\vec{0}}}
\def\k{{\vec{k}}}

\def\k{{\bf k}}

\def\kps{{k_D^2}}

\DeclareRobustCommand{\SkipTocEntry}[4]{}

\def\d{{\rm d}}

\def\beq{\begin{equation}}
\def\eeq{\end{equation}}
\def\d{\partial}

\def\d{\partial}

\def\ba#1\ea{\begin{align}#1\end{align}}
\def\bg#1\eg{\begin{gather}#1\end{gather}}
\newcommand{\bseq}{\begin{subequations}}
\newcommand{\eseq}{\end{subequations}}

\DeclareSymbolFont{extraup}{U}{zavm}{m}{n}
\DeclareMathSymbol{\varheart}{\mathalpha}{extraup}{86}
\DeclareMathSymbol{\vardiamond}{\mathalpha}{extraup}{87}

\def\({\left(}
\def\){\right)}
\def\[{\left[}
\def\]{\right]}

\begin{document}

\begin{titlepage}

\setcounter{page}{1} \baselineskip=15.5pt \thispagestyle{empty}

\vbox{\baselineskip14pt
%\hbox{hep-th/0000000}
}
{~~~~~~~~~~~~~~~~~~~~~~~~~~~~~~~~~~~~
~~~~~~~~~~~~~~~~~~~~~~~~~~~~~~~~~~
~~~~~~~~~~~ }

\bigskip\

\vspace{2cm}
\begin{center}
{\fontsize{19}{36}\selectfont  \sc
Long-Range Nonlocality in Six-Point String Scattering:  simulation of black hole infallers
%Long-Range Nonlocality 
%via Longitudinal \\ String Spreading at Six Points
}
\end{center}

\vspace{0.6cm}

\begin{center}
{\fontsize{13}{30}\selectfont  Matthew Dodelson$^{1}$ and Eva Silverstein$^{1,2}$}
\end{center}

%\vspace{0.2cm}

\begin{center}
\vskip 8pt

\textsl{
\emph{$^1$Stanford Institute for Theoretical Physics, Stanford University, Stanford, CA 94306}}

%\vskip 7pt
%\textsl{ \emph{$^2$Kavli Institute for Theoretical Physics, University of California, Santa Barbara, CA 93106}}

%\vskip 7pt
%\textsl{ \emph{$^2$SLAC National Accelerator Laboratory, 2575 Sand Hill, Menlo Park, CA 94025}}

\vskip 7pt
\textsl{ \emph{$^2$Kavli Institute for Particle Astrophysics and Cosmology, Stanford, CA 94025}}

\end{center}

\vspace{0.5cm}
\hrule \vspace{0.1cm}
{ \noindent \textbf{Abstract} We set up a tree-level six point scattering process in which two strings
% $C$ and $1'$ 
are separated longitudinally such that they could only interact directly via a non-local spreading effect such as that predicted by light cone gauge calculations and the Gross-Mende saddle point.     
One string, the  `detector',  is produced at a finite time with energy $E$ by an auxiliary $2\to 2$ sub-process, with kinematics such that it has sufficient resolution to detect the longitudinal spreading of an additional incoming string, the `source'.  
%The spreading calculation specifically predicts that they could interact despite the fact that their centers are separated longitudinally by a scale of order $E\alpha'$ where $E$ is the energy of $1'$. 
We test this hypothesis in a gauge-invariant S-matrix calculation convolved with an appropriate wavepacket peaked at a separation $X$ between the central trajectories of the source and produced detector.
The amplitude exhibits support for scattering at the predicted longitudinal separation $X\sim\alpha' E$, in sharp contrast to the analogous quantum field theory amplitude (whose support manifestly traces out a tail of the position-space wavefunction).       
The effect arises in a regime in which the string amplitude is not obtained as a convergent sum of such QFT amplitudes, and has larger amplitude than similar QFT models (with the same auxiliary four point amplitude).  
In a linear dilaton background, the amplitude depends on the string coupling as expected if the scattering is not simply occuring on the wavepacket tail in string theory.  
%in sharp contrast to tree-level quantum field theory in a similar kinematic regime, whose amplitude as a function of $X$ traces out the tail of the wavefunction.  
%Part of the amplitude in both string theory and quantum field theory comes from the residues of poles -- which contains no such spreading effect in either theory -- but additional contributions behave differently in the two theories.        
% with its peak support shifted early by this amount.  
%In an appropriate kinematic regime, the worldsheet OPE leads to a factorized Regge form for the amplitude which precisely corresponds to the decomposition into the auxilliary four-point process and the detector-source scattering.  We support this interpretation by showing that the size of the amplitude is inconsistent with 
%Gauge invariance in open string theory precludes alternative interactions between the incoming strings.
%The spreading interaction range degrades as predicted when deformed so that the resolution of the detector string is reduced.   
%We contrast this with the behavior of tree-level quantum field theory with similar kinematics. 
This manifests the  scale of longitudinal spreading in a gauge-invariant S-matrix amplitude, in a calculable process with significant amplitude.  It simulates a key feature of the dynamics of time-translated horizon infallers.  
\vspace{0.3cm}
 \hrule

\vspace{0.6cm}}
\end{titlepage}

\tableofcontents

\section{Introduction and setup}

Light cone \cite{lennyspreading}\ and saddle point \cite{grossmende}\ calculations strongly suggest a significant degree of longitudinal nonlocality in string theory in the regime of large center of mass energy compared to the string scale, $E\gg 1/\sqrt{\alpha'}$.  
As reviewed in \cite{BHpaper,Smatrixpaper}, both analyses \cite{lennyspreading, grossmende}\ lead to the relations\footnote{The Gross-Mende saddle point is slightly larger than the longitudinal spreading scale, $\Delta X^+\, \Delta X^-\simeq \alpha'^2 k_{\perp}^2$ \cite{Smatrixpaper, SCH}. This leads to a $k_\perp^2$-dependent suppression factor in the four-point function \cite{Smatrixpaper}.}
\begin{align}\label{deltax}
(\Delta X_\perp)^2 &\simeq \alpha'\log\frac{E^2}{k_\perp^2}\\
\Delta X^+\, \Delta X^- &\simeq\alpha'\hspace{5 mm} \text{for }k_\perp^2>1/\alpha'
\end{align}
for relativistic $2\to 2$ scattering, where $k_\perp$ denotes the transverse momentum transfer in the process.\footnote{The relevant decomposition into light cone and transverse directions is determined uniquely by requiring that $k_\perp$ is shared equally among the scattered strings, defining the `brick wall frame' as in \cite{bpst}.}  As the resolution in light cone time $\Delta X^-\simeq k_\perp^2/E$ improves at large center of mass energy, the nonlocality in the $X^+$ direction increases.  A computation in \cite{bpst}, combining light cone and saddle point techniques, explicitly derives the cutoff on the mode number of worldsheet vacuum fluctuations which corresponds to the finite light-cone time resolution,\footnote{In the appendix, we clarify the distinction between this detector-dependent cutoff and the subtraction of the power law divergence in a similar sum over modes that enters into the string mass spectrum in light cone gauge.  We thank Milind Shyani and Ying Zhao for discussions.} as explained recently in \cite{BHpaper}.
More broadly, the UV softness of string amplitudes in itself implies some spreading of probability in position space.  

%This alone does not determine the direction or extent of the spreading, so the light cone and saddle point predictions are more quantitatively useful, but this basic intuition is also worth keeping in mind.   

However, the quantities computed in \cite{lennyspreading}\ are not manifestly gauge-invariant, and the embedding coordinates in \cite{grossmende}\ are evaluated on complex saddle points and hence are not directly applicable to the real geometry of the scattering process.  Therefore it is of interest to understand the role of the relations (\ref{deltax}) in real, gauge-invariant quantities.  This has been discussed in a number of very interesting works such as \cite{ACV}\cite{locality}\cite{JoeBHcomp}\cite{ggm}, but the results on the basic question of long-range interaction were inconclusive.  In \cite{Smatrixpaper}, we began a detailed analysis of this using wavepackets convolved against four and five point tree-level string scattering amplitudes, exhibiting features -- such as the deflection of the traced-back trajectory of a scattered string wavepacket before its center of mass would collide -- which are difficult to explain purely by the transverse spreading $(\Delta X_\perp)^2$.   Excluding purely transverse non-locality is important, and the arguments in \cite{Smatrixpaper}\ are subtle, giving only indirect indications of the long range $E/k_\perp^2\le \alpha' E$ predicted by (\ref{deltax}).   

More generally, it is crucial to take into account the possibility that fluctuations in the central positions of strings could account for effects that would otherwise be ascribed to the longitudinal spreading.\footnote{In \cite{Smatrixpaper}, this was conditional on an assumption that the incoming strings deflect directly into the corresponding outgoing ones in Regge kinematics, and that scattering can occur at the peak trajectory -- the most probable central value among position wavepackets.}
% (which as we will describe below, corresponds to scattering at a saddle point value in momentum).        
In this work, we will find that in an appropriate six point scattering process, the large scale of longitudinal spreading directly emerges at the level of the peak wavepacket separation of the relevant strings, in a kinematic regime we identify explicitly, improving substantially on \cite{Smatrixpaper}.  In this regime, moreover, a simple tree level quantum field theory analogue model would not interact as strongly at this separation, provided that we identify its four-point couplings with the amplitudes of certain sub-processes in the string theory (one of which is Regge-soft).   In order to test the spreading interpretation further, we summarize a result of \cite{LDpaper}\ in a weakly varying dilaton background, which yields string coupling dependence consistent with scattering at this peak wavepacket separation.    As a result, the longitudinal spreading prediction survives a substantial test.

Let us summarize the setup in more detail.  We will analyze a $3\to 3$ scattering process which directly exhibits the expected scale $\alpha' E$ of longitudinal non-locality.  A four-point sub-process produces a string at finite time (say $T=0$) which is kinematically set up to be an effective detector of the longitudinal spreading of a third incoming source string.   
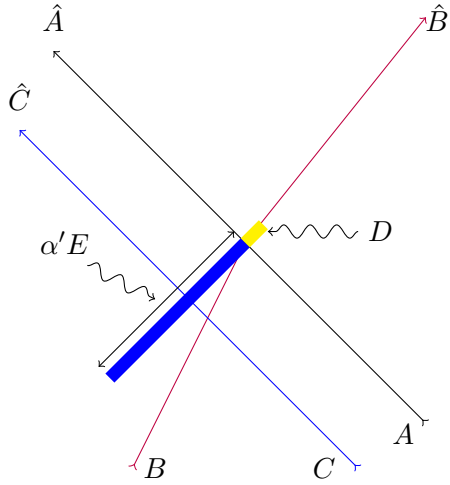
\begin{figure}[htbp]
\begin{center}
\begin{tikzpicture}[scale=3]
 \draw[>->,color=purple] (-.5,-1) -- (0,0) -- (.8,1);
 \draw[>->,color=blue] (.5,-1) --  (-1,.5);
 \draw[>->] (.8,-.8) -- (-.85,.85);
 \draw (-.85,1) node {$\hat{A}$};
 \draw (.85,1) node {$\hat{B}$};
  \draw (-1,.65) node {$\hat{C}$};
  \draw (.35,-1) node {$C$};
  \draw (-.4,-1) node {$B$};
  \draw (.7,-.85) node {$A$};
  \draw (-.8,0) node {$\alpha' E$};
  \draw[snake it,->] (-.7,-.1) -- (-.4,-.25);
  \draw[<->] (-.65,-.55) -- (-.05,.05); 
\fill[color=blue] (-.58,-.62) -- (.02,-.02) -- (-.02,.02) -- (-.62,-.58);
\fill[color=yellow] (.02,-.02) -- (.1,.06) -- (.06,.1) -- (-.02,.02);
\draw[snake it,->] (.5,.05) -- (.1,.05);
\draw (.6,.05) node {$D$};
 \end{tikzpicture}
\end{center}
\caption{ \label{sixfig}
The setup for our process, showing trajectories of interest which have support in the string amplitude.  The central trajectories of strings $A$ and $B$ collide at $T=X=Y=0$, producing outgoing strings $\hat{A}$ and $\hat{B}$.  We introduce the third incoming string $C$, with kinematics such that string $\hat{B}$ (and the closely related off-shell string $D$) has optimal light cone time resolution to detect the longitudinal spreading of strings $C$ and $\hat{C}$ (shown in blue) as in (\ref{k1pplus}).   To simplify the interpretation, we exclude a direct $2\to 2$ interaction $BC\to (B+C+\hat C)+\hat C$ by computing the amplitude for vertex operator orderings  such as $A\hat{A}C\hat{C}B\hat{B}$ which preclude such a sub-process.} 
% The amplitude for the process decomposes via a worldsheet OPE into the $AB\to 1'3$ and $C1'\to 2 1$ four-point amplitudes times the $1'$ Pomeron propagator.  It exhibits support for  
%It has a phase which implies peak support for the 
%$C\to 2$ trajectories propagating through the interaction region a time 
%$T_{C*}$
%of order $\alpha' E$  before the center of mass collision of $A$ and $B$, independently of the value of $\phi$, consistently with the longitudinal spreading predictions of \cite{lennyspreading, grossmende}.  We will refer to this process in the kinematic regime well described by $1'$ exchange as `case $1'$'.  
\end{figure}
According to the prediction of (\ref{deltax}), 
this third incomer could interact with the putative detector even if its center crosses the interaction region earlier, by a timescale of up to order $\alpha' E$.  With separated Gaussian wavepackets for the trajectories of $A$ and $C$, with peak $X=X_C-X_A$, longitudinal spreading would produce contributions supported at $X$ of order $\alpha' E$, a dynamical scale that is insensitive to the shape of the wavepacket.  This is in contrast to the corresponding process in quantum field theory, for which scattering only occurs for particle $C$ delayed with respect to $A$; in that case any contribution with $X<0$ is ascribable to scattering a tail of the wavefunction in position space.  

%The amplitude we will compute below decomposes via a worldsheet OPE into factors describing precisely the predicted dyanamics.  

By analyzing the amplitude for scattering between localized external states, we perform a quantitative test of the prediction that the third incomer interacts with the putative detector at a separation in the $X^+$ direction of order $\alpha' E$.\footnote{Below we will be more precise about the factors multiplying $\alpha' E$ in the interaction scales arising in various regimes.}   
With Gaussian wavepackets for the trajectories of $A$ and $C$, with peak $X=X_C-X_A$ and position-space width $1/\sigma \ll E\alpha'$, we find a string theory amplitude $A(X)$ which exhibits a peak of support early in $X$ by the predicted amount.  
We take a relatively thin width $\sigma$ in momentum space in order to optimize the detector kinematics, which leads to somewhat coarse (but still sufficient) resolution in position space.   
In sharp contrast, QFT comparison models instead produce a distribution with width $\sim 1/\sigma$ at small $X$, with Gaussian-suppressed support at large $X$ consistent with scattering on the tail of the wavepackets.  
%We show explicitly how the QFT result for $A(X)$ at sufficiently large $|X|$ can be ascribed to wavepacket tails.  
If we identify the four-point sub-process amplitudes in tree-level QFT and string theory, then
string theory produces parametrically stronger scattering at $X\sim E\alpha'$, and a further
analysis using a linear dilaton background fits with this picture of the scattering geometry \cite{LDpaper}.
%that this result does not fit with it scattering purely on the tail of the wavepackets, in part using
%a dilaton background as a tracer of the string interactions \cite{LDpaper}.  
%In an appropriate kinematic regime, the corresponding calculation in string theory contains contributions shifted early compared to this tail contribution, and also spread out at the scale $\sim\alpha' E$.   
This result arises in a regime in which the string theory amplitude is not obtained as a convergent sum over QFT propagators (which we compare and contrast to a special regime with such an expansion).

%  Similar results hold with respect to the wavefunction of string B, as we explain in detail below.   

%These results are summarized in Figures \ref{AQFT}-\ref{ASTtriangle}.  
 
%with and and a phase peak support at $T_*\sim -2\pi \alpha' E$.  
%This range of longitudinal non-locality remains robust against deformations that preserve the predicted scale, and  appropriately declines under deformations that degrade the predicted sensitivity of the detector.   Alternate transverse spreading-induced interactions  or rare fluctuations can be excluded using compactly supported wavepackets, and in some cases simply by virtue of the overall size of the amplitude, which is parameterically  too large to admit such interpretations in a kinematic regime we derive below.
%These results come from an analysis of two versions of this process, each relevant in a particular window of kinematic parameters of the same amplitude calculation, as described in figures \ref{sixfig}\ and \ref{sixfigb}.  

The predicted long-range longitudinal non-locality could have far-reaching implications, becoming particularly interesting when applied to horizon physics \cite{BHpaper,backdraft}.  In that context, the trajectories of test objects time-translated relative to each other by $\Delta t$ develop an exponentially large center of mass energy squared $s\propto e^{\Delta t/2r_s}$ in the near-horizon region, where $r_s$ is the curvature radius in the geometry.
This region, a Rindler patch of flat spacetime, is given by
\be\label{Rindler}
ds^2\approx -2dX^+\, dX^- +d X_\perp^2  ~~~~~~ {\rm for} ~~~  X^+X^- \ll r_s^2, ~~~~ X_\perp^2\ll r_s^2.
\ee
%where $r_s$ is the curvature radius of the geometry.  
The large near-horizon energy arises even for weakly curved geometries, with the evolution in the geometry gradually building up a large relative boost.  The two trajectories cross the horizon displaced in the $X^+$ direction by an amount that grows linearly with the center of mass energy.  

In this situation, non-locality that grows in a way commensurate with the center of mass energy can lead to interaction between the two infallers \cite{backdraft}.  In \cite{BHpaper}, we analyzed this explicitly for the non-locality introduced by longitudinal string spreading, finding a wide window of parameters where this occurs, given (\ref{deltax}).  This would not occur for weakly interacting particles at the same order, and we characterized this as a breakdown of effective quantum field theory in horizon physics.  (However, it is an interesting question whether this would occur just as well for field theories such as QCD which exhibit Regge behavior and may have a string-theoretic formulation, something we can investigate using holographic large-N gauge theories \cite{Byungwoo}.)  

\begin{figure}[htbp]
\begin{center}
\begin{tikzpicture}[scale=2.7]
\draw (1,.85) node {$\hat{B}$};
\draw (.6,-.9) node {$C$};
\draw (-.9,.65) node {$\hat{C}$};
\draw (0,.85) node {$\hat{A}$};
\draw (1,-.1) node {$A$};
\draw (-1,-.8) node {$B$};
\draw[>->,color=blue] (.5,-1) -- (-1,.5);
\fill[color=yellow] (.55,.45) -- (.45,.55) -- (.55,.65) -- (.65,.55);
\draw[>->,dashed] (1,0) -- (0,1);
\draw[->] (.5,.5) -- (1,1);
\draw[dashed,>-] (-1,-1) -- (.5,.5);
\draw (-.1,-.3) -- (.55,.35);
\draw[snake it,->] (1.2,.5) -- (.7,.55);
\draw (-.1,-.3) -- (-.2,-.2);
\draw (.55,.35) -- (.45,.45);
\draw (.5,-.2) node {$\Delta X^+$};
\draw (1.3,.5) node {$D$};
\end{tikzpicture}
\end{center}
\caption{ \label{rindler} The six-point process naturally simulates horizon physics by producing a detector string $D$ at finite longitudinal separation $\Delta X^+$ from the source string $C$. From this perspective strings $A,B,$ and $\hat{A}$ (drawn as dashed lines) are auxiliary external legs, whose only purpose is to collide and produce $D$.}
\end{figure}
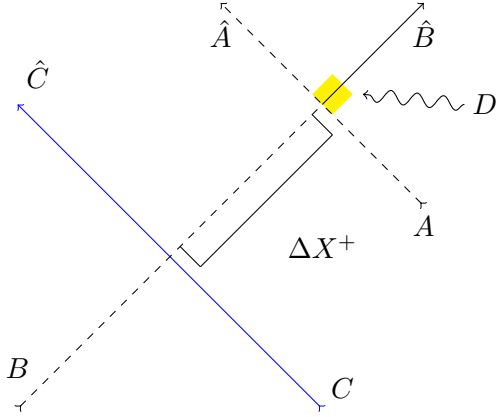

In fact, the process described by our flat spacetime six-point function provides a calculable simulation of the configuration of early and late infallers in horizon physics.  In that problem, the two strings' centers never cross paths in the near horizon region; as just noted they are widely separated in $X^+$, and appear in the near horizon region as if they had already moved past each other.  That means that in the Rindler patch, they are configured just like the source and detector strings in our six-point process, as shown in Figure \ref{rindler}.\footnote{We thank D. Marolf for early discussions of a setup like this.}  It will be very interesting to apply this to articulate more explicit predictions for observables in black hole physics and cosmology.  

\subsection{Quantum field theory comparison models}

We will explicitly compare and contrast the behavior of the string theory amplitude just set up with two tree-level field theory models, whose amplitudes appear below in (\ref{AQFT0}-\ref{AQFT1}).  The simplest is a six-point contact interaction, with a coupling $\lambda_6\sim g_\text{s}^4$; we will refer to this model as QFT0.  The second, QFT1, is a model with four point interactions $\lambda_{AB\hat{A}D}\propto g_\text{s}^2$ and $\lambda_{CD\hat{C}\hat{B}}\propto g_\text{s}^2$, with $D$ a massless particle.  The six point scattering in this theory includes massless $D$ exchange, and we will focus on this model in making our comparisons.

%One could also consider a QFT model with some additional massive bosons, spaced as in string theory with $m_n^2\sim n/\alpha'$ for a finite range of integer $n$.  However, in the regime we will be interested in, the softness of the string theory amplitude arises from the contributions of higher spin massive particles, with alternating signs for their residues.  This is not straightforward to mock up in a self-contained theory of point particles since higher spin fields couple only via irrelevant operators.  Of course higher spin massive particles can arise as composites at low energies in a self-contained QFT, but such particles may exhibit spreading effects associated with their partons.\footnote{See for example \cite{SpinsStrings}\ for a recent discussion of the relation between higher spins and string theory.}  We plan to investigate spreading in large-N gauge theories in future work \cite{Byungwoo}.  But in this paper, we will take our control group to be the simple tree-level renormalizable QFT models just mentioned.       

%if we avoid having fields with only irrelevant interactions, we should restrict these fields to spins $\le 1$.    
%Finally, we can also consider a model we will call QFT2 with a finite number of additional massive particles, $1'_n$, with a mass spectrum spaced as in string theory, $m_n^2\sim n/\alpha'$.  
%This last model is the most similar to tree-level string theory, but of course it still lacks the UV softness of the string amplitudes.  We will comment further on this below.          

\section{Kinematics, predicted detector resolution, and pole structure}

In order to set up the process described in the previous section, we consider $3\to 3$ string scattering.   For simplicity we work in kinematics for which the incoming and outgoing strings move in three dimensions $(T, X, Y)$.  In general for relativistic string motion the wavefunctions are linear combinations of vertex operators with momenta 
\bea\label{fullkin}
k_a&=&(\omega_a, p_a, q_a), ~~~~ a=A, B, C \nonumber \\
k_j &=&(-\omega_j, -p_j, q_j), ~~~~ j=\hat{A},\hat{B},\hat{C}
\eea
where
\be\label{omegadef}
\omega_I=+\sqrt{p_I^2+q_I^2}, ~~~~ I=A-C, \hat{A}-\hat{C}
\ee
%We will shortly specify a wavepacket given by linear combination of on-shell momentum eigenstates, 
Since we are working in a Minkowski spacetime background, the scattering amplitude contains an energy-momentum conserving delta function.  In the process of interest, the strings are moving mainly in the $X$ direction.  By working with appropriate wavepackets and incorporating the delta function in the amplitude, we will be able to consistently restrict the momenta to this regime while still localizing the strings sufficiently in position space to test for longitudinal interaction.     

We will work with on-shell wavefunctions (linear combinations of momentum eigenstate vertex operators) -- that are pure momentum eigenstates for the outgoing strings $\hat{B}, \hat{C}$, and $\hat{A}$.  We must also specify wavefunctions for the incoming strings $A$, $B$, and $C$.  For reasons that will become clear below, in the longitudinal ($X$) direction we will work with Gaussians of string-scale width for $A$ and $C$ (with central values in position space separated by a parameter $X$) and for $B$ we will specify a wavefunction supported only within a very small range of $\tilde p_B$  of order $E$.   Here and below, we use a tilde to denote momenta that are integrated over in the wavepackets that we will explicitly set up below.   

Define the total energy-momentum as
\be\label{Ktot}
k_{\text{tot}}=-({k}_{\hat A}+{k}_{\hat B}+{k}_{\hat C})=(\omega_{\text{tot}}, P, Q).
\ee
This is constant, fixed by the choice of momentum eigenstates for strings $\hat{A},\hat{B},\hat{C}$.  
We will work in a frame where the total transverse momentum vanishes ($Q=0$), and $P$ is of order $-E$.  We are interested in a regime in which $A$ and $C$ are moving toward negative $x$ with momenta $p_A, p_C\sim -E$, and $B$ is moving toward positive $X$ with momentum $p_B\sim E$, as depicted in Figure \ref{sixfig}.   

We will find it sufficient, and simplest, to work at fixed transverse momenta 
\be\label{Qs}
q_B=0, ~~~~ q_A=-q_C\equiv q 
\ee
and will consider separately the cases $q=0$ and $q\ne 0$.  

We solve the longitudinal ($x$-direction) momentum conservation condition with 
\be\label{pcons}
\tilde p_A = P-\tilde p_B-\tilde p_C
\ee        
Moreover, for all our wavepackets, once we impose energy conservation $\tilde p_B$ will vary, if at all, within a small enough range that the variation of $A$ and $C$ essentially cancel:  $\delta \tilde p_A\simeq -\delta \tilde p_C$.  Thus $\delta \tilde p_C$ is conjugate to the 
longitudinal spatial separation $\tilde X$ between $C$ and $A$.  

Next let us consider the energy conserving delta function
\be\label{enmom}
\delta\left(f(\tilde p_B, \tilde p_C))\right),
\ee
with
\bea\label{fdef}
 f(\tilde p_B, \tilde p_C) &=& \omega_A(\tilde p_B, \tilde p_C) +\omega_B(\tilde p_B)+\omega_C(\tilde p_C)-\omega_{\text{tot}}\nonumber \\
 &=& \sqrt{(P-\tilde p_B-\tilde p_C)^2+q^2}+|\tilde p_B|+  \sqrt{\tilde p_C^2+q^2}-\omega_{\text{tot}}
\eea
The delta function sets this function to zero.  

First consider the case $q=0$.  Energy conservation fixes $\tilde p_B$
\be\label{pBqzero}
\tilde p_B=\frac{1}{2}(\omega_{\text{tot}}+P) ~~~~~~~~~~~~~~~~(q=0)
\ee  
for
\be\label{PCrangeqzero}
\frac{1}{2}(P-\omega_{\text{tot}})<\tilde p_C< 0 ~~~~~~~~~~~~(q=0)
\ee
For $q=0$, if the trajectories of $A$ and $C$ are separated in the longitudinal plane at one time, they never meet.   We will introduce Gaussian wavepackets with support well within the endpoints of the range (\ref{PCrangeqzero}).  This case of $q=0$ is sufficient for our simplest test of spreading below in \S\ref{qzero}.

Next let us consider $q\ne 0$.  In this case, there is a meeting point of the $A$ and $C$ trajectories (as depicted below in figure \ref{Tmeetfig}).  
Let us consider (\ref{fdef}) as a function of $\tilde p_C$ at fixed $\tilde p_B$.  It has a minimum at $\tilde p_C=(P-\tilde p_B)/2$.  For sufficiently large $\tilde p_B$, this minimum is above zero and there is no solution.  For sufficiently small $\tilde p_B$, there are 2 solutions.  In the regime where $q\ll |p|$ for all strings, this is given to good approximation by
\be\label{pCpB}
\tilde p_C^{\pm}\equiv \frac{1}{2}\left(P-\tilde p_B\pm\sqrt{(\tilde p_B-P)^2-\frac{2(\tilde p_B-P)q^2}{2(\omega_{\text{tot}}+P-2\tilde p_B)}} ~ \right)
\ee
At one special value of $\tilde p_B$ there is a single solution to $f=0$, at which $\tilde p_A=\tilde p_C=(P-\tilde p_B)/2$.  Losing the solution as we increase $\tilde p_B$ through this point has a simple interpretation:  it corresponds to a kinematic configuration such that $A$ and $C$ cannot meet at a point if they cross the $T=0$ plane at separate positions in $x$, since they are moving at the same speed in the $x$ direction.  Thus there is no support for a pointlike interaction between them.  For $q=0$ we will work with a wavepacket for $B$ with nonzero support only for a small range of $\tilde p_B$.  The amplitude, including the energy conserving delta function, will then restrict the range of $\tilde p_C$ to be sufficiently small that we can work in a simple kinematic regime.      

In all of our calculations below,  the wavepackets will have significant support only for momenta satisfying $\omega_I\sim E\gg |q_I|$ in which the kinematics simplifies to the form
\begin{align}\label{Eq}
k_A&=(E_A+q_A^2/(2E_A),-E_A,q_A)\\
k_B&=(E_B+q_B^2/(2E_B),E_B,q_B)\\
k_C&=(E_C+q_C^2/(2E_C),-E_C,q_C)\\
{k}_{\hat A}&=(-E_{\hat A}-q_{\hat A}^2/(2E_{\hat A}),E_{\hat A},q_{\hat A})\\
{k}_{\hat B}&=(-E_{\hat B}-q_{\hat B}^2/(2E_{\hat B}),-E_{\hat B},q_{\hat B})\\
{k}_{\hat C}&=(-E_{\hat C}-q_{\hat C}^2/(2E_{\hat C}),E_{\hat C},q_{\hat C}),
\end{align}
constrained by the energy-momentum conserving delta function, as just discussed.  
The restricted support in momentum space will entail a position-space tail, whose effects we will include in our analysis.  
%where 
%\begin{align*}
%E_{\hat B}=E_B+\frac{q_A^2}{4E_A}+\frac{q_B^2}{4E_B}+\frac{q_C^2}{4E_C}-\frac{q_{\hat B}^2}{4E_B}-\frac{q_{\hat C}^2}{4E_{\hat C}}-\frac{(q_A+q_B+q_C+q_{\hat B}+q_{\hat C})^2}{4E_{\hat A}}.
%\end{align*}

The amplitude depends on kinematic invariants
\be\label{KIJs}
K_{IJ}=2\alpha' k_I\cdot k_J
\ee
which provide a useful, albeit redundant, parameterization.  
These are fixed (of order $q^2$) between strings moving in the same direction, and large in magnitude (of order $\pm E^2$) for strings moving in opposite directions.   The intermediate string $D$ described above,
\be\label{konep}
k_{D}=k_C+k_{\hat B}+k_{\hat C}=-(k_A+k_B+k_{\hat A})
\ee 
and specifically the behavior of $k_{D}^2$ as a function of $E_C$, will play an important role.   
Some of the invariants we will refer to in our analysis are explicitly
\begin{align}\label{someKIJs}
K_{A\hat{A}}&=\alpha'\frac{(q_AE_{\hat A}-(q_A+q_B+q_C+q_{\hat B}+q_{\hat C})E_A)^2}{E_AE_{\hat A}}\\
K_{C\hat{C}}&=\alpha'\frac{(q_CE_{\hat C}+q_{\hat C}E_C)^2}{E_C E_{\hat C}}\\
k_{D}^2&=\frac{K_{C\hat{C}}}{\alpha'}+4E_{\hat B}(E_C-E_{\hat C})+\frac{(E_Cq_{\hat B}+E_{\hat B}q_C)^2}{E_{\hat B}E_C}-\frac{(E_{\hat C}q_{\hat B}-E_{\hat B}q_{\hat C})^2}{E_{\hat B}E_{\hat C}} 
\end{align} 
%It will also be useful to record the behavior of the quantities $K_{IJ}=2\alpha' k_I\cdot k_J$ which enter directly into the string amplitudes.  This is a redundant description, but useful in the calculation.  They are given in Appendix \ref{sec:KIJs}.  In our analysis below, we will work with kinematic invariants sufficiently large that the amplitude can be reliably calculated by saddle point.    

\subsection{Detector and spreading prediction}\label{detectorprediction}

The string worldsheet topology and vertex operator ordering allows for a process in which $A$ and $B$ join to produce Strings $D$ and $\hat{A}$, with $D$ joining $C$ to produce $\hat{B}$ and $\hat{C}$.  In our analysis we will not {\it assume} this sequence of events in spacetime, but in this subsection we will briefly review the prediction \cite{lennyspreading, BHpaper}\ for a spreading-induced interaction between $C$ and $D$ in such a process.   

If we view string $D$ as a detector of string $C$'s spreading, then for $\alpha' k_D^+ k_D^-\ge  O(1)$ the light cone prediction is\footnote{See specifically Section 2.3\ of \cite{BHpaper}\ for the suppression of the spreading radius for a timelike detector trajectory, corresponding to $\alpha' k_D^+k_D^-\ge 1$ in (\ref{refinedprediction}).  For $\alpha'k_D^+k_D^-<1$, the light cone spreading prediction is $\Delta X^+_{\text{spreading}}\sim \alpha'E$.}
\cite{lennyspreading, BHpaper}\
\be\label{refinedprediction}
\Delta X^+_{\text{spreading}} \sim \frac{ k_{D}^+}{k_D^+ k_D^-}\sim \frac{E_{\hat B}}{k_D^+k_D^-}.
\ee  
The last estimate here is applicable in the regime $|E_C-E_{\hat C}|\ll E$, for which
%String $1'$ will behave as a Pomeron for sufficiently small $|k_D^2|$ (albeit still $\gg 1/\alpha'$), arising from an OPE limit in which the vertex operators for strings 1, 2, and $C$ approach each other. 
String $D$ and its continuation into string $\hat{B}$ have light cone energy
\be\label{k1pplus}
-k_{D}^+\simeq -k_{\hat B}^+\simeq \sqrt{2}E_{\hat B}.
\ee
The sign here corresponds to the fact that string $\hat{B}$ is outgoing, while string $D$ is an incoming leg of the process $CD\to \hat{C}\hat{B}$ in which it is a putative detector of the spreading of string $C$.
According to the predictions of longitudinal spreading \cite{lennyspreading,BHpaper,Smatrixpaper,bpst}, this implies sufficient light-cone time resolution for string $D$ or $\hat{B}$ to detect string $C$, with a predicted range bounded above by $\alpha' E$ (at high energy with fixed momentum transfer).   
%This resolution is optimal regardless of the value of $\phi$, predicting that the extent of the spreading should not depend on that angle.  We will test this explicitly in the amplitude below, and will find this to be the case.  

It is important to note, as in \cite{BHpaper, Smatrixpaper}\ that the predicted interaction between the string and the detector is causal:  in the free single-string state, the mode sum which determines the mean square size of the string is quadratically divergent, cut off by the energy resolution of the detector.  A time advance, in which interaction occurs without the centers of the strings meeting, is not acausal in this context.   In the next subsection we briefly review the manifestation of causality in the S-matrix, and its limitations.  

\subsection{Pole structure, causality, and the S-matrix}\label{sec:poles}

In tree-level quantum field theory, there is no spreading phenomenon in light cone gauge analogous to that in string theory, and causality demands that point particles scatter without time advances.  
In that case, one can  work with gauge-invariant local operators which commute outside the light-cone to establish causality and hence the absence of time advances. 

In string theory, we work with the S-matrix because we do not have such simple local observables.
The manifestation of causality in the S-matrix is encoded in the $i\epsilon$ prescription at the poles,
in that the Fourier transform of a simple pole in momentum space gives a step function in position space. 
The limitations on this arise because the Fourier transform is not exactly what arises in computing S-matrix amplitudes, since energies are positive and the amplitude includes the energy-momentum conserving delta function which is not analytic.
But since the $i\epsilon$ prescription is the same in string theory and QFT \cite{JoeBook}\cite{Edepsilon}, it is important to understand the implications of S-matrix data alone for scattering advances as a function of the center of mass of each string. 

At the level of the momentum space amplitude, string theory does not generally decompose into a convergent sum of particle propogators.  This will play a role in our results below.  
%Even though string theory has the same $i\epsilon$ prescription 
%Even when there is a convergent sum of this kind, there are infinite sequences of poles and the amplitudes are softer in the UV than the individual terms.   
 The standard limitations on the information contained in S-matrix amplitudes will also enter our analysis.  In general the connection between analyticity and causality is not completely sharp in the S-matrix \cite{Smatrixbook}\cite{AllanNima}.    
 Because the energies of the particles are positive, one does not obtain the full Fourier transform, and hence we do not precisely get a step function in time.\footnote{We expand on this via toy integrals in the appendix.}  In some circumstances, this is a small effect.  But assessing that for a given scattering problem requires analyzing it in detail with explicit wavepackets, taking into account the behavior of the momentum space amplitude over the range of momenta supported.  
%One must take into account that the full momentum space amplitude is not analytic in the momenta because of the positivity of the energies $\omega$ appearing in the energy-momentum conserving delta function.  This, and the positivity of the energy, introduces endpoints into the momentum integrals that affect the Fourier transform of the amplitude.  

A related limitation is the uncertainty principle: imperfect knowledge of the the relative momenta of scatterers can complicate the determination of their range of interaction.
Consider more specifically the process we are setting up in this work.
The trajectories for A and C on the peak of the wavepackets we will set up in the generic case $q_A,q_C\not=0$ are depicted below in figure \ref{Tmeetfig}\ in \S 5.
For $p_A-p_C<0$ as depicted there, scattering at $T=0$ would require longitudinal non-locality; conversely if 
$p_A-p_C>0$ scattering at $T=0$ is delayed, not requiring non-locality. 
We will use appropriate wavepackets to restrict the support of $p_A-p_C$ to be negative, as depicted.  For a class of momentum ranges, the resulting amplitudes exhibit a sharp distinction between quantum field theory and string theory.   
%Moreover, in appropriate kinemaic regimes the string theory amplitude does not arise as a sum convergent over simple QFT propagators, and there is additional structure in the relevant direction ($p_C$) in momentum space.  

In S-matrix amplitudes with wavepackets, the scattering may appear advanced as a function of the peak position in the wavefunction, but this may be consistent with purely delayed scattering on its tail.  We will present results that are consistent with this interpretation in our tree-level QFT comparison models, whose amplitudes track the tails of the position space wavefunctions and exhibit strong dependence on their parameters.  In string theory, we will find rather different behavior which is consistent with the longitudinal spreading prediction.  
In a regime where the amplitude is not obtained as a convergent sum over particle propagators, we will see that its amplitude is larger than the corresponding QFT one, provided that we identify the four-point sub-process amplitudes in QFT and string theory.   The possibility that the scattering proceeds via longitudinal spreading (as opposed to only occurring on the wavepacket tail) passes a further test in \cite{LDpaper}\ that we will summarize.

\section{Wavepackets}\label{sec:wavepacket}

In order to understand aspects of the position space geometry of the scattering process, it is useful to convolve the momentum-space string amplitude with nontrivial wavepackets for the strings $A$, $B$, and $C$.  That is, we consider linear combinations of on-shell vertex operators for the incoming strings.  

Before discussing our wavepackets in detail, let us briefly review how they fit into the total probability for scattering, taking into account relativistic normalization factors.\footnote{See e.g. chapter 4 of \cite{PeskinSchroeder}\ for the relevant background and conventions.}   We will set up wavepackets for incomers $A, B,$ and $C$ normalized as  
\be\label{Psinormgen}
\int  \frac{d^3\k}{(2\pi)^3}|\Psi(\k)|^2 =1.
\ee
Then the probability for $A$, $B$, and $C$, to scatter into $\hat A, \hat B, \hat C$ within a momentum range $d^3 \k_{\hat A} d^3 \k_{\hat B} d^3 \k_{\hat C}$ of outgoing momenta $\k_{\hat A}, \k_{\hat B}, \k_{\hat C}$ is given by
\be\label{Pdiff}
P = \prod_{j=\hat A, \hat B, \hat C} \frac{d^3\k_{j}}{(2\pi)^3}\frac{1}{2 \omega_{j}}| \langle \k_{\hat A}, \k_{\hat B}, \k_{\hat C}|S|\phi_A \phi_B\phi_C\rangle |^2
\ee
where 
\be\label{phidef}
|\phi_A\phi_B\phi_C\rangle = \prod_{a=A, B, C}\int \frac{d^3\k_a}{(2\pi)^3\sqrt{2 \omega_a}}\Psi_A(\k_A)\Psi_B(\k_B)\Psi_C(\k_C) |\k_A, \k_B, \k_C\rangle
\ee
In the kinematic regime which will interest us, the factors of $\omega_I$ in the denominator will not vary substantially with the $\k_I$ supported by the wavepackets.  As a result, we can pull out those factors and focus on the behavior of the momentum space amplitude convolved with the wavepackets $\Psi_I(\k_I)$, keeping in mind that the final result for the amplitude includes the inverse powers of $\omega$.  
We note that these factors are common to string theory and quantum field theory.  

These factors must also be included in considering the implications for black hole physics.\footnote{We thank D. Stanford for a discussion of this point.} In that setup, however, the incoming energies $\omega$ which enter into the relativistic normalization in the asymptotically flat region of the black hole geometry are much smaller than the center of mass energy of time-translated infallers in the near horizon region \cite{BHpaper}.     

\subsection{Wavepacket widths}

Before getting into any details, let us comment on the width of the wavepackets.  The light cone gauge spreading prediction described above in \S\ref{detectorprediction}\ is given in terms of the momentum of the detector $D$, so we will work with wavepackets that are supported in a window of momenta for which longitudinal spreading is expected from that point of view.   
At the same time, of course, we wish to resolve aspects of the geometry of the process in position space so we cannot work with pure momentum eigenstates.  We will see that we can use wavepackets that localize sufficiently in momentum and position space to satisfy both of these requirements.          

Let us spell this out more explicitly, since it is a somewhat subtle aspect of our analysis.   We will include wavepackets for $A$ and $C$, whose relative position is conjugate to the momentum variable $\delta E_{C}\sim -\delta E_A$ (as discussed above, string $A$ will solve the longitudinal momentum conservation delta function,  absorbing the change $\delta E_C$ in $C$'s longitudinal momentum). 

The required resolution in position space is not too stringent in itself:  we are interested in resolving the relative positions of $A$ and $C$ within the large predicted spreading scale $\sim E\alpha'$.  In momentum space,
however,  requiring that the wavepackets have their support reasonably close to the optimal detector kinematics leads to a nontrivial condition.  In particular, let us require $k_D^2$ to stay of order $q^2$ in magnitude, since according to (\ref{refinedprediction}) the spreading would degrade strongly otherwise.   We can translate that to a limit on the support of $\delta E_C$ using (\ref{someKIJs})
\be\label{kDvariation}
k_D^2 = 4 E_{\hat B}\delta E_C +\dots.
\ee
Restricting the support of $\kps$ to within of order $q^2$ restricts $\delta E_C$ to within of order $q^2/E$.  This is a strong constraint on the momenta.  Fortunately, it is still within the broad position space resolution we need, but the corresponding wavepackets are quite wide in position space compared to string scale.    
Below, we will find the most interesting results for wavepackets of this kind -- of width $\sim q/E$ in momentum space  (with the order $q$ parameter being specifically $\sqrt{-\eta}\equiv \sqrt{-k_B\cdot k_{\hat B}+k_C\cdot k_{\hat C}}$).  We will comment further on the role of this parameter below.      

\subsection{Gaussian wavepackets for $A$ and $C$}

For $A$ and $C$, we will work with Gaussian wavepackets,  centered on trajectories with momentum $p\sim -E$, as in the setup of Figure \ref{sixfig}.  
In momentum space the wavepackets take the form
\be\label{wavepacketPGauss}
\Psi_{X, p}(\tilde p)=\frac{1}{\pi^{1/4}\sigma^{1/2}}\exp (-i \tilde p X) \exp({-(\tilde p-p)^2/2\sigma^2})
\ee
where $\tilde p$ is a canonical momentum variable, and the subscripts are parameters of the wavefunction.
The position space wavefunction is 
\be\label{posGauss}
\Psi_{X, p}(\tilde X, T)=\int \frac{d\tilde p}{\pi^{1/4}\sigma^{1/2}} e^{i\sqrt{\tilde p^2+q^2}T+i\tilde p(\tilde X-X)}e^{-(\tilde p-p)^2/2\sigma^2}
\ee
At $T=0$ this is a Gaussian $\propto e^{-(\tilde X-X)^2/2\sigma^2}$ centered at $\tilde X=X$ with width $1/\sigma$.   At other times, it is approximately Gaussian centered on the trajectory $T=-(\tilde X-X)$.  The evolved position space wavepacket also has a contribution with a power-law tail, but this is suppressed by $e^{-p^2/2\sigma^2}$. 
%We can, for example, take the width to be of string scale:  $\sigma^2\sim 1/\alpha'$; this strongly suppresses the additional term in our regime with $p\sim -E$, while localizing the wavepackets in position space well below the spreading scale $\sim \alpha' E$.   
We will center string $A$ such that at the peak of its wavepacket, it crosses $T=0$ at $X_A=0$, and send C in with a peak trajectory passing through $X\equiv X_C$ at $T=0$.      

\subsection{Wavepackets for string $B$}

Let us next discuss wavepackets for string $B$.  
In the case $q=0$, energy conservation fixes $\tilde p_B$ (\ref{pBqzero}), and we can work with any wavepacket for $B$ with support at that value.   The remainder of this section is not essential for the reader interested in the simplest test of spreading in $q=0$ kinematics, which is contained in \S\ref{qzero}\ below.  

\subsubsection{$B$ wavepackets used in $q\ne 0$ analysis}

For our $q\ne 0$ analysis below, we will consider wavefunctions which are sharply supported between two nearby values of $\tilde p_B$.  One example of this is simply the step function
\be\label{wavepacketPsharp}
\Psi_{X_B, p_{B,\text{min}}, p_{B, \text{max}}}(\tilde p_B)=\frac{1}{\sqrt{\Delta p_B}}\exp (-i \tilde p_B X_B) \theta ({p_{B,\text{max}}}-\tilde p_B) \theta (\tilde p_B-p_{B, \text{min}})
\ee
where we defined
\be\label{DelpB}
\Delta p_B=p_{B,\text{max}}-p_{B,\text{min}}
\ee
Here $\tilde p_B$ is the canonical momentum variable, and the subscripts are parameters of the wavefunction.  In our setup, $p_{B, \text{min}}$ and $p_{B, \text{max}}$ are both of order $E$.  

This wavepacket for B is sharply cut off in momentum space, which will lead to a power law tail in position space.   This may seem like a counterintuitive choice, given that we would like to constrain the position space geometry. 
However, as we will discuss in detail below (see Figure \ref{Tmeetfig}\ and eqn (\ref{Tmeet})), the resolution we need for B's trajectory is much weaker than that for A and C.
The utility of the sharp cutoff in $\tilde p_B$ is that this, in combination with the energy-momentum conservation in the amplitude, sharply ensures that (i) the strings are moving in the intended direction, and (ii) the range of momentum integrated over is small enough that $\delta \tilde p_C$ is absorbed by $\delta \tilde p_A$ to good approximation.  This simplifies the analysis and interpretation, leaving enough structure in position space to strongly distinguish the behavior of string theory and tree-level quantum field theory.  

If we take B to have zero transverse momentum (an eigenstate with $\tilde q_B=0$), the corresponding position space wavefunction is
\be\label{wavepacketXsharp}
\Psi_{X, p_{B,\text{min}}, p_{B,\text{max}}}(\tilde X_B, T) = i\frac{ e^{i p_{B,\text{min}} (T-(\tilde X_B-X_B))}-e^{i {p_{B,\text{max}}} (T-({\tilde X_B}-{X_B}))}}{\sqrt{2 \pi\Delta p_B } (T-(\tilde X_B-X_B))}
\ee  
Again $X_B$ is the peak of the wavefunction, not to be confused with that of a position eigenstate.  
The corresponding probability in position space is
\be\label{PsiXsq}
|\Psi|^2=\frac{2}{\pi}\frac{\sin^2(\frac{\Delta p_B}{2}(T-(\tilde X_B-X_B)))}{\Delta p_B(T-(\tilde X_B-X_B))^2}
\ee
This is nearly constant for $|T-(\tilde X_B-X_B)|\ll 1/\Delta p_B$ but drops rapidly beyond that scale.  We will put $X_B=0$ in our analysis, so that the peak of this wavepacket describes B hitting A at $X=T=0$.  

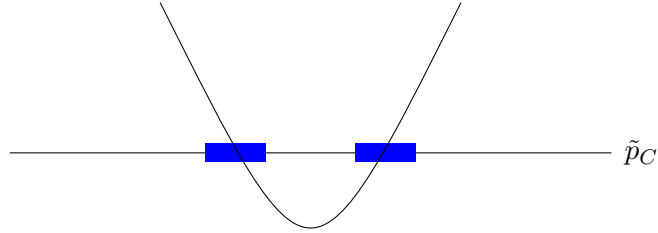
\begin{figure}[htbp]
\begin{center}
\begin{tikzpicture}[scale=4]
\draw (-1,0) -- (1,0);
\fill[blue] (-.35,.03) -- (-.15,.03)--(-.15,-.03) -- (-.35,-.03)--(-.35,.03);
\fill[blue] (.35,.03) -- (.15,.03)--(.15,-.03) -- (.35,-.03)--(.35,.03);
    \draw (-.5,.5) .. controls (0,-.5) .. (.5,.5);
\draw (0,1) node{$\sum_a \tilde{\omega}_a-\omega_{\text{tot}}$};
\draw (1.1,0) node{$\tilde{p}_C$};
\end{tikzpicture}
\end{center}
\caption{The argument of the energy conserving delta function as a function of $\tilde p_C$, for a given $\tilde p_B$.  The light intervals indicate the corresponding ranges of solutions to energy conservation for $\tilde p_C$, given a small range of $\tilde p_B$ supported by our wavepacket.  This small range of $\tilde p_C$ enforced by energy conservation restrict our calculations to a simple regime of the momentum-space amplitude, while still localizing the longitudinal separation of A and C using Gaussian wavepackets with good resolution ($1/\sigma\ll \alpha' E$).}
\label{ranges}
\end{figure}

A variant of this which is useful is to take a triangular region of support between $p_{B,\text{min}}$ and $p_{B, \text{max}}$,
\bea\label{PsiTriangle}
\Psi_{\text{tri}}(\tilde p_B) &=& \sqrt{\frac{3}{\Delta p_B}}\left(\frac{2}{\Delta p_B}(\tilde p_B-p_{B, 0})+1\right) ~~~~ p_{B,\text{min}}<\tilde p_B<p_{B,0}\nonumber \\
\Psi_{\text{tri}}(\tilde p_B) &=& \sqrt{\frac{3}{\Delta p_B}}\left(-\frac{2}{\Delta p_B}(\tilde p_B-p_{B, 0})+1\right) ~~~~ p_{B,0}<\tilde p_B<p_{B,\text{max}} 
\eea
where we defined
\be\label{pBtot}
p_{B,0}=\frac{p_{B,\text{min}}+p_{B, \text{max}}}{2}
\ee
and we work at $X_B=0$.  The Fourier transform to position space gives in this case
\be\label{PsiTriangleX}
\tilde \Psi_{\text{tri}}(\tilde X_B)=\frac{8\sqrt{3\Delta p_B}}{\sqrt{2\pi}}e^{-ip_{B0}(\tilde X_B-T)}\frac{\sin^2(\frac{\Delta p_B (\tilde X_B-T)}{4})}{(\Delta p_B (\tilde X_B-T))^2}
\ee
This has a tail $\propto 1/(\tilde X_B-T)^2$, and a finite variance in position and momentum space.  

We will analyze the amplitudes for QFT and string theory convolved with these wavepackets.  It will be useful for interpreting the potential contribution from wavefunction tails to check the sensitivity of these amplitudes to the width $\sigma$ of the Gaussians (\ref{posGauss}) and the difference between the step function and triangular wavefunctions for B.  

As mentioned above, we will work with $p_{B,\text{max}}, p_{B,\text{min}}$ such that the resulting range of $p_C$ values allowed by energy-momentum conservation (\ref{pCpB}) satisfies
\be\label{deltap}
\Delta p_C\ll \tilde p_C
\ee
for all $\tilde p_C$ contributing to the amplitude.  The required range of $p_B$ can be obtained by differentiating the energy conservation condition $\sum_a \tilde \omega_a=\omega_{\text{tot}}$, which gives
\be\label{dpbdpc}
\frac{d\tilde p_C}{d\tilde p_B} = \frac{\frac{\tilde p_A}{\tilde \omega_A}-1}{\frac{\tilde p_C}{\tilde\omega_C}-\frac{\tilde p_A}{\tilde\omega_A}}\sim - \frac{E^3}{q^2(\tilde p_C-\tilde p_A)}
%-\frac{2 E}{\tilde p_C-\tilde p_A}
\ee
The difference $\tilde p_C-\tilde p_A$  is the separation between the two solutions to energy conservation  (\ref{pCpB}).    Within each range of solutions (Figure \ref{ranges}), we have
\be\label{DelpCpB}
\Delta p_C\sim \frac{-E^3}{q^2(p_C - p_A)}\Delta p_B
\ee

%so that in a wide range of $X$ including the spreading radius $X\sim E$, the wavefunction has a power law tail.   
%The reason for this is that our kinematics regime applies as long as $\delta E_C\ll E_C\sim E$.  This easily leaves room for the maximal $\delta E_C$ up to which we integrate (which is $E_{C, max}-E_{C, min}=\Delta p$ here), to satisfy (\ref{deltap}).  
 
 As a result, the integral is only over $\tilde p_C$ in the regime $\tilde p=\tilde p_C-\tilde p_A \simeq -(\tilde E_C-\tilde E_A)$.  That is, for the above kinematics with $E_I\gg |q_I|$, we can satisfy energy-momentum conservation to good approximation by absorbing a deformation $\delta\tilde E_C$ by $\delta \tilde E_A\simeq -\delta\tilde E_C$.     
This momentum variable is conjugate to the separation $\tilde X_C-\tilde X_A$ between the two strings in position space at time $T=0$.  To sum up, we choose $p_{B,\text{min}}, p_{B,\text{max}}$ such that over the full range of support in momentum space, we have the hierarchy
\be\label{pdiff}
|\delta E_C| \simeq |\delta E_A|\ll E_A\simeq E_C.
\ee
This restricted range for $\tilde p_A$ and $\tilde p_C$ is dictated by the {\it amplitude} along with the wavepacket for $B$, not the wavefunctions for $A$ and $C$.    

%As a result, the Fourier transform of the amplitude has a broad distribution in position space even at $T=0$.  

%The corresponding position space wavefunction has a tail away from T=0 -- the past light cones of $\tilde X_A$ and $\tilde X_C$ at T=0.  Can this tail explain a nonzero contribution with $\tilde X_C<\tilde X_A$ at $T=0$ consistently with the tiny range (\ref{pcrange})?  In this setup, $p_B$ is fixed, so $B$'s trajectory can be anywhere in X.   But A is lightlike in the X-T plane, so although C has some transverse motion as $\tilde q_C$ varies,
%there is no overlap of A, B, and C, before T=0.    But there is support for trajectories which do overlap at a very late time $\propto -\tilde X_{AC}E/q_M$ consistently with going through T=0 in the `advanced' order.
%If we even mildly localize B and/or 1, we should suppress such a contribution.  Or maybe we can precisely match this late intersection to a tail in QFT as it is...   

\section{Momentum space amplitude}\label{sec:amplitude}

In this section, we introduce the momentum space string amplitude and describe some of its key features within our kinematic regime of interest.  We will be interested in two open string orderings, and two kinematic regimes. We will compare and contrast string theory with QFT, with the sign of one of the kinematic variables playing an interesting role.  In one of the regimes, the amplitude can be expanded into a convergent sum over QFT propagators $\frac{1}{\kps+{n_{D}}-i\epsilon}$.  In the other regime, the analogous expansion in propagators comes dressed with additional dependence on the detector kinematics $\kps$.  This will enable us to distinguish the behavior of string theory in the latter regime from that of QFT, which scatters on the tail of the wavepacket for $X<0$.

Before entering into the details, let us summarize the result of this section.  
In our regime of interest, the relevant piece of the amplitude will take the form
\be\label{Afourform}
A^{(1)}_4 A^{(2)}_4 B(\alpha'\kps, \eta),
\ee
with $\eta=B\hat B-C\hat C$ non-integer.  (We will discuss the case of integer $\eta$ below as well; it has a different pole structure.)
Here the first two factors are the amplitude for the auxiliary four point process in our setup and the four point amplitude describing the process $CD\to \hat C\hat B$.  The last factor varies strongly with the momentum variable $\tilde p_C$ conjugate to the separation $\tilde X$ between A and C, via its dependence on $\kps$.  When $\eta>0$, this Beta function can be expanded into a convergent sum over $\kps$ propagators, while for $\eta<0$ this is not the case.   

\subsection{Open string orderings and gauge invariance}

We will work with open strings, focusing on particular orderings  $A\hat{A}C\hat{C}\hat{B}B$ and $A\hat{A}B\hat{B}C\hat{C}$.  In each of these cases, strings $B$ and $C$ cannot simply scatter early into strings ${B}+C+\hat C$ and $\hat{C}$ as that would violate gauge invariance.   This is useful because such scattering would introduce an extra ambiguity of interpretation:  even if the scattering is supported for an early central trajectory of $C$ relative to $A$, that could proceed simply via $BC\to \hat C (B+C+\hat C)$ followed by $(B+C+\hat C)A\to \hat B\hat A$.   This is important because the internal $B+C+\hat C$ exchange could generate a spread amplitude in $X=X_C-X_A$,  the separation between strings $A$ and $C$ in our setup.  As mentioned above, this separation is conjugate to $\tilde p_C$, on which  $\kps$ varies strongly.  The kinematic variable  $(k_B+k_C+k_{\hat C})^2$ also varies strongly with $\tilde p_C$.  So spreading in a case with $B+C+\hat C$ exchange would be more difficult to test given this more prosaic interpretation of an early, local four point interaction involving $B$ and $C$.      

Gauge invariance in the orderings we consider do allow an early 5 point interaction $CB\to \hat B \hat C (A+\hat A)$ followed by a 3 point interaction $A(A+\hat A)\to \hat A$.  A local field theory with constant 3 and 5 point vertices joined by an $A\hat A$ propagator would not produce spreading in $X$, as $A\hat A$ does not vary strongly with $\tilde p_C$.   String spreading could introduce an interaction between $\hat A$ and $D$ at large $X$.    

%But this does not produce spreading in $X$, as $A\hat A$ does not vary strongly with $\tilde p_C$.

\subsection{Simplification near a pole}

We would like to work with a complete expression for the amplitude that we convolve with a wavepacket.  At five points, the integral over vertex operator positions yields a closed form expression in terms of Beta functions and a $_3F_2$ hypergeometric function.  At six points this is not the case, but we can obtain a complete expession if we work sufficiently near a pole, factorizing the amplitude into a 3 point function times a five point function.   We will first work this out for bosonic open string tachyon scattering in the relativistic regime $E\gg 1/\sqrt{\alpha'}$, and then describe the generalization to the superstring, for which the external states and the leading poles are massless.         

In the bosonic case, factorizing near the leading pole in $A+\hat{A}$ is a good approximation for
\be\label{KA3small}
1+K_{A\hat{A}}\ll \frac{1}{\log(\alpha' E^2)}
\ee
as can be seen as follows from the generalized OPE analysis in \cite{bpst}.  Sufficiently near the pole, the integral over worldsheet vertex operator positions $y_I$ is dominated by the regime in which the vertex operators for String $A$ and String $\hat{A}$ approach each other, $y_{A\hat{A}}\to 0$.  In this regime, including the leading correction to the OPE gives \cite{bpst}
\be\label{genOPE}
e^{i k_A X(y_A)} e^{i k_{\hat A} X(\hat{y}_A)} \sim e^{i (k_A+ k_{\hat A})X(y_A)+i k_A y_{\hat{A}A} \partial_y X(y_A)}
\ee    
Contracting the correction term in the exponent  with other vertex operators in the amplitude, one finds that it is of order $y_{A\hat{A}} K_{AI}$, where $I$ indexes the other vertex operators.  The leading $K_{AI}$ that contribute are of order $\alpha' E^2$ (for $I=\hat{B}, B$).   
This leads to an integral of the form
\be\label{yA3int}
\int d\log(y_{A\hat{A}}) e^{i y_{A\hat{A}}\alpha' E^2} e^{(1+K_{A\hat{A}})\log(y_{A\hat{A}})}
\ee
%where we have expanded in $K_{A\hat{A}}$ since our goal is to determine when it is valid to neglect the higher terms in this series.
%The exponential in (\ref{yA3int}) oscillates rapidly unless  $y_{A3}<1/E^2\alpha'$.  
The dominant contribution to this integral is at $y_{A\hat{A}}$ of order $1/(\alpha' E^2)$ (beyond which it is suppressed by the oscillatory factor).  Plugging back in and expanding in $K_{A\hat{A}}$ gives a series of the form $\sum c_n (1+K_{A\hat{A}})\log(\alpha' E^2)$.  The condition for this to be a valid expansion is (\ref{KA3small}).

We must respect this condition consistently with the linear combination of momentum eigenstates we take for our wavepacket.
In particular, by varying the first line of (\ref{someKIJs}) with respect to $\delta E_A\simeq -\delta E_C$, we find that since we keep $|\delta E_A|\simeq |\delta E_C|\ll E_A\simeq E_C$ (\ref{pdiff}), the variation of $K_{A\hat{A}}$ is negligible.  

Given this, we can derive a closed form expression for the amplitude \cite{BP, superstringamp}, as we will discuss in the following subsections.  We should emphasize that we will not work right {\it on}\ the pole, but will keep a small but finite distance away from it.  We describe this kinematic regime in detail in Appendix B.  

\subsection{Simplified amplitude from the vertex operator integral}

As we will discuss shortly, near one pole the amplitude is given explicitly in terms of Beta functions and hypergeometric functions \cite{BP, superstringamp}.  In this section we will derive the simplified amplitude 
(\ref{Afourform}) in our kinematic regime via the vertex operator integrals.    
Let work for simplicity in the bosonic theory,  and define
\begin{align}\label{IJdef}
IJ=1+2\alpha' k_I\cdot k_J=1+K_{IJ}
\end{align}
In this section, we will take these kinematic invariants large (aside from $A\hat A\ll 1$), neglecting order one offsets that arise in the exact amplitude.

For the ordering $A\hat{A}B\hat{B}C\hat{C}$, the integral over vertex operator positions can be written as\footnote{See e.g. \cite{JoeBook}\ for the relevant background on tree-level string scattering and the integral expression for the momentum space amplitude.} 
\begin{align}\label{yint}
\frac{1}{A\hat A}\int_{0}^{1}dy_{\hat{C}}\int_{0}^{y_{\hat{C}}}dy_{A\hat{A}}\, y_{A\hat{A}}^{\alpha' k_D^2}\, y_{\hat{C}}^{B\hat{C}}(1-y_{A\hat{A}})^{A\hat{A}C}(1-y_{\hat{C}})^{C\hat{C}}(y_{\hat{C}}-y_{A\hat{A}})^{A\hat{A}\hat{C}}.
\end{align}
where we have set $y_B=0, y_C=1,$ and $y_{\hat B}=\infty$.  

We will estimate the $y_{\hat C}$ integral in two regimes of the $y_{A\hat A}$ integral.  
In our kinematics, $B\hat C\sim 4 E^2\alpha'$, which is very large.  As a result, the factor $y_{\hat{C}}^{B\hat{C}}$ implies that the $y_{\hat C}$ integral is dominated by $y_{\hat{C}}=1$.  In more detail, the saddle point equation for $y_{\hat C}$ is
\be\label{saddlehatC}
\frac{B\hat C}{y_{\hat C}}-\frac{C\hat C}{1-y_{\hat C}}+\frac{A\hat A \hat C}{y_{\hat C}-y_{A\hat A}}=0
\ee
In our kinematics (described in detail in appendix B), the numerator of the third term is parametrically smaller than the second, which is parametrically smaller than the first.   
When $y_{A\hat A}$ is not close to $y_{\hat C}$, the third term is negligible and this equation has a solution 
\be\label{Firstysoln}
y_{\hat C}\simeq \frac{1}{1+C\hat C/B\hat C}
\ee
This reduces the integral to
%For $y_A\ll y_C$,
\begin{align}
\frac{1}{A\hat A}B\hat{C}^{-C\hat{C}}\int_{0}^{1}dy_{A\hat{A}}\, y_{A\hat{A}}^{\alpha' k_D^2}\, (1-y_{A\hat{A}})^{A\hat A C+A\hat A \hat C}.
\end{align}
The final exponent can be simplified,
\begin{align}
C A\hat A + A\hat A \hat C=(k_A+k_{\hat{A}}+k_C+k_{\hat{C}})^2-C\hat{C}-A\hat A=B\hat{B}-C\hat{C}-A\hat A=\eta-A\hat A,
\end{align}
Given $A\hat A\ll 1$, what remains is the integral expression for $B(\kps\alpha', \eta)$ times two four-point amplitudes, one of which is near a pole and one in a Regge regime.  This is the contribution of interest in our application, as it varies strongly with variations in $\delta p_C\sim -\delta p_A$, the variable conjugate to the longitudinal separation of A and C.  We will comment on the contour integral for $B(\kps\alpha', \eta)$ further below in \S\ref{contour}\ in a more specific kinematic regime of interest; its dominant contribution indeed comes from $y_{A\hat A}$ far from $1\sim y_{\hat C}$.     

The remaining contribution comes from the region
$x=y_{A\hat{A}}/y_{\hat{C}}\sim 1$ in the integral.  This leads to weak dependence on $\delta p_C\sim -\delta p_A$, since the factor depending on $\kps\alpha'$ in (\ref{yint}) is now close to 1.  This extra term is given explicitly below in (\ref{twoterm})-(\ref{twotermapprox}).  

%This gives, to good approximation,
%\begin{align*}
%\int_{0}^{1}dy_{\hat{C}}\int_{0}^{1}dx\,  y_{\hat{C}}^{B\hat{C}}(1-xy_{\hat{C}})^{A\hat{A}C}(1-y_{\hat{C}})^{C\hat{C}}(1-x)^{A\hat{A}\hat{C}}=B\hat{C}^{-B\hat{B}}B(A\hat{A}\hat{C},-\eta).
%\end{align*}

\subsection{The closed form string theory amplitude for two orderings}

Let us now enter into further details of the momentum space string amplitude.  We will start with the bosonic theory.  
At the end of this section, we briefly describe the small shifts (removing the tachyon pole) arising in the superstring case.  Finally,  we will comment on the size of the amplitude in different versions of our basic process depending on the identification of the auxiliary process.

\subsubsection{Ordering $A\hat{A}C\hat{C}\hat{B}B$}\label{formone}

One useful form of this amplitude is
\be\label{A3C21Bfull}
A_{\text{ST}}=\frac{1}{A\hat{A}} B(\alpha' k_D^2-1, B\hat{B})B(C\hat{C}, \hat{B}\hat{C})_3F_2(1-B\hat{C},\alpha' k_D^2-1, C\hat{C}; \alpha' k_D^2-1+B\hat{B}, \hat{B}\hat{C}+C\hat{C}; 1)
\ee
with the standard $i\epsilon$ prescription, where we have used (\ref{IJdef}).
This is equal to the following sum
\be\label{A3C21B}
\frac{1}{A\hat{A}}\sum_{n=0}^{\infty} \frac{(1-B\hat{C})_n}{n!}B(\alpha' k_D^2-1+n, B\hat{B})B(C\hat{C}+n, \hat{B}\hat{C})
\ee
We can collect the factors in the $n$th term in the sum as
\be\label{summandfactors}
\frac{\Gamma(\hat{B}\hat{C})}{\Gamma(1-B\hat{C})}\frac{\Gamma(1-B\hat{C}+n)}{\Gamma(\hat{B}\hat{C}+n+C\hat{C})}\frac{\Gamma(C\hat{C}+n)}{n!}\frac{\Gamma(\alpha'k_D^2-1+n)\Gamma(B\hat{B})}{\Gamma(\alpha'k_D^2-1+n+B\hat{B})}
\ee
%The first factor is order 1 in our kinematics.  The next factor depends on $n$ very weakly; we could in fact choose $12+C\hat{C}=-B2$ (up to tiny deformations of $C\hat{C}$ in our integral over $\tilde E_C$).  
%There is then no net $(-1)^n$ and these first two factors are order 1.  If we take our momentum range over positive $k_D^2$, the remaining factors are all positive.  The last factor
%decays faster than $(k_D^2+n)^{-B\hat{B}}$.
The sum here converges as long as
\be\label{oneconverge}
B\hat{B}\hat{C}\equiv B\hat{B}+B\hat{C}+\hat{B}\hat{C}>1. 
\ee

For comparison with QFT, it is interesting to expand this in terms of  $\kps$ propagators by expanding $B(\alpha' k_D^2-1+n, B\hat{B})$ in terms of its poles.  We work in the regime where the sum over $n$ converges as in (\ref{oneconverge}). 
This gives
\be\label{propagatorsum}
\frac{\Gamma(\hat{B}\hat{C})}{\Gamma(1-B\hat{C})}\sum_m \frac{(1-B\hat{B})_m}{m!}\sum_n \frac{\Gamma(1-B\hat{C}+n)}{\Gamma(\hat{B}\hat{C}+n+C\hat{C})}\frac{\Gamma(C\hat{C}+n)}{n!} \frac{1}{k_{D,0}^2+(n+m)/\alpha'-i\epsilon}
\ee
If we perform the sum over $m$ first, it converges (giving back the original $B(\alpha' k_D^2-1+n, B\hat{B})$).   
However, the sum over $n$ does not generally converge if performed first; that requires $B\hat{C}+\hat{B}\hat{C}$ sufficiently large.

For special choices of kinematics, this amplitude simplifies further.  One useful example is to apply Saalschutz's theorem
\be\label{Saals}
_3F_2(a, b, -N; d, 1+a+b-d-N; 1)=\frac{(d-a)_N(d-b)_N}{(d)_N (d-a-b)_N}
\ee 
with
\be\label{ids}
N=B\hat{C}-1, ~~~ B\hat{B}\hat{C}=2, ~~~a=\alpha'\kps-1, ~~~b=C\hat{C}, ~~~ d=\alpha'\kps-1+B\hat{B} 
\ee
(with $N$ a positive integer).    

We can write this simplified amplitude in two useful forms.  First define
\be\label{etadef}
\eta \equiv B\hat{B}-C\hat{C}
\ee
The sign of this variable will play an interesting role in the dynamics.
In terms of this we can us (\ref{Saals}) to write the amplitude as
\be\label{FirstAprops}
\frac{B(C\hat{C}, \hat{B}\hat{C})}{A\hat{A}}B(\alpha'\kps-1, B\hat{B})\frac{\Gamma(\alpha'\kps-1+B\hat{B})\Gamma(\eta)}{\Gamma(B\hat{B})\Gamma(\alpha'\kps+\eta)}\times\frac{\Gamma(\alpha'\kps-1+\eta+N)\Gamma(B\hat{B}+N)}{\Gamma(\alpha'\kps-1+B\hat{B}+N)\Gamma(\eta+N)}
\ee
This further simplifies to
\be\label{fullsimp}
\frac{B(C\hat{C}, \hat{B}\hat{C})}{A\hat{A}}B(\alpha'\kps-1, \eta)\times\frac{\Gamma(\alpha'\kps-1+\eta+N)\Gamma(B\hat{B}+N)}{\Gamma(\alpha'\kps-1+B\hat{B}+N)\Gamma(\eta+N)}
\ee
Within the kinematic regime of our setup, $N\sim B\hat{C}\sim E^2\alpha'$ is much greater than the other variables in the arguments of the $\Gamma$ functions, so the last factor in (\ref{FirstAprops}-\ref{fullsimp}) is approximately 1.  

In this regime, the amplitude is approximately given by
\be\label{sumovern}
\frac{B(C\hat{C}, \hat{B}\hat{C})}{A\hat{A}} B(\alpha' k_D^2-1, \eta)
\ee
For $\eta>0$, the last Beta function has a convergent expansion in terms of massive $D$ propagators.    
 For $\eta<0$ it does not, but the function $B(\alpha'\kps-1, B\hat{B})$ in the equivalent form (\ref{FirstAprops}) can be expanded in propagators, dressed by additional $\kps$ dependence.
  
\subsubsection{Ordering $A\hat{A}B\hat{B}C\hat{C}$}\label{A3B1C2}

For the ordering $A\hat{A}B\hat{B}C\hat{C}$, we obtain a similar form of the amplitude without as many special kinematic choices. It is simplest to write the five-point function as the sum of two terms,
\begin{align}\label{twoterm}
A&=B(C\hat{B},B\hat{B})B(A\hat{C}\hat{A},-\eta){_3F_2}(B\hat{B},A\hat{C}\hat{A},\alpha'k_D^2-1+\eta,1+\eta,C\hat{B}+B\hat{B}; 1)\notag\\
&+B(C\hat{B},C\hat{C})B(\alpha' k_D^2-1,\eta){_3F_2}(C\hat{C},\alpha' k_D^2-1,A\hat{C}\hat{A}-\eta,1-\eta,C\hat{B}+C\hat{C}; 1)
\end{align}
If $\eta$ is not an integer, the hypergeometric functions reduce to 1 and the answer is approximately
\begin{align}\label{twotermapprox}
A=C\hat{B}^{-B\hat{B}}B(A\hat{C}\hat{A},-\eta)+C\hat{B}^{-C\hat{C}}B(\alpha' k_D^2-1,\eta).
\end{align}
The second term is similar to the simple form (\ref{sumovern}) above.  The first term will not vary strongly as we integrate over the momentum variable $\tilde p_C$ conjugate to the separation between $A$ and $C$.  

In this ordering, the case where $\eta$ is a negative integer works out as follows. The hypergeometric function on the first line of (\ref{twoterm}) has poles at negative integers $\eta$, while the poles in the second line of (\ref{twoterm}) come from the beta function out in front. The remaining finite part comes from expanding the $B(C\hat{B},B\hat{B})$ around negative integer $\eta$,
\begin{align}
-\log(C\hat{B})\sum_{n=0}^{\infty}\frac{(-1)^n}{n!}\frac{\Gamma(B\hat{B}+n)\Gamma(A\hat{C}\hat{A}+n)\Gamma(\alpha'k_D^2-1+\eta+n)}{\Gamma(B\hat{B})\Gamma(A\hat{C}\hat{A})\Gamma(\alpha' k_D^2-1+\eta)\Gamma(1+\eta+n)}(C\hat{B})^{-B\hat{B}-n}.
\end{align}
The first surviving term is at $n=-\eta$,
\begin{align}\label{loginteger}
&-\frac{(-1)^\eta}{(-\eta)!}\log(C\hat{B})\frac{\Gamma(C\hat{C})\Gamma(A\hat{C}\hat{A}-\eta)}{\Gamma(B\hat{B})\Gamma(A\hat{C}\hat{A})}\frac{\Gamma(\alpha' k_D^2-1)}{\Gamma(\alpha' k_D^2-1+\eta)}C\hat{B}^{-C\hat{C}}.
\end{align}

\subsubsection{Integer $\eta$}

Let us briefly comment on the special case of integer $\eta$.  
For positive integer $\eta$, the function $B(\alpha'\kps, \eta)$ reduces to a finite sum of propagators, and  
for negative integer $\eta$, the function $B(\alpha'\kps, \eta)$ reduces to a polynomial in $\alpha'\kps$ of order $-\eta$.   However, the full amplitude contains a logarithmic factor (\ref{loginteger}) which also depends on $p_C$.  

The amplitude (\ref{loginteger}) is interesting, but physically distinct from that of our main setup aimed at simulating black hole infallers:  it does not exhibit $\kps$ poles corresponding to  production of the putative detector.  We will focus on non-integer $\eta<0$, although some of the analysis below goes through in either case.  The integer $\eta<0$ case may be useful for separating issues using the absence of $\kps$ poles in that case.

\subsubsection{Superstring}

\indent In order to avoid dealing with tachyon kinematics, let us briefly mention the analog of our amplitude for massless vector open superstrings.
% First consider a slight variant of (\ref{A3C\hat{C}1Bfull}),
%\begin{align}
%\frac{1}{K_{A\hat{A}}}B(K_{12},K_{B\hat{B}})B(K_{AC}+K_{C3},K_{C\hat{C}}-K_{B\hat{B}})+\frac{1}{K_{A\hat{A}}}B(K_{12},K_{C\hat{C}})B(\alpha'k_D^2,K_{B\hat{B}}-K_{C\hat{C}}).
%\end{align}
There, from extra structure in the vertex operators for the massless external states, one obtains a sum of terms with two or three of the kinematic invariants $IJ$  (\ref{IJdef}) shifted by $-1$, which among other things removes the tachyon pole in $\kps$.  The amplitude also contains a kinematic factor \cite{superstringamp}\ which is a function of the external momenta and polarization vectors, and is independent of $\alpha'$. It will not substantially affect the shape of the amplitude in position space.

%An example of such a term is 
%\begin{align}\label{supertwoterms}
%&\frac{K(\epsilon_I,k_I)}{K_{A\hat{A}}}\left(B(K_{12},K_{B\hat{B}}-1)B(K_{AC}+K_{C3},K_{C\hat{C}}-K_{B\hat{B}})\right.\notag\\
%&\hspace{20 mm}\left.+B(K_{12},K_{C\hat{C}}-1)B(k_D^2\alpha',K_{B\hat{B}}-K_{C\hat{C}})\right).
%\end{align}
%The kinematic factor $K(\epsilon_I,k_I)$ is a function of the external momenta and polarization vectors, and is independent of $\alpha'$. It will not affect the shape of the amplitude in position space.
%with residues
%\be\label{jres}
%\ee

\subsubsection{The size of the amplitude}\label{sizesec}

For the application to black hole physics, the $AB\hat{A}$ system is auxiliary, as explained in the introduction. This means that we should strip off the four-point amplitude corresponding to the $AB\hat{A}D$ subprocess in order to estimate the size of the effect in the black hole.  In the form of the amplitude we have analyzed here, this means we strip off the $K_{A\hat A}$ pole,  which leaves us with a Regge-suppressed interaction from the first factor in (\ref{FirstAprops}-\ref{fullsimp}).
% Neglecting the overall kinematic factor $K(\epsilon_I,k_I)$, the second term in (\ref{supertwoterms}) then becomes
%\begin{align}
%B(K_{12},K_{C\hat{C}}-1)B(\alpha' k_D^2,K_{B\hat{B}}-K_{C\hat{C}}).
%\end{align}
%For large $K_{C\hat{C}}\ll K_{12}$, this is suppressed by a large negative power of $E$, due to the usual Regge suppression factor  $K_{12}^{-K_{C\hat{C}}}$.   
(It is less suppressed than in hard scattering, but still soft.)  

To potentially obtain a more dramatic effect, we can instead work in a kinematic regime where the auxiliary process is Regge suppressed.  This can be obtained in two ways.  One would be to work with $K_{C\hat{C}}$ rather than $K_{A\hat{A}}$ near a pole, so that the $AB\to D\hat{A}$ sub-process is Regge suppressed.   

However, we can obtain the equivalent situation simply by time-reversing the above setup, considering $\hat{A},\hat{B}$, and $\hat{C}$ as incoming strings.  Then the auxiliary process is $\hat{B}\hat{C}\to D C$ and the predicted spreading-induced interaction is $\hat{A} D\to BC$.  In both our orderings $A\hat{A}C\hat{C}\hat{B}B$ and $A\hat{A}B\hat{B}C\hat{C}$, gauge invariance again prevents an early interaction between strings $\hat{B}$ and $\hat{A}$ in this time reversed version.  In this version,  the Regge suppressed factor $B(C\hat{C}, \hat{B}\hat{C})\sim (-C\hat{C}/\hat{B}\hat{C})^{C\hat{C}}$ is auxiliary.  

It is important to note that although $K_{A\hat{A}}\ll 1$, the amplitude is not strictly on the $A\hat{A}$ pole.  In the appendix we elaborate on the relevant aspects of our kinematic regime.  Specifically, eqn (\ref{B1C2}) shows that $\eta\ne 0$ requires $K_{A\hat{A}}\sim \delta q^2\alpha'\ne 0$.    

%$K_{C\hat{C}}$ is near the massless pole. Proceeding in the same way as above, the analog of the second term in (\ref{supertwoterms}) is
%\begin{align}
%\frac{1}{K_{C\hat{C}}}B(K_{13},K_{A\hat{A}}-1)B(\alpha'k_D^2,K_{B\hat{B}}-K_{A\hat{A}}).
%\end{align}
%Here $B(K_{12},K_{A\hat{A}}-1)$ is the four-point amplitude for the AB31$'$ subprocess. Stripping this factor off yields an unsuppressed scattering amplitude for the process of interest, which may have important applications to the black hole situation.

Finally, we note that there may be effects in curved spacetime which affect the range of spreading.  A linear dilaton gradient has an interesting calculable effect discussed in \cite{LDpaper}, and something similar may arise in curved spacetime.  As in \cite{BHpaper}, one may most directly relate flat spacetime results to those in a black hole in the near-horizon Rindler region, putting the strings in their light cone ground state there.  Having made this choice of intermediate state, the geometry may have a nontrivial effect as we evolve the state back through the outside geometry of the black hole.  The effect of the curved geometry and the form of our scattering states evolved back toward the boundary is also important to understand in the context of AdS/CFT.          

\subsection{QFT comparison amplitudes}

It will be useful to compare and contrast this with the tree-level quantum field theories defined above, with the following amplitudes.  First, the six point contact interaction alone gives 
\be\label{AQFT0}
A_{\text{QFT}0}\sim \lambda_6 \delta^{(3)}(\sum \tilde k_I)
\ee
Four point couplings with a massless $D$ particle has the amplitude
\be\label{AQFT1}
A_{\text{QFT}1}\sim \lambda_4^{(1)}\lambda_4^{(2)} \frac{1}{k_D^2-i\epsilon}\delta^{(3)}(\sum \tilde k_I)
\ee
%Finally, we add in additional massive poles, giving
%\be\label{AQFT2}
%{\cal A}_{\text{QFT}2}\sim g_\text{s}^4 \sum_{n=0}^N \frac{1}{k_D^2+n/\alpha'-i\epsilon}\delta^{(3)}(\sum \tilde k_I)
%\ee
%for some finite $N$.  
%We will find that a key difference in behavior between string theory and quantum field theory descends from the softness of string amplitudes as we vary $E_C$ toward larger values of $k_D^2$.   This distinction is manifest in comparing (\ref{transitions}) and (\ref{AQFT0}-\ref{AQFT1}).  
%The softness in momentum space in the string theory case will translate into spreading of probability in position space in contrast to the behavior of these QFT models.  
We could consider more general QFT models with a finite number of poles at various mass scales.

In string theory, the amplitude has richer structure as a function of $\kps$.  As noted above, it is not always expressable as a convergent sum over QFT propagators.  This will lead to the key difference in their behavior.  (Even when there is a convergent sum over propagators, it involves an infinite sum over higher spin intermediate states, leading to softer behavior than any QFT truncation.)

\subsection{Note on pole structure and open string gauge symmetry}

It is worth emphasizing an interesting subtlety in our analysis.   As explained at the beginning of this section, we work with open string orderings precluding the possibility of an early $BC$ collision producing an intermediate $B+C+\hat C$ string plus $\hat C$.  If we had allowed that, then the analogue of our simplified form of the amplitude
 \be\label{Afourformagain}
A^{(1)}_4 A^{(2)}_4 B(\alpha'\kps, \eta),
\ee
would be (e.g. for ordering $A\hat A BC\hat C\hat B$)
\be\label{Afourformagain}
A^{(1)}_4 A^{(2)}_4 B(\alpha'\kps, \alpha'(k_B+k_C+k_{\hat C})^2),
\ee
This has poles in  $B+C+\hat C$ as well as in $\kps$.  Implementing the standard $i\epsilon$ prescription, these sets of poles are on opposite sides of the axis in the complex $p_C$ plane.

In going from that ordering to our orderings of interest, we lose the  $B+C+\hat C$ poles.  The remaining poles in $\kps$ are on one side of the axis within the kinematic regime of momenta in our setup, the same side as the pole in our QFT1 comparison model.   In the QFT model, this pole structure is tied to the purely delayed interactions.   

As discussed above in \S\ref{sec:poles}, this alone does not imply locality. What we will find below is that for $\eta>0$, the amplitude can be written as a convergent sum over particle propagators (a finite sum if $\eta$ is integer) and it leads to results consistent with local scattering, with early support as a function of the peak separation $X=X_C-X_A$ of the position space wavefunction indistinguishable from scattering on the tail of the wavefunction.   In contrast, for $\eta<0$, the amplitude (\ref{Afourformagain}) has a rich structure in momentum space.  It grows like $|\kps|^{-\eta}$ at large $|\kps|$ within a broad window of $p_C$ in our regime.  For a window of small $\kps$, it decomposes into two terms with strongly varying phases that generate strong support for early (and late) interactions at a scale $|X_C-X_A|\sim E\alpha'$.            

\subsection{Combining the orderings}

A complete calculation includes all the orderings of open strings.  These contain different sets of poles, and generically do not cancel out.  For example, the two orderings we evaluated above have a common factor $B(\kps, \eta)$, but a different pole structure in $\hat B\hat C$ in one of the auxiliary factors.     

\section{Results in QFT and String Theory}

In this section, we integrate the wavepackets in Section \ref{sec:wavepacket}\ against the string  (\ref{A3C21Bfull}) and tree level field theory (\ref{AQFT0}-\ref{AQFT1}) amplitudes in Section \ref{sec:amplitude}.   We work in transverse momentum eigenstates, evaluating the amplitude on $\tilde q_A=-\tilde q_C=q, \tilde q_B=0$.    

We will start by focusing on the case with $q= 0$, and then generalize to $q\ne 0$.  To begin with we will analyze wavepackets with support over a particular range of $\tilde p_C$ for which the momentum-space amplitude contains the structure relevant for spreading.  In \S\ref{wide}\ we will generalize to a larger family of wavepackets.               

\subsection{The $q=0$ case}\label{qzero}

The geometry and kinematics is particularly simple if we set $q_A=q_C=0$.  In that case, the central trajectories never meet in position space.  In momentum space the structure is also simplified, as discussed above in (\ref{pBqzero}-\ref{PCrangeqzero}). Energy conservation sets $p_B$ to a specific value at which  $\tilde p_C$ (and $\tilde p_A=P-p_B-\tilde p_C$) can vary widely while still satisfying energy-momentum conservation.  In this section, we will analyze Gaussian wavepackets for $A$ and $C$ that constrain the momentum $\tilde p_C$ within an interesting regime for which the momentum space amplitude (\ref{Afourform}) exhibits a rich structure that will play a key role in our test of string spreading.   

For this case, we simply use Gaussian wavepackets for $A$ and $C$, with width $\sigma$.
After convolving the amplitude with the wavepackets and imposing energy-momentum conservation, the amplitude takes the form
\bea\label{genint}
A(X)_{p_C, \sigma_0}\sim \int \frac{d\tilde p_C}{\sigma_0} e^{-(\tilde p_C-p_C)^2/2\sigma_0^2}e^{-i\tilde p_C X} \hat{{\cal A}}(\tilde p_C) \nonumber\\
\eea
where 
\be\label{sigmaeff}
\frac{1}{\sigma_0^2}=\frac{2}{\sigma^2}
\ee
and $\hat{\cal A}$ is the momentum-space amplitude with the energy-momentum conserving delta function stripped off.  

Let us work with the ordering of \S\ref{A3B1C2}, for which the form of the amplitude takes the very simple form (\ref{twotermapprox}) to good approximation for general $q$, including $q=0$.\footnote{This is not the case for our derivation above of the simplification of the the $A\hat{A}C\hat{C}\hat{B}B$ ordering, whose reduction to the form (\ref{sumovern}) involved taking $B\hat{B}\hat{C}=2$); from (\ref{B12kin}) this requires $q\ne 0$.  }  The second term in in this amplitude varies strongly with $\kps$.  We will also specify  
\be\label{etasign}
\eta<0,
\ee
which leads to a simple structure relevant for our test of longitudinal spreading.  
In the range 
\be\label{kintervalm}
0<\alpha' k_D^2<-\eta
\ee
it is useful to express this as
\begin{align}
&\frac{B(C\hat{C}, \hat{B}\hat{C})}{A\hat{A}}\frac{\sin\pi(\alpha' k_D^2+\eta) \Gamma(\alpha' k_D^2)\Gamma(-\alpha' k_D^2-\eta)}{\Gamma(-\eta)\sin\pi\eta}\notag\\
&=A_4^{(1)}A_4^{(2)}\times \frac{\sin\pi(\alpha' k_D^2+\eta) \Gamma(\alpha' k_D^2)\Gamma(-\alpha' k_D^2-\eta)}{\Gamma(-\eta)\sin\pi\eta}.
\end{align}
This has several important features.  The sinusoidal factor expands into two phases $e^{\pm i\pi (\eta+\kps)}\sim e^{\pm i\pi \eta}e^{\pm i (4 E_{\hat B} \delta p_C+\dots)}$.  These vary strongly with $p_C$:  in each term this phase combines with the $e^{-i p_C X}$ factor to shift the support in $X$ by a large scale $\pm 4\pi E_{\hat B}\alpha'$.   To assess the effect of the other factors, we start by noting that
the function $\Gamma(\alpha'\kps)\Gamma(-\eta-\alpha'\kps)$ has a minimum at $\alpha'\kps=-\eta/2$.  The width of this minimum is of order
\be\label{DelEamp}
\Delta_A \kps \alpha' \sim \sqrt{-\eta}\Rightarrow \Delta_A p_C\sim \frac{\sqrt{-\eta}}{4 E_{\hat B}\alpha'}
\ee
where the subscript $A$ indicates that this is the width of the extremum in the amplitude (as opposed to the wavepacket).  We will use a Gaussian wavepacket which introduces a maximum in the integrand at this point.

In particular, we specify the center and width of Gaussian wavepackets for A and C as follows.   We will use the same prescription in string theory and QFT so that we can directly compare the shape and size of the resulting amplitude $A(X)$ in the QFT1 model and string theory.
For simplicity, let us center our Gaussian wavepackets at $\alpha' k_D^2=-\eta/2$. 
Let us further specify a sufficiently small width for the wavepackets in momentum space that the Gaussian maximum dominates over the amplitude minimum at $\alpha' k_D^2=-\eta/2$, producing a net maximum.
Including all numerical factors, if we define
\be\label{c1def}
\sigma_0\equiv \frac{\sqrt{c_\sigma (-\eta)}}{4 E_{\hat B}\alpha'}, ~~~~ c_\sigma <\frac{1}{4}
\ee
in terms of a numerical factor $c_\sigma$, then the integrand has a net maximum at $\alpha'\kps=-\eta/2$. 

We obtain a stronger condition if we require that this be a global maximum (where for $\kps<0$ we replace the poles by half their residues).   At $\kps=0$ or $-\eta$, the wavepacket suppression is $\sim e^{\eta/8c_\sigma}$ and the amplitude is unsuppressed; this is smaller than the amplitude $2^\eta$ at the maximum of the wavepacket as long as  
\be\label{csigstrong}
c_\sigma < \frac{1}{8\log(2)}
\ee
%\be\label{sigboundqzero}
%\frac{1}{\sigma_0} > \frac{4 E_{\hat B}\alpha'}{\sqrt{-\eta}} \gg 1.
%\ee
Imposing the above conditions, we can still satisfy 
\be\label{sigres}
\frac{1}{\sigma_0}\ll X_*=4\pi E_{\hat B}\alpha'.
\ee
These choices simplify our analysis, while being sufficient to distinguish string theory from QFT.  The narrow width in momentum space was anticipated in our discussion at the beginning of \S\ref{sec:wavepacket}:  it ensures that support for $\kps$ is limited to a window for which longitudinal spreading is expected according to (\ref{refinedprediction}).  

We can now put this together with all amplitude factors and integrate the amplitude against the wavepacket, expanding the $\sin\pi(\alpha'\kps+\eta)$ into its two phase terms.  This gives one term with the central trajectory of C advanced\footnote{Again we emphasize that the interaction itself is consistent with causality:  the interaction between the {\it endpoints} of string $C$ and the detector $D$ may be purely delayed.} (centered at  $X=-X_*=-4\pi E_{\hat B}\alpha'$)
\be\label{resultm}
e^{-\frac{(X+X_*)^2\sigma_0^2}{2(1-4 c_\sigma)}}\frac{B(C\hat{C}, \hat{B}\hat{C})}{A\hat{A}} 2^\eta,
\ee
as well as a similar delayed term centered at $+4\pi E_{\hat B}\alpha'$.

Let us finally compare this to the QFT1 model with propagator $\frac{1}{\kps-i\epsilon}$, which gives approximately
\be\label{QFT1qzero}
e^{-X^2\sigma_0^2/2}\frac{1}{A\hat{A}} \frac{\lambda_{CD\hat C \hat B}}{(-\eta/2)}
\ee
Here there are no phase terms and the Gaussian is centered at $X=0$, consistently with QFT scattering on the tail of the Gaussian wavepacket.  This shape is very different from the string theory case, which as just noted is peaked at $X=\pm 4\pi E_{\hat B}\alpha'$.  

Let us also compare the magnitude of the amplitude of the two theories at
$X\sim \pm X_*= \pm 4 \pi E\alpha'$.   As discussed above in \S\ref{sizesec}, the appropriate comparison involves stripping off the auxiliary process amplitude, which we may take to be $\lambda_{CD\hat C \hat B}\equiv B(C\hat{C}, \hat{B}\hat{C})$.  The remaining QFT amplitude is
\be\label{QFTsuppression}
\frac{1}{A\hat{A}}\frac{1}{(-\eta/2)}\exp(-X_*^2\sigma_0^2/2)\sim \frac{1}{A\hat{A}}\frac{1}{(-\eta/2)}\exp(-16\pi^2 E^2\alpha'^2\sigma_0^2/2)
\ee
This is $ > \frac{1}{-(\eta/2)}e^{\pi^2\eta} $ (the price we paid for forcing the Gaussian width small enough to focus on the simple extremum of the amplitude), but is still exponentially suppressed.
%From (\ref{sigboundqzero}) we can write

In terms of (\ref{c1def}), the ratio of the magnitudes of the string and QFT amplitudes at $X\sim -4\pi E_{\hat B}\alpha'$ is
\be\label{sizeratioqzero}
\frac{A_{\text{QFT}}}{A_{\text{ST}}}\simeq \frac{1}{(-\eta/2)} \left(\frac{2}{e^{\pi^2 c_\sigma/2}}\right)^{-\eta}  
\ee
Altogether,
% noting that $2/e^{\pi^2}$ is of order $10^{-4}$, 
this exhibits a window $\frac{2\log(2)}{\pi^2}<c_\sigma<\frac{1}{8\log(2)}$ in which the QFT result is parametrically suppressed (as we increase $-\eta$) compared to string theory.  
%This restricted range of $c_\sigma$ here is still consistent with the Gaussian wavepacket times the amplitude exhibiting a maximum at $\kps_0=-\eta/2$, toward the lower end of this range of $c_\sigma$ ($\frac{2\log(2)}{\pi^2}<c_\sigma< \frac{1}{4}$).    
Thus the size as well as the shape of the string theory amplitude indicates physics beyond this tree level QFT model, even within the simplified regime of parameters defined above.   Again, we should emphasize that this comparison involves stripping off a Regge-soft four point amplitude in string theory (identifying it with the corresponding four point coupling in the QFT1 model).    

The shape and amplitude of the string theory result suggests that it is not scattering on the tail, but this could be subtle.
One could ask whether it is scattering on the tail of the wavefunction with only delayed interactions with respect to the central positions of C and A, at a momentum with enhanced amplitude compensating for the suppression on the tail of the wavefunction.  The momentum space amplitude, including the residues of the poles for non-integer $\eta$, grows like $|\alpha'\kps|^{-\eta}$ for $|\alpha'\kps|\gg -\eta$.  But the Gaussian wavepacket $\propto e^{-\delta\tilde p_C^2/2\sigma_0^2}$ strongly suppresses the contributions from this region, so they cannot compete with the amplitude given in (\ref{resultm}) at $X=-X_*$. 

One way to assess this is to introduce a varying dilaton background as a tracer of the interactions \cite{LDpaper}.  We can introduce a linear dilaton background $\phi \sim V\cdot X=-V^- X^+$ in the $X^+$ direction (for a wide range of $X^+$) to distinguish purely delayed scattering from advanced scattering (in terms of the central positions of $A$ and $C$).
The string coupling in this region behaves as
\be\label{gs}
g_s(X^+) = g_0 e^{V\cdot X} = g_0 e^{-V^- X^+} = g_O^2
\ee  
where $g_O$ is the open string coupling.  
The advanced scenario depicted in figure \ref{sixfig}\ involves $C+\hat C$ splitting off early, which implies a 
relative factor of $g_O(\Delta X^+_*)$ compared to scattering at the origin, with $\Delta X^+_*=X_*/\sqrt{2}=2\sqrt{2}\pi E_{\hat B}\alpha'$.  We analyze this in \cite{LDpaper}, finding agreement with this prediction.  Moreover, the linear dilaton affects the spreading prediction in a calculable way, degrading it for sufficiently large $V^-p^{+}_D \sim \sqrt{2}V^-E_{\hat{B}}$.  The scattering amplitude in the linear dilaton background confirms this prediction as well.

%          This simply requires collecting all relative factors in the amplitude away from the peak in position and momentum by the excursions
%\be\label{excursion}
%\delta X=4\pi E_{\hat B}\alpha' , ~~~~ \delta p_C\equiv \kappa_C \frac{(-\eta/2)}{4 E_{\hat B}}
%\ee
%These excursions incur Gaussian penalties in position and momentum from the wavepacket  with $\sigma$ given by (\ref{c1def}), along with the enhancement from the momentum space amplitude.  The net relative factor is given by
%\be\label{netfactor}
%\exp\left(\eta[c_\sigma \pi^2+\frac{\kappa_C^2}{8 c_\sigma}-\log|1+\kappa_C|]\right)
%\ee
%The first term in the exponent is the tail suppression in position, the second the suppression in momentum, and the third the enhancement from the amplitude (recall that $\eta<0$ here).  For $c_\sigma$ in our range $\frac{2\log(2)}{\pi^2} \lesssim c_\sigma < 1$ the factor in square brackets is positive, meaning that the net effect of going out on the tail is always parametrically suppressed.   Integrating over such contributions, the relative size of the tail contribution remains parametrically suppressed.  So we can exclude a tail interpretation of this result.  
%Indeed, one finds that the advanced term in the string theory result is reproduced for wavepackets compactly supported over an appropriate range of $\tilde X<0$.  

\subsubsection{Contour integral, light cone evolution, and interaction scales}\label{contour}

In this subsection, we will note an interesting feature of the vertex operator integral generating the factor $B(\alpha'\kps, \eta)$ in our amplitude.  This can be written
\be\label{Betaint}
\int_0^1 dy \,y^{-1+\alpha' k_D^2} (1-y)^{-1+\eta}
\ee
The upper end of this integral is modified in the full amplitude; there are no poles in $\eta$.  

For simplicity, let us work with our amplitude between the poles of the function (\ref{Betaint}), at negative half-integer values of $\eta=(2k+1)/2$ (for integer $k$).  We will also take $\alpha' \kps, -\eta$ and $-\alpha' k_D^2-\eta$ to be large and positive, as is the case near the center of the wavepacket above.  In this regime, the Beta function reduces to 
\be\label{Bsimple}
B(\alpha' \kps, \eta) \simeq \cos(\pi \alpha' k_D^2)B(\alpha'k_{D}^2,-\alpha' k_D^2-\eta)
\ee

The integrand of (\ref{Betaint}) has a maximum at
\be\label{ysaddle}
y_s =e^{\pm i\pi}\left| \frac{\alpha' k_D^2}{\alpha' k_D^2+\eta}\right|.
\ee
The integrand has a cut which we can take to run from $-\infty$ to 0, and a square root branch cut from $y=1$ to $y=\infty$.  (The latter is a square root branch cut because we took $\eta$ half-integral.)
Now integrate from 0 to 1 along the following contour.  We split it into two pieces: the first is $1/2$ times the integral from $y=0$ to $-\infty$ above the first cut, around the upper half plane to $+\infty$, and back to $y=1$ above the second cut.  The second piece is $1/2$ times the reflection of the first across the real line (i.e. going below the cuts).   Since $\alpha' \kps+\eta\ll -1$, the integrand is highly suppressed at infinity. The two contours between $y=1$ and $y=\infty$ cancel each other, since we obtain a factor of $-1$ crossing the square root branch cut.  The contours between $y=0$ and $y=-\infty$ get contributions from the saddles (\ref{ysaddle}).  Altogether we get 
\be\label{Betasimplecase}
\frac{1}{2}(e^{i\pi \alpha' k_D^2}+e^{-i\pi \alpha' k_D^2})|y_s|^{\alpha' k_D^2}(1+|y_s|)^{\eta}
\ee     
times the width of the saddle.  In the regime defined above, this is a good approximation to the amplitude.

It is interesting to express this as follows.  Let us write $y=e^{-i\tau}$ and $X^+=\alpha' p_B^+\tau$, as in light cone gauge quantization (now taking the light cone `time' variable in the $X^+$ rather than the $X^-$ direction).  We can write (\ref{Betaint}) as
\be\label{Xplusform}
\int_\gamma dX^+ \left(1-e^{-i X^+/(\alpha'p_B^+)}\right)^\eta e^{-i X^+\kps/p_B^+},
\ee      
with the contour $\gamma$ for $X^+$ corresponding to the one explained above for $y=e^{-i X^+/(\alpha'p_B^+)}$.  
The saddles are generically at the complex values
\be\label{Xplussaddles}
X_s^+=\pm \alpha'  \pi p_B^++i\alpha' p_B^+\log |y_s|.
\ee
However, if we specialize to $k_D^2\sim -\eta/2$, where we centered the wavepackets above, then $|y_s|\sim 1$, so this is a purely real saddle, corresponding to Lorentzian time evolution. The last factor in the integrand in (\ref{Xplusform}) is then 
\be\label{kDfactor}
e^{-2iX_s^+(p_{C+\hat{C}}^-+\ldots)} =e^{\mp 2\pi i\alpha' p_B^+(p^-_{C+\hat{C}}+\ldots)}
\ee            
where the $\ldots$ denote terms that vary weakly with $E_C$. This could be viewed as evolving string $C+\hat C$ for a time $\Delta X^+= \pm 2 \pi \alpha' p_B^+$, or 
\begin{align*}
\Delta X=\sqrt{2}\Delta X^+=\pm4\pi \alpha' E_B=\pm X_*.
\end{align*}
  This interpretation fits quantitatively with the possibility that the effect derived above is not on the tail of the wavepackets, but occurs at the peak value $\pm X_*$.  Although it is not possible to read off real-time physics from S matrix amplitudes, it may be possible to extract gauge-invariant information from time differences such as the one just noted, perhaps in combination with a background field such as the linear dilaton described above.  
In the latter analysis \cite{LDpaper}, a single power of the open string coupling $g_O$ evaluated at the peak (early) trajectory of $C$.  This is consistent with an early 3 point interaction in which $C\to C+\hat C$.  That lines up with the evolution along $X^+$ just described.                

\subsubsection{More specific comparison to light cone prediction}

To finish this analysis, let us briefly comment on the comparison of this result to the refined light cone predictions reviewed above in \S\ref{detectorprediction}.  The result (\ref{resultm}) fits with (\ref{refinedprediction}) given  a distribution of the form
\be\label{lightconecomparison}
\exp\left(-\text{const}\times \frac{\Delta X^+}{\Delta X^+_{\text{spreading}}}\right)\sim \exp\left(-\text{const}\times  \frac{E_{\hat B}\alpha'}{E_{\hat B}/k_{D0}^+k_{D0}^-}\right)\sim 2^\eta
\ee
where we used that the leading contribution to (\ref{resultm}) came from $\kps_0=-\eta/2$, and that as explained in our appendix $B$, the detector has negligible transverse motion: $k_{D\perp} =q_D$ is small, of order $\sqrt{K_{A\hat A}}\sim \delta q$.  Such a distribution linear in the exponent for the longitudinal direction is suggested by the linear dependence of the worldsheet action $S\sim \frac{1}{\alpha'}\int \partial X^+\partial X^-$ on $X^+$, in contrast to the Gaussian distribution for transverse spreading arising from $S\sim \frac{1}{\alpha'}\int (\partial X_\perp)^2$. See \cite{BHpaper, Smatrixpaper}\ for more discussion of this point.  

More generally, it would be interesting to better understand the role of $\eta$ vis a vis the light cone gauge calculations.   From the point of view of the light cone gauge calculations reviewed in \S\ref{detectorprediction}, the spreading estimate (\ref{refinedprediction}) is given in terms of the momentum of the detector.   This is based on a simple estimate of the resolution required, refined by the $1/k_D^+k_D^-$ factor as discussed in \cite{BHpaper}.   But the state of the detector $D$ has more parameters (in principle an infinite number).   As we have seen, the structure of the amplitude is strongly dependent on the sign of $\eta$:  in particular, this determines whether it can be written as a convergent sum of propagators $1/(\kps+n-i\epsilon)$.  It is possible that additonal detector parameters (such as $\eta$) enter into the efficiency of the detection of string spreading. 
This would be interesting to explore further.

%In the next section, we will consider more general Gaussian wavepackets with support over a wider range of momenta, hence better localized in position.   

\subsection{Generalization to $q\ne 0$}

Next, we will analyze the case with $\tilde q_A=-\tilde q_C=q, \tilde q_B=0$.  The geometry of the $A$ and $C$ trajectories is more involved here, in that they meet at a finite (but very late) time.  For this analysis, we will localize string $B$ using a wavepacket (\ref{wavepacketPsharp}) that sharply restricts the range of $\tilde p_B$, leading via energy-momentum conservation to a sharp restriction on the range of $\tilde p_C$.  This entails a broad position-space tail for $B$, which we study in detail below.   For the $A$ and $C$ wavepackets, we may use Gaussians of width $\sigma$, for example taking $\sigma$ of order string scale.

\subsubsection{Momentum basis}

Performing the integral over $\tilde p_A$ absorbs the longitudinal momentum conserving delta function.  For the square wavepacket for string B (\ref{wavepacketPsharp}), this gives 
\bea\label{genint}
A(X)_{p_A, p_{B,\text{min}}, p_{B,\text{max}}, p_C, q, \sigma}\equiv \int d\tilde p_C \int_{p_{B,\text{min}}}^{p_{B,\text{max}}} & \frac{d\tilde p_B}{\sqrt{\Delta p_B}} &  e^{-(P-\tilde p_B-\tilde p_C-p_A)^2/2\sigma^2}e^{-(\tilde p_C-p_C)^2/2\sigma^2}e^{-i\tilde p_C X}\nonumber \\
& & \times ~~ \delta(\sum\tilde\omega_a-\omega_{\text{tot}}) ~ \hat{A}(\tilde p_C) 
\eea
where again $X$ is the peak value of the separation between $A$ and $C$, $p_A$ is the peak value of $A$'s momentum, and so on; $\hat{A}$ denotes the amplitude with the energy-momentum conserving delta function stripped off.  Below, we will also present results for the case of the triangular wavepacket for string B (\ref{PsiTriangle}).     
In all of our cases of interest, $\hat A$ varies significantly only in the $\tilde p_C$ direction in the range of momenta supported by the wavepackets and energy-momentum conservation.  

%They are in Gaussian wavepackets, with tails suppressed by $\exp(-X^2\sigma^2/2)$.  

Next we perform the integral over $\tilde p_B$, absorbing the energy-conserving delta function (\ref{enmom}).  Denoting these solutions $\tilde p_{B*}(\tilde p_C)$, and suppressing the parameters aside from $X$, we are left with
\be\label{psiA}
A(X)=\int \frac{d  \tilde p_C}{|\partial_{\tilde p_B} f[\tilde p_{B*}(\tilde p_C)]|\sqrt{\Delta p_B}} \theta(\tilde p_{B*}(\tilde p_C)-p_{B,\text{min}})\theta(p_{B,\text{max}}-\tilde p_{B*}(\tilde p_C))  e^{-(\tilde p_C-p_C)^2/\sigma^2}e^{-i\tilde p_C X}
\hat { A}(\tilde p_C)
\ee
where $f$ was defined in (\ref{fdef}).  
Given $p_{B,\text{min}}$ and $p_{B,\text{max}}$, we have two ranges of support for $\tilde p_C$ as depicted in Figure \ref{ranges}; we can use the former to tune the latter, keeping all contributions $|\tilde p_C-p_C|\ll |p_C|$. 

The derivative $f'$ varies weakly with $\tilde p_C$:
\be\label{fderiv}
\partial_{\tilde p_B} f[\tilde p_{B*}(\tilde p_C)]=-\frac{\tilde p_A}{\tilde \omega_A}+\frac{\tilde p_B}{\tilde \omega_B}|_{p_{B*}}\simeq 2
%\frac{1}{|\partial_{\tilde p_B} f(\tilde p_{B*}(\tilde p_C)|}=
\ee
where in the last step we used that the integral is restricted to the regime we have discussed where $-\tilde p_A\sim \tilde p_B\sim \tilde \omega_I\sim E$. 
We are left with a relatively simple integral over a pair of tunable, finite ranges of $\tilde p_C$.   We will evaluate this numerically.  For simplicity we consider a single range of $\tilde p_C$, since each behaves similarly, and we treat $f'$ as a constant (\ref{fderiv}).

Let us work in a range of $\tilde p_C$ which does not overlap with the $D$ poles, but starts close to the massless pole and extends further into the regime of spacelike $k_{D}$.  
This gives very distinct results in string theory as compared to all the QFT comparison models, as we will see.    

Before moving to that, we will further map out the geometry of the process, setting up an equivalent calculation of $A(X)$ in a position basis.  This will be very useful in assessing the contribution from the tail of the B wavepacket.  

\subsubsection{Position basis}

\begin{figure}[htbp]
\begin{center}
\begin{tikzpicture}[scale=3]
\draw (-1,-.8) -- (1,-.8);
\draw (-1,.05) -- (1,.05);
\draw (1.3,-.8) node {$T=0$};
\draw (1.4,.05) node {$T=T_{\text{meet}}$};
\draw[color=red] (.2,-1.1) -- (-.5,1);
\draw[color=blue] (.9,-1.1) -- (-.9,.8);
\draw[color=red] (.25,-1.15) node {$C$};
\draw[color=blue] (.95,-1.15) node {$A$};
\draw[color=red] (-.55,1.05) node {$C$};
\draw[color=blue] (-.95,.85) node {$A$};
\draw[|->] (.1,.15) -- (.1,1); 
\draw (.3,.5) node {$\text{delays}$};
 \end{tikzpicture}
\end{center}
\caption{At the very late time $T_{\text{meet}}\sim E^3 X/(q^2(p_C-p_A)$ (\ref{Tmeet}), the central trajectories of strings A and C meet.   After this time, the interaction is purely delayed (not requiring any nonlocality). This region is accessible to the strings out on a highly suppressed power law tail of the wavefunction $\Psi_B$, as discussed in the text.  String B propagates to the right in this picture, and the B wavefunction is peaked such that B meets A at $T=0$.}
\label{Tmeetfig}
\end{figure}
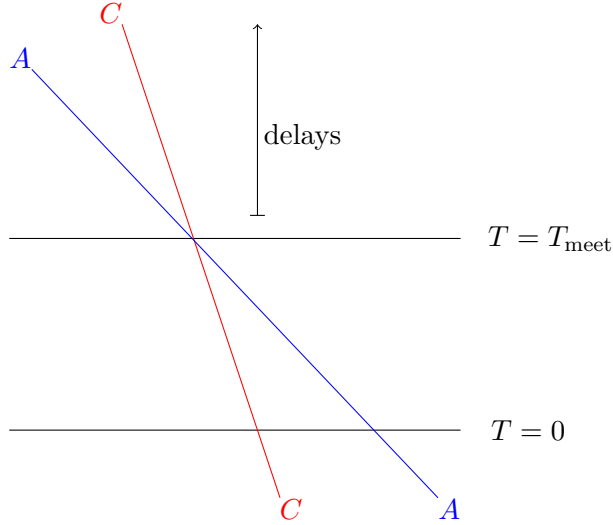

As depicted in Figure \ref{Tmeetfig}, for $p_C\ne p_A$ the trajectories of A and C meet at a very late, but finite, timescale:
\be\label{Tmeet}
T_{\text{meet}}=\frac{2}{q^2}X \frac{p_A^2 p_C^2}{p_C^2-p_A^2}
\ee
In our integral, the values of $p_A$ and $p_C$ that are supported are such that this quantity is of order $E^3 X/(q^2 (p_A-p_C))$, given $1/\sqrt{\alpha'}\sim |\Delta p_C|\ll |p_C|\sim E$.   If $q=0$, the trajectories never meet as we discussed above in subsection \S\ref{qzero}.

This point of direct intersection of the trajectories, and the subsequent region where C is delayed relative to A, is accessible on the tail of 
the position space $\Psi_B$ wavefunction (\ref{wavepacketXsharp}) or (\ref{PsiTriangle}), which is a power law.   Since the wavepackets localizing A and C are Gaussian with width $1/\sigma\sim\sqrt{\alpha'}$, if we assume local interactions then the tail of $\Psi_B$ takes over at a value of $X$ satisfying
\be\label{Xcrossover}
e^{-X^2\sigma^2/2}\sim \frac{1}{{\Delta p_B} T_{\text{meet}}}= \frac{1}{X}\frac{q^2 |p_A^2-p_C^2|}{ p_A^2 p_C^2\sqrt{2\pi}\Delta p_B}
\ee
for the square wavepacket; they meet further out in $X$ for the triangular wavepacket.

%Plugging in the numbers used in the above results, this fits with the crossover scale in the QFT plots of $A(X)$. 
%Any interaction that is local well within the scale $\alpha' E$ tracks the Gaussian until this crossover scale, at which point it takes advantage of the power law tail from $\Psi_B$. 
%The result for string theory does not track the Gaussian tail, and only begins to decrease significantly outside the longitudinal spreading scale $\sim E/(k_D^2+(1/\alpha'))$.  This is consistent with it scattering at the peak of the Gaussian, via longitudinal spreading, out to this scale, outside of which it then scatters on the tail of $\Psi_B$.   

It will be useful to compute $A(X)$ (\ref{genint}) also in a position basis.  Specifically, we can write it as
\be\label{AXX}
A(X)=\int d\tilde X_B \tilde\Psi_B(\tilde X_B) g(\tilde X_B; X)
\ee
where 
\be\label{gdef}
g(\tilde X_B; X) \simeq \int d\tilde p_C ~ \psi_{C, X}(\tilde p_C) ~ e^{-i \tilde p_{B*}(\tilde p_C)\tilde X_B} \hat{A}(\tilde p_{B*}(\tilde p_C), \tilde p_C)
\ee
where as above, $\tilde p_{B*}(\tilde p_C)$ denotes a solution to the energy conserving delta function, and $\hat A$ is the momentum-space amplitude with the energy-momentum conserving delta functions stripped off.  
This is given explicitly by
\be\label{pBstar}
\tilde p_{B*}=\frac{1}{2}\frac{q^2+(P-\tilde p_C)^2-(\tilde\omega_C-\omega_{\text{tot}})^2}{\tilde\omega_C-\omega_{\text{tot}}+P-\tilde p_C}
\ee
(as long as $\tilde p_B\ge 0$).  In this form, we can determine the range of $\tilde X_B$ that contributes.

The integral (\ref{AXX}) is explicitly
\be\label{AXXfull}
A(X)=\int d\tilde X_B \frac{\sin(\Delta p_B\tilde X_B/2)}{\sqrt{\Delta p_B}\tilde X_B} \int d\tilde p_C \, e^{-i X\tilde p_C}e^{-i\tilde X_B(\tilde p_{B*}(\tilde p_C)-p_{B0})} e^{-\delta\tilde p_C^2/(2\sigma^2)}\hat A(\tilde p_C) 
\ee
say for our first version (\ref{wavepacketXsharp}) of the B wavepacket (a square step in momentum space).

\subsubsection{QFT model results}

\begin{figure}[htbp]
\begin{center}
\includegraphics[width=4.5cm]{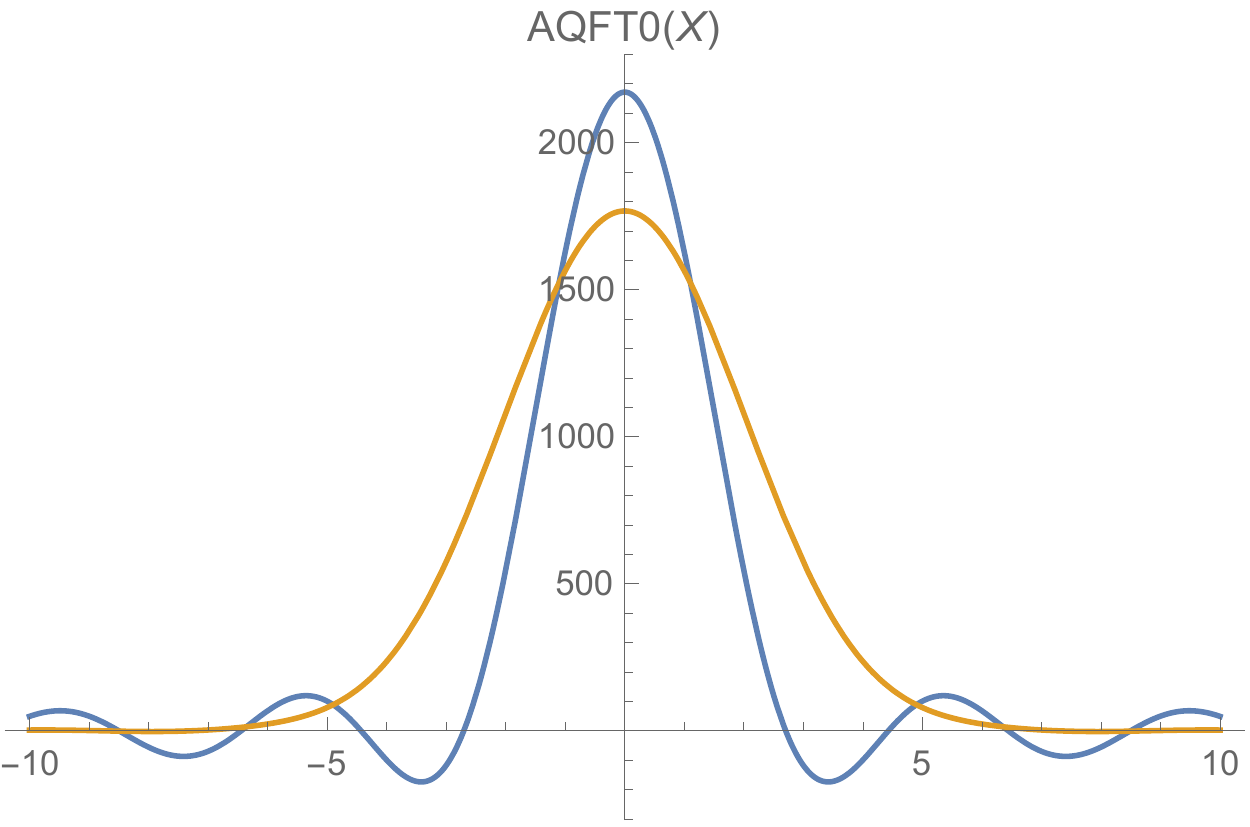}\includegraphics[width=4.5cm]{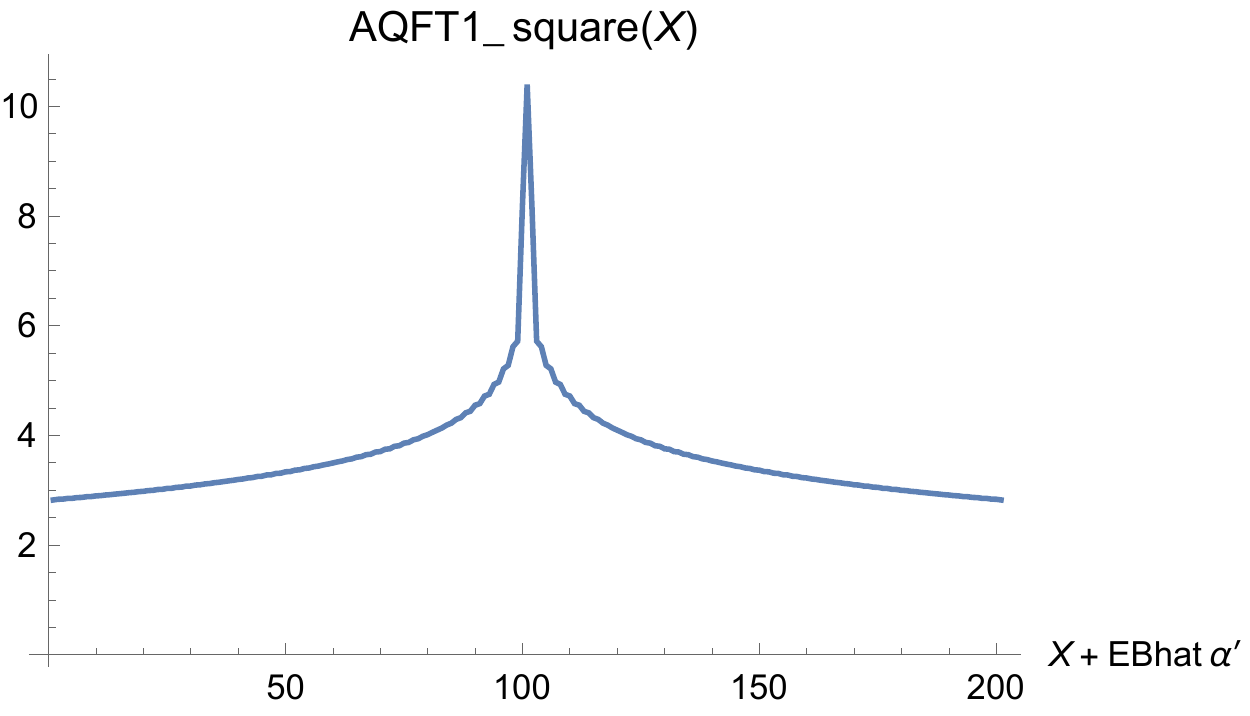}\includegraphics[width=4.5cm]{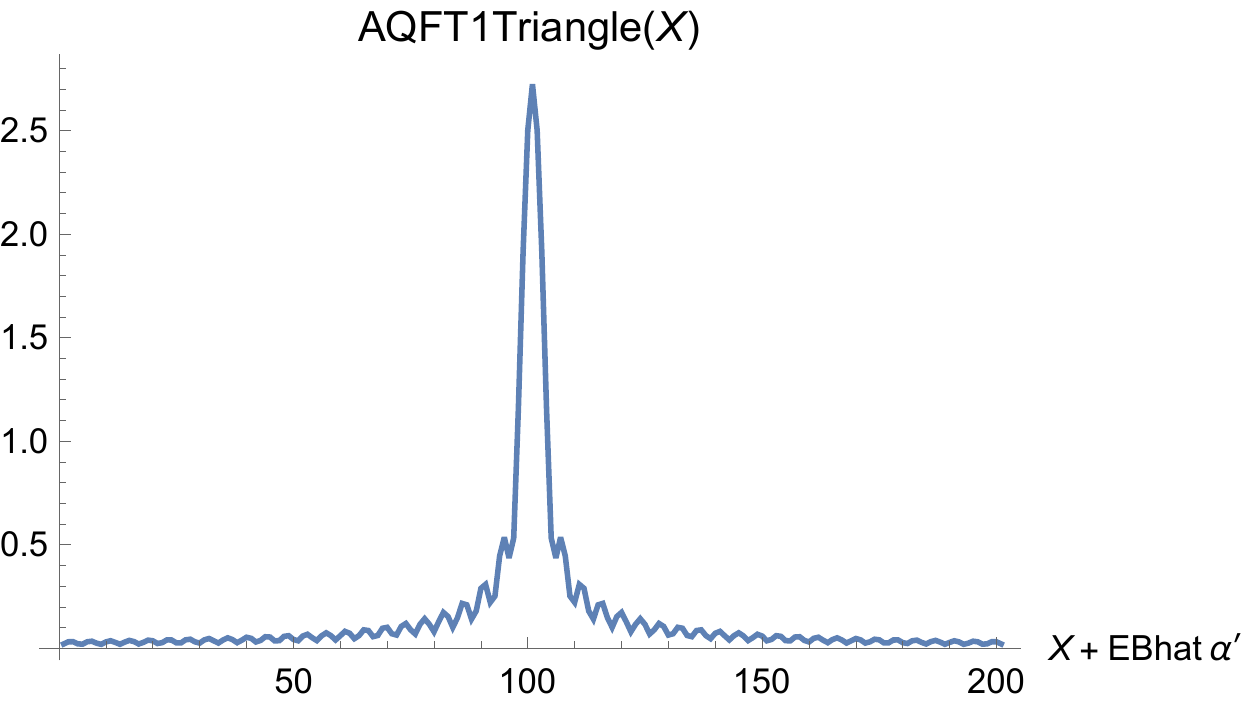}
\end{center}
\caption{The shape of the distribution $A(X)$ for the two QFT comparison models.  The first two plots, B is in the square wave packet for B described in the text.  First, we plot the six point contact interaction (our $QFT_0$ comparison model), plotted for $\sigma=1/\sqrt{\alpha'}$ in blue and $\sigma=1/(2\sqrt{\alpha'})$ in yellow.  Its width is $1/\sigma$, the width of the wavepacket, and does not extend out to $\sim \alpha' E$.  We plot the numerical integral for the other example with a single massless $D$ pole (QFT1) for $\sigma = 1/\sqrt{\alpha'}$.  The parameters are within the regime described above, with $E_{\hat B}$ of order 100.  In the third plot, we replace the step wavepacket by the triangular step $\Psi_{B, \text{tri}}$ described in the main text.  This wavepacket has a more suppressed power law tail in position space. Comparing this to the second plot in Figure (\ref{AQFT}), we see that $A(X)$ changes consistently with the system scattering on this reduced tail once it takes over from the Gaussian tail of the A and C wavefunctions. }
\label{AQFT}
\end{figure}

The QFT0 model is a simple contact interaction, and its amplitude $A(X)$ directly tracks the tail of the wavepacket, as in figure \ref{AQFT}\ below.

Let us next calculate $A(X)$ for the QFT1 model, which will provide a very useful model to compare and contrast with string theory in various regimes.   Let us start by analyzing this explicitly in the regime where the range $\Delta p_C$ is somewhat smaller than the Gaussian width $\sigma$, so the latter can be neglected to good approximation.

Given this, the momentum basis expression for $A(X)$ in the QFT1 model is given to good approximation by the integral (using integration variable $y_C=-\delta\tilde p_C-\Delta p_C/2$)
\bea\label{QFT1simple}
A(X) &\simeq& \frac{e^{-i p_{C0}X}}{\sqrt{\Delta p_B} 4E_{\hat B}\alpha'} \int_0^{\Delta p_C}d y_C\frac{1}{y_C + \frac{{k_D^2}_{\text{min}}}{4 E_{\hat B}\alpha'}}e^{-i X y_C} \nonumber \\
&=& \frac{1}{4 E\alpha'\sqrt{\Delta p_B}}e^{i X (-\frac{{k_D^2}_{\text{min}}}{4E_{\hat B}}-p_{C0})}\left(\Gamma( 0, i X \frac{{k_D^2}_{\text{min}}}{4E_{\hat B}})-\Gamma(0, i X(\frac{{k_D^2}_{\text{min}}}{4E_{\hat B}}+\Delta p_C))\right)
\eea
where ${k_D^2}_{min}$ is the minimal value of $k_D^2$ in the supported range of momenta.  

In figure \ref{AQFT}\ we plot $A(X)$ for the QFT models (including there the dependence on $\sigma$ computed numerically). The essential physics is contained in the analytic result (\ref{QFT1simple}).  It is peaked at the origin and falls off away from there.  This suggests that its support for $X<0$ is accounted for by scattering on the tail of the wavepacket, as expected from causality.

{\bf{$\Psi_B$ tail contributions}}

Indeed, we can explicitly reproduce this formula from the tail part of the $\Psi_B$ wavefunction, using the position basis representation of $A(X)$.      
Starting from (\ref{AXXfull}), we can perform the integral in various orders.  
Let us first Fourier transform the amplitude $\hat A(\tilde p_C)$, giving a step function. 
%For the string theory case, to obtain the Fourier transform we first compute
%\be\label{FTB}
%\int_{-\infty}^\infty d\delta\tilde p_C e^{-i\delta\tilde p_C\tilde X}B(-4 E_{\hat B}[\delta\tilde p_C-\frac{\Delta p_C}{2}]+{k_D^2}_{min}-i\epsilon, \eta)
%\ee
%This gives
%\be\label{simplerB}
%\frac{1}{4 E_{\hat B}} e^{-i \tilde X( \frac{{k_D^2}_{min}}{4 E_{\hat B}}+\frac{\Delta p_C}{2})}\int_{-\infty}^\infty d\kappa e^{-i\kappa\tilde X/4E}B(-\kappa-i\epsilon, \eta)
%\ee
%Summing over the residues of the poles gives 
%\be\label{final}
%\frac{1}{4 E_{\hat B}} e^{-i \tilde X( \frac{{k_D^2}_{min}}{4 E_{\hat B}}+\frac{\Delta p_C}{2})} (1-e^{-i \tilde X/4E})^{\eta-1}\theta(\tilde X)
%\ee
Then inverting the Fourier transform, this leads to the position space expression for $A(X)$ in the QFT1 model:
\be\label{ggenn}
e^{-i p_{C0}X}\int \frac{d\tilde X}{4 E_{\hat B}\alpha'} \theta(\tilde X)f\left(\tilde{X}\right)\int d\tilde X_B \frac{\sin \frac{\Delta p_B\tilde X_B}{2}}{\sqrt{\Delta p_B}\tilde X_B}\int d\delta\tilde p_C\,  e^{-\delta\tilde p_C^2/2\sigma^2} e^{i\delta\tilde p_C(\tilde X-X)}e^{-i(\tilde p_{B*}-p_{B0})\tilde X_B}
\ee
where 
%in the string theory case
%\be\label{hatA}
%f\left(\tilde{X}\right)=e^{i\tilde{X}(\frac{-(\eta-1)}{8 E_{\hat B}\alpha'}-\frac{{k_D^2}_{min}}{4 E_{\hat B}}-\frac{\Delta p_C}{2})}\sin\left(\frac{\tilde X}{8\alpha'  E_{\hat B}}\right)^{\eta-1} 2^{\eta-1} ~~~~~~ {\text{(string theory)}}
%\ee
%In the QFT1 model, 
\be\label{fQFT}
f(\tilde X)=e^{-i\tilde{X}(\frac{{k_D^2}_{\text{min}}}{4 E_{\hat B}}+\frac{\Delta p_C}{2})}~~~~~~{\text{(QFT1)}}
\ee
%In the following two subsetions, we will study this integral in two regimes:  the tail, with $\tilde X_B<-2 T_{\text{meet}}$, and the central region with $|\tilde X_B| < 1/\Delta p_B$.  

We will first analyze the tail, i.e. the contribution from $\tilde X_B<-2T_{\text{meet}}$. 
For this purpose, it is useful to perform the $\delta\tilde p_C$ integral starting from (\ref{ggenn}).  First, we Taylor expand $\tilde p_{B*}$ to second order
in $\delta\tilde p_C$, finding
\bea\label{pBderivs}
p_B' &\equiv& \frac{d\tilde{p}_B}{d\tilde{p}_C}|_{p_{B*}}=\frac{q^2}{4}\left(\frac{1}{p_A^2}-\frac{1}{p_C^2}\right)\approx \frac{q^2(p_C-p_A)}{2p_C^3}=\frac{q^2(p_A-p_C)}{2E^3}=\frac{X}{2T_{\text{meet}}} \nonumber \\
p_B'' &\equiv& \frac{d^2\tilde{p}_B}{d^2\tilde{p}_C}|_{p_{B*}}=\frac{q^2}{p_C^3}\simeq -\frac{q^2}{E^3},
\eea
with higher derivatives negligible in our kinematics.  
At this order, the $\delta \tilde p_C$ integral is now just a Gaussian.  Evaluating that leads to
\be\label{gnext}
\int \frac{d\tilde X}{4 E_{\hat B}\alpha'} \theta(\tilde X)\tilde{\hat A}(\tilde{X})e^{ip_{C0}(\tilde X-X)}\int_{\text{tail}} d\tilde X_B \frac{\sin \frac{\Delta p_B\tilde X_B}{2}}{\sqrt{\Delta p_B}\tilde X_B} \frac{\exp\left(-\frac{\frac{1}{2}(\tilde X-X-p_B'\tilde X_B)^2}{\frac{1}{\sigma^2}+i p_B'' \tilde X_B}\right)}{\sqrt{\frac{1}{\sigma^2}+i p_B'' \tilde X_B}}
\ee
The variance in the denominator of the exponent here is
\be\label{downstairs}
\frac{1}{\sigma^2}+i p_B'' \tilde X_B \simeq \frac{1}{\sigma^2}+i \frac{X}{(p_A-p_C)}\frac{ \tilde X_B}{T_{\text{meet}}}
\ee
Since we are integrating $\tilde X_B$ out on the tail, the second term here is larger in magnitude than $X/(p_A-p_C)$.  This is much larger than $1/\sigma^2$ in our calculation of $A(X)$.    

Next, expand the $\sin(\tilde X_B\Delta p_B/2)$ factor into its two phase terms.  
From the Gaussian factor and one term in this expansion of the $\sin$ factor, along with the $\Delta p_C$ part of the phase in (\ref{fQFT}), we obtain
\be\label{Bstructure}
e^{\frac{i}{2}(\beta \tilde X_B+\frac{(\tilde X-X)^2}{p_B'' \tilde X_B})}e^{-i\frac{\tilde X}{2}(\Delta p_C+2\frac{p_B'}{p_B''})}
\ee
where
\be\label{betadef}
\beta=\pm\Delta p_B +\frac{p_B'^2}{p_B''}
\ee
From (\ref{Bstructure}) we derive resonant contributions to the $\tilde X_B$ integral 
\be\label{XBres}
X_{B*}=\pm \frac{(\tilde X-X)}{\sqrt{\beta p_B''}}
\ee
with width
\be\label{XBwidth}
\Delta\tilde X_{B*}\simeq \frac{\sqrt{p_B''} \tilde X_{B*}^{3/2}}{\tilde X-X}
\ee
Since our tail contribution has an endpoint of integration, we should check if the width is smaller than the distance to this endpoint.   We find that 
the resonance is well within the contour of integration if we integrate over a somewhat larger range of $\tilde X_B$ (still very large in magnitude), say $\tilde X_B <-T_{\text{meet}}$ as opposed to $-2 T_{\text{meet}}$.  This larger integral includes the purely delayed tail.  \footnote{Alternatively, we could work just on the tail $\tilde X_B<-2 T_{\text{meet}}$ and obtain a portion of the width of the resonance for some range of $\tilde X$.}

Note that the $\pm$ in (\ref{betadef}) and (\ref{XBres}) are independent:  we have two resonances, and within each resonance there are two terms from expanding the $\sin$ into phases.  
Plugging back in, the phase becomes
\be\label{XTform}
e^{i\tilde X(\pm\sqrt{\beta/p_B''}-\Delta p_C/2- p_B'/p_B'')}
\ee
Next let us work in the regime $\Delta p_B/2\ll |p_B'^2/p_B''|$, which corresponds to (\ref{hierarchyDelta}) below.  Given that, we can expand the square root in the $\Delta p_B$ term.  Then (\ref{XTform}) becomes
\be\label{approxXT}
e^{i\tilde X(\pm p_B'/p_B'' \pm \Delta p_B/(2 p_B')- \Delta p_C/2- p_B'/p_B'')}
\ee
From this, we note that there is a term for which the signs are such that this exponent will cancel.  
In the resulting integral over $\tilde X$, the remaining phase is $\exp(-i\tilde X{k_D^2}_{\text{min}}/4E_{\hat B})$.  

Plugging in the resonance and width appropriately, and including all parametric factors, the two resonant contributions to the $\tilde X_B$ integral are 
\be\label{res1}
\text{Res}_1=\frac{e^{-i(p_{C0}+\frac{\Delta p_C}{2}+\frac{{k_D^2}_{\text{min}}}{4 E_{\hat B}})X}}{4 E_{\hat B}\alpha'\sqrt{\Delta p_B}}\int_0^\infty \frac{d\tilde X}{\tilde X-X}e^{-i(\tilde X-X)\frac{{k_D^2}_{\text{min}}}{4 E_{\hat B}}}\left( 1-e^{-i\Delta p_C(\tilde X-X)}\right)
%\left(1-e^{-i\frac{\tilde X}{4 E_{\hat B}\alpha'}}\right)^{\eta-1}
\ee
and
\be\label{res2}
\text{Res}_2=\frac{e^{-i(p_{C0}+\frac{\Delta p_C}{2}+\frac{{k_D^2}_{\text{min}}}{4 E_{\hat B}}+2\frac{p_B'}{p_B''})X}}{4 E_{\hat B}\alpha'\sqrt{\Delta p_B}}\int_0^\infty \frac{d\tilde X}{\tilde X-X}e^{-i(\tilde X-X)(\frac{{k_D^2}_{\text{min}}}{4 E_{\hat B}}+2\frac{p_B'}{p_B''})}\left( 1-e^{-i\Delta p_C(\tilde X-X)}\right)
%\left(1-e^{-i\frac{\tilde X}{4 E_{\hat B}\alpha'}}\right)^{\eta-1}
\ee
The first resonance is on the tail $\tilde X_B<-2T_{\text{meet}}$.  
%In the QFT1 model, the resonances are given by the corresponding formulas without the final factor in the integrand.  Let us start with that case.   

Using the integral 
\be\label{betaint}
\int_{x}^\infty \frac{dw}{w} e^{i\beta w}=\Gamma(0, -i\beta x)
\ee
valid for $x>0$, we find that the first resonance reproduces the result for $A(X)$ calculated in momentum space above in (\ref{QFT1simple}), valid in the regime outside of the width of the Gaussian wavepacket.  The second resonance gives a phase times
\be\label{Res2QFT}
\frac{1}{4 E\alpha'\sqrt{\Delta p_B}}\left(\Gamma( 0, i X (\frac{{k_D^2}_{\text{min}}}{4E_{\hat B}}+2p_B'/p_B''))-\Gamma(0, i X(\frac{{k_D^2}_{\text{min}}}{4E_{\hat B}}+2p_B'/p_B''+\Delta p_C))\right)
\ee
The hierarchy we took above, which amounts to 
\be\label{hierarchyDelta}
\Delta p_C\ll \frac{p_B'}{p_B''} \sim |p_A-p_C|
\ee
means that we can expand the second argument in $\Delta p_C$.  As a result, the second resonance gives a subdominant contribution.   
Altogether, this reproduces our momentum basis result for $A(X)$ in the QFT1 model, from a contribution on the B wavepacket tail at large negative $\tilde X_B$.     
 
 \smallskip
 
{\bf QFT1 summary}
             
\smallskip

%$\bullet$ {\bf QFT (Figures \ref{AQFT}-\ref{AQFT1triangle}):}  As a function of the peak $X$ of the wavefunction for the separation of $C$ and $A$, both QFT comparison models are strongly suppressed away from the origin.  The width of the profile $A(X)$ increases as we decrease the width $\sigma$ of the momentum-space wavepackets, consistent with scattering on the tail of this wavefunction out to a value of $X$ where the tail of the B wavefunction takes over.  When we change the B wavefunction from a square step to a triangular step in momentum space, the QFT amplitude responds by matching onto the reduced tail in position space. Altogether, the shape of $A(X)$ in QFT is sensitive to the wavepacket tails, but insensitive to the minimal value of $k_D^2$.   

We have computed $A(X)$ quantitatively in both a momentum and position basis, obtaining results consistent with the expectation that the QFT models are scattering on the tail for $X<0$.   In figure \ref{AQFT}, we plot $A(X)$ for the QFT comparison models.

%\begin{figure}[htbp]
%\begin{center}
%\includegraphics[width=7cm]{QFT1triangle.pdf}
%%\includegraphics[width=5cm]{AQFTone.pdf}
%%\includegraphics[width=5cm]{AQFTtwo.pdf}
%\end{center}
%\caption{The shape of $|A(X)|$ for QFT1, replacing the step wavepacket by the triangular step $\Psi_{B, \text{tri}}$ described in the main text.  This wavepacket has a more suppressed power law tail in position space. Comparing this to the second plot in Figure (\ref{AQFT}), we see that $A(X)$ changes consistently with the system scattering on this reduced tail once it takes over from the Gaussian tail of the A and C wavefunctions.}
%\label{AQFT1triangle}
%\end{figure}
%Having worked out this case in detail, we are now in a position to generalize to string theory in appropriate kinematic regimes.  

\subsubsection{String Theory with $q\ne 0$}

Next, we analyze $A(X)$ in string theory in two interesting kinematic regimes.  In order to clarify the distinction between string theory and QFT, we focus on the expression for the amplitude as a sum over $\kps$ propagators.  

First, we treat the case with $\eta=B\hat{B}-C\hat{C}>0$.  In that case, $A(X)$ is spread out to the extent predicted by light cone calculations.  However, the amplitude (\ref{fullsimp}) has a convergent expansion in terms of QFT propagators, for each of which we may ascribe scattering at early values of $X$ to the tail of the wavefunction. 
 
Next, we treat $\eta<0$.  In that case, we find stronger spreading of $A(X)$, and there is no such convergent expansion in terms of $\kps$ propagators.  However, we can express it in terms of propagators dressed with additional $\kps$ dependence, explicitly showing how the analogous tail contribution in position space fails to account for $A(X)$ (as it did for the QFT propagators).  

%   In (\ref{niceform}), this includes additional $\kps$ dependence accompanying each propagator, which will yield distinct behavior in string theory.  (The other form of the amplitude discussed above in (\ref{propagatorsum}) leads to a sum over propagators without additional structure, but that sum converges in a different kinematic regime.)

%\subsubsection{$\eta<0$ advanced interaction}

\smallskip

{\bf {$\eta>0$ and a cautionary tail}}\label{caution}

\smallskip

Let us briefly compare this result with the amplitude in a different regime, including a subspace with a convergent sum over QFT propagators.   For this, we work with (\ref{fullsimp}), and in the regime $\eta>0$.

\begin{figure}[htbp]
\begin{center}
\includegraphics[width=7cm]{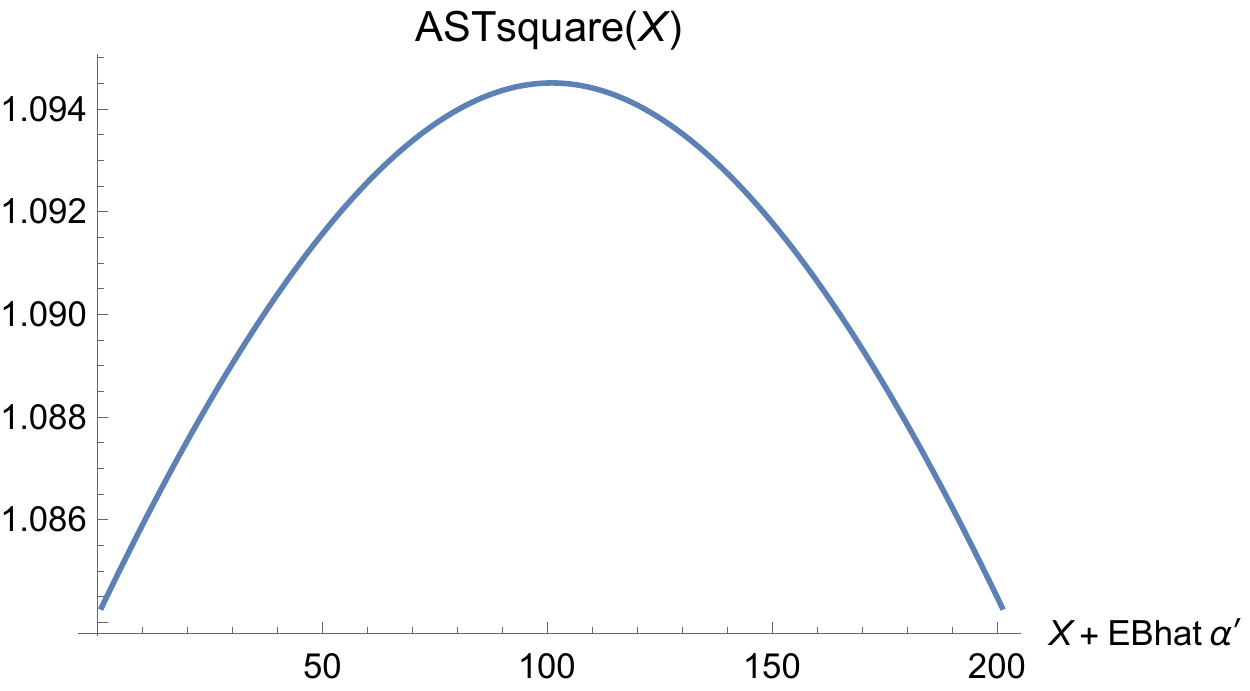} \includegraphics[width=7cm]{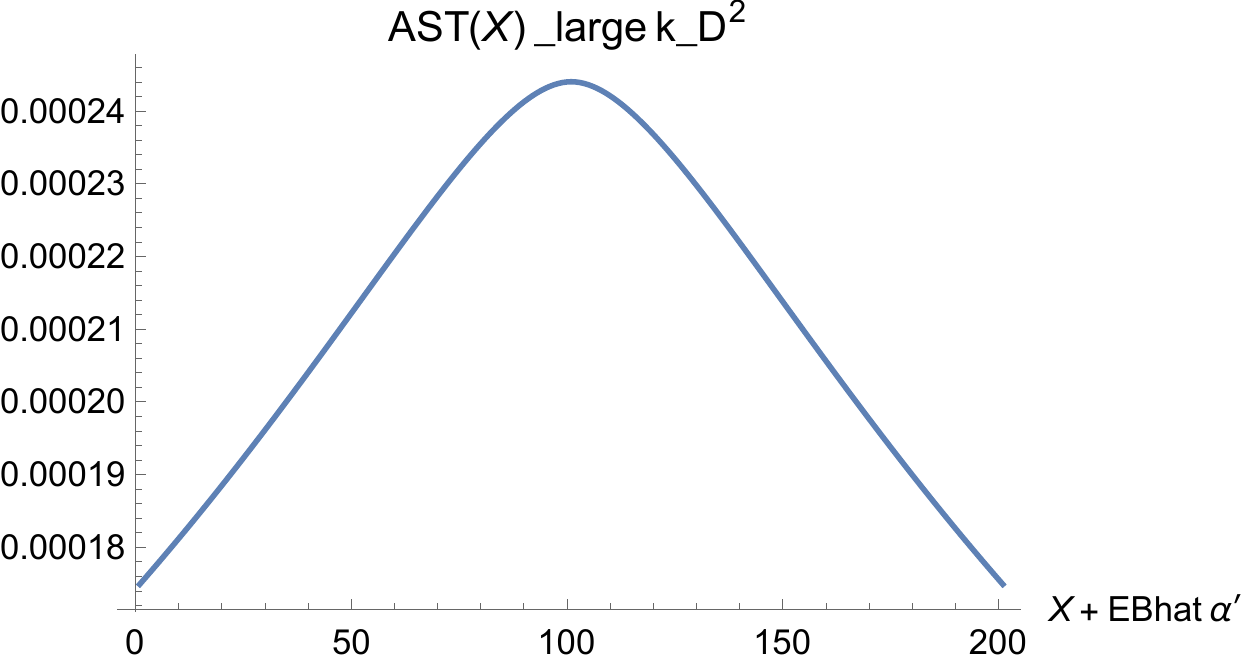}
\end{center}
\caption{The shape of the distribution $|A(X)|$ for tree-level string theory  in the regime of \S\ref{formone}\ and \S\ref{caution}, plotted for $\sigma=1/\sqrt{\alpha'}$ and two values of the minimal $k_D^2$ ($1/(10\alpha')$ and $10/\alpha'$).
%and for $\sigma=1/2\sqrt{\alpha'}$.  
Its width does not increase as we decrease $\sigma$, in contrast to QFT,  but it does depend on $E/k_D^2$ as expected from (\ref{refinedprediction}).  Nonetheless, as described in the text, there is a subset of kinematic parameter space in which this shape arises from a convergent sum of of QFT propagators.}
\label{AST}
\end{figure}
\begin{figure}[htbp]
\begin{center}
\includegraphics[width=6cm]{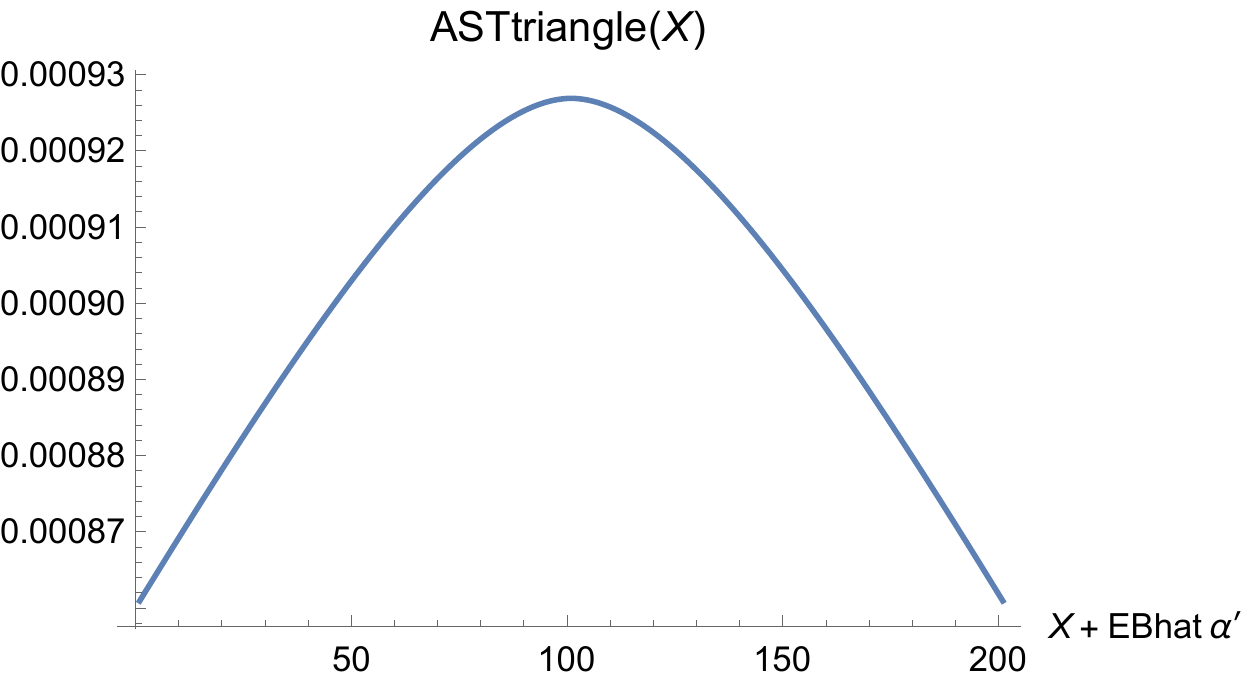}
\end{center}
\caption{The shape of $|A(X)|$ for string theory in the regime of \S\ref{formone}\ and \S\ref{caution}, replacing the step wavepacket by the triangular step $\Psi_{B, \text{tri}}$ described in the main text.  This wavepacket has a more suppressed power law tail in position space. Comparing this to the plots in Figure (\ref{AST}), we see that within the spreading range $X\sim \alpha' E$ depicted, the shape of $A(X)$ is relatively insensitive to the modified B wavefunction tail.  (The size of the amplitude is suppressed, reflecting the fact that the support in momentum space is weighted toward larger values of $k_D^2$ in the triangular step wavefunction.). Nonetheless, as described in the text, there is a subset of kinematic parameter space in which this shape arises from a convergent sum of of QFT propagators. }
\label{ASTtriangle}
\end{figure}  

The shape of $A(X)$ is plotted in figures \ref{AST}-\ref{ASTtriangle}.    
As a function of $X$, the peak wavepacket separation between A and C, it is spread out at the level predicted for spreading-induced interactions.   This includes the predicted dependence on $\kps$ and relative insensitivity to wavepacket parameters. 

However, at least in the subset of the kinematic parameter space we are in, this may be a red herring.  The Beta function factor $B(\kps\alpha', \eta)$ (in the superstring generalization) has a convergent expansion in terms of $\kps$ propagators.  As such, it may be possible to ascribe the scattering at $X<0$ entirely to the tail in this case.  %A caveat to that interpretation, however, is that the reduction to (\ref{sumovern}) based on (\ref{specialpoint}) does not strictly apply to our convolution with the wavepacket since $C\hat{C}$ varies as we integrate over $\tilde p_C$.   
So our analysis in this regime does not lead to a clear identification of the spread shape for $A(X)$ to longitudinal spreading induced interaction, despite the appearance of the predicted scales.  
This example is a cautionary tale against over-interpreting $A(X)$ by itself, although it still illustrates an interesting difference operationally between QFT and string theory in the same setup.  

%In any case, the regime analyzed in the previus subsection more cleanly indicates non-tail contributions via a direct shift and spreading of $A(X)$ compared to the contributions of QFT $\kps$ propagators. 

Since the effect proposed in \cite{lennyspreading}\ does not admit an immediate effective field theory description, it makes sense to search for S-matrix evidence for it from amplitudes not obtained as a convergent sum over QFT propagators.  We turn to that next.

\smallskip
  
{\bf {$\eta<0$ and a stronger test of Longitudinal String Spreading}}    

\smallskip  

Let us start from the form (\ref{FirstAprops}) of the amplitude, and expand the $B(\kps, B\hat{B})$ in terms of propagators, giving
\be\label{propexpB1}
\frac{B(C\hat{C}, \hat{B}\hat{C})}{A\hat{A}}\sum_{n_D}\frac{(1-B\hat{B})_{n_{D}}}{{n_{D}} !}\frac{1}{\kps+{{n_D/\alpha'}}-i\epsilon}\sin\pi(\alpha'\kps+\eta)\Gamma(-\alpha'\kps-\eta)\Gamma(\alpha'\kps+B\hat{B})\frac{\Gamma(\eta)}{\Gamma(B\hat{B})}
\ee
Working in the regime $B\hat{B}\gg -\eta\gg 1$ reduces this to
\be\label{propexpB1simp}
\frac{B(C\hat{C}, \hat{B}\hat{C})}{A\hat{A}}\sum_{n_{D}} \frac{(1-B\hat{B})_{n_{D}}}{{n_{D}} !}\frac{1}{\kps+{{n_D/\alpha'}}-i\epsilon}\sin\pi(\alpha'\kps+\eta)\left(\frac{B\hat{B}}{-\eta}\right)^{\alpha'\kps}
\ee

This exhibits nontrivial $\kps$ dependence multiplying each propagator $\frac{1}{k_D^2+{n_{D}/\alpha'}-i\epsilon}$, so we can make a direct comparison with QFT term by term.   Let us expand the $\sin\pi(\alpha'\kps+\eta)$ into its two phase terms, restricting our momentum interval $\Delta p_C$ to the regime $0\lesssim \alpha'\kps \ll -\eta$.

Since $\kps$ is approximately linear in $p_C$, we get the result for a QFT model with a propagator  (e.g. for ${n_{D}}=0$ the QFT1 model in (\ref{QFT1simple})) but with the replacement
\be\label{Xshift}
X\to X \pm 4\pi E_{\hat B}\alpha'+ 4 i E_{\hat B}\alpha' \log\left(\frac{B\hat{B}}{-\eta}\right) 
\ee
where the $\pm$ depends on which term of the expanded $\sin$ function we consider.

As a result, the extra exponential dependence on $\kps$ yields two terms.  In each, the support of the amplitude is spread out and shifted (either early or late) by a scale $\sim E\alpha'$.  
%\be\label{Xmaxn}
%\Delta X\sim 4 E_{\hat B} \alpha'\log(\frac{B2}{1-B\hat{B}2})
%\ee
As a result, the calculation of the tail contribution $\tilde X_B<-2T_{\text{meet}}$ to $A(X)$ in our position basis in this case does not account for the amplitude:  the dominant resonance is shifted early relative to the tail (and the support in $\tilde X_B$ is also spread out as a result of the imaginary part of the shift (\ref{Xshift})).  

\begin{figure}[htbp]
\begin{center}
\includegraphics[width=8cm]{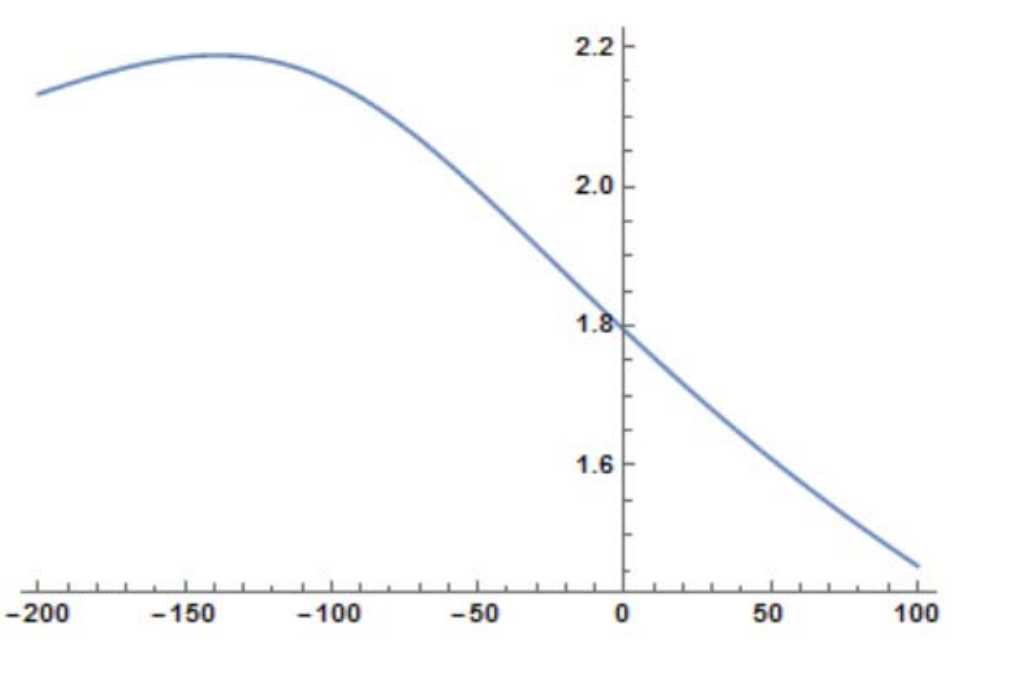}
\end{center}
\caption{The string theory amplitude near the ${n_{D}}=0$ pole has a contribution shifted and spread out by order $\sim E\alpha'$ compared to the QFT1 model in the kinematic regime described in the text.  Given that the latter is on the tail, the former is not.  Here we depict one term of the expansion of the $\sin\pi(\kps+\eta)$ factor in the amplitude in terms of phases; there is a similar term shifted to larger $X$ by $\sim 4\pi E_{\hat B}$ and similarly spread out. }
\label{ShiftShape}
\end{figure}

Figure \ref{ShiftShape}\ illustrates the contrast.  The string theory result is spread by the predicted amount $\sim E\alpha'$ compared to the QFT1 model with amplitude  $\frac{1}{k_D^2-i\epsilon}$  (which scatters on the tail of the wavepacket).

This analysis has focused on the terms in the deformed propagator expansion.  We should make sure that the effect does not cancel out among the propagators.  We have numerically calculated $A(X)$, revealing that the advanced and delayed contributions at the wide scale $\sim 4\pi E_{\hat B}\alpha'$ survive in the amplitude summed over ${n_{D}}$.

\begin{figure}[htbp]
\begin{center}
\includegraphics[width=5cm]{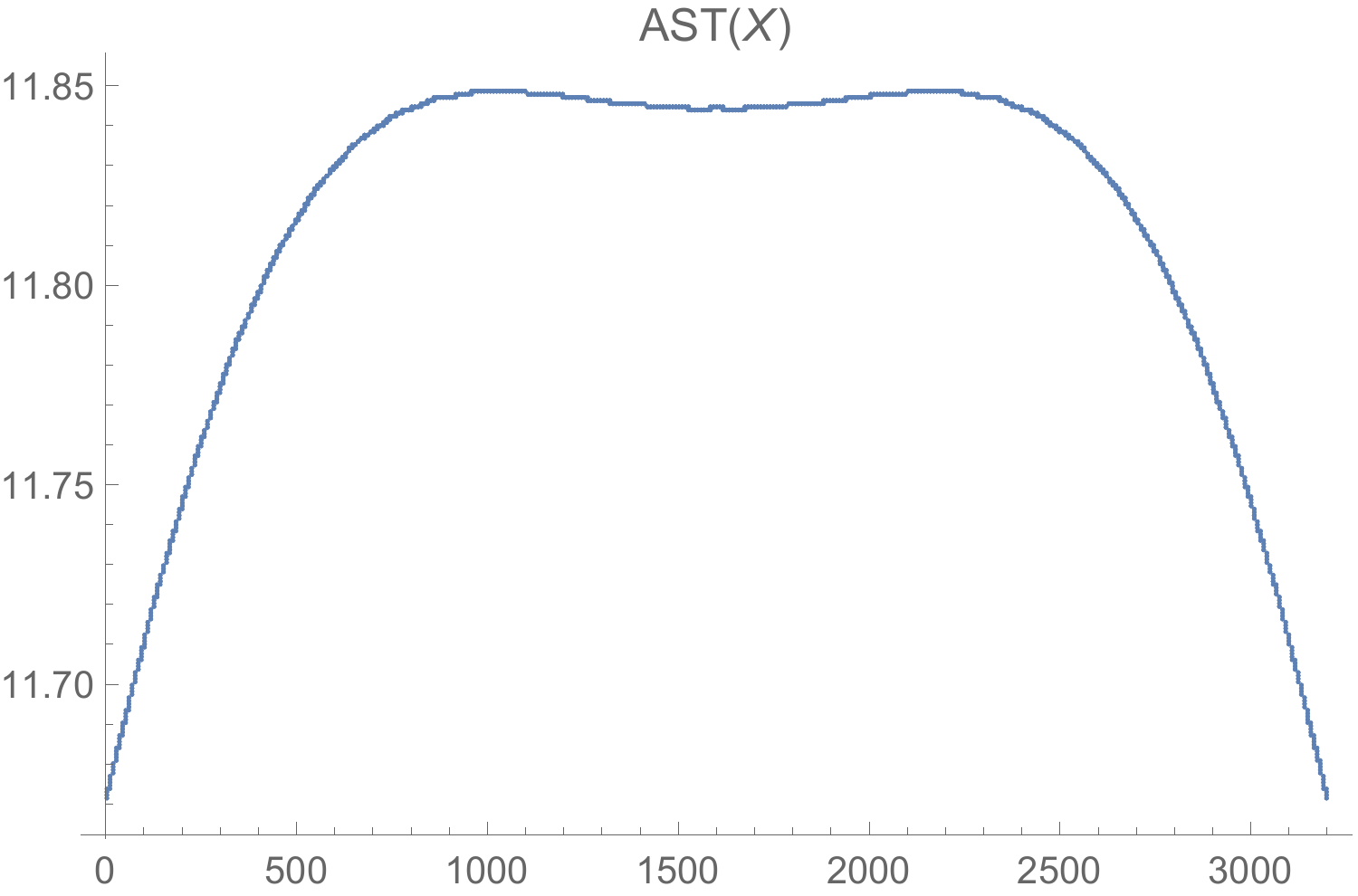} \includegraphics[width=5cm]{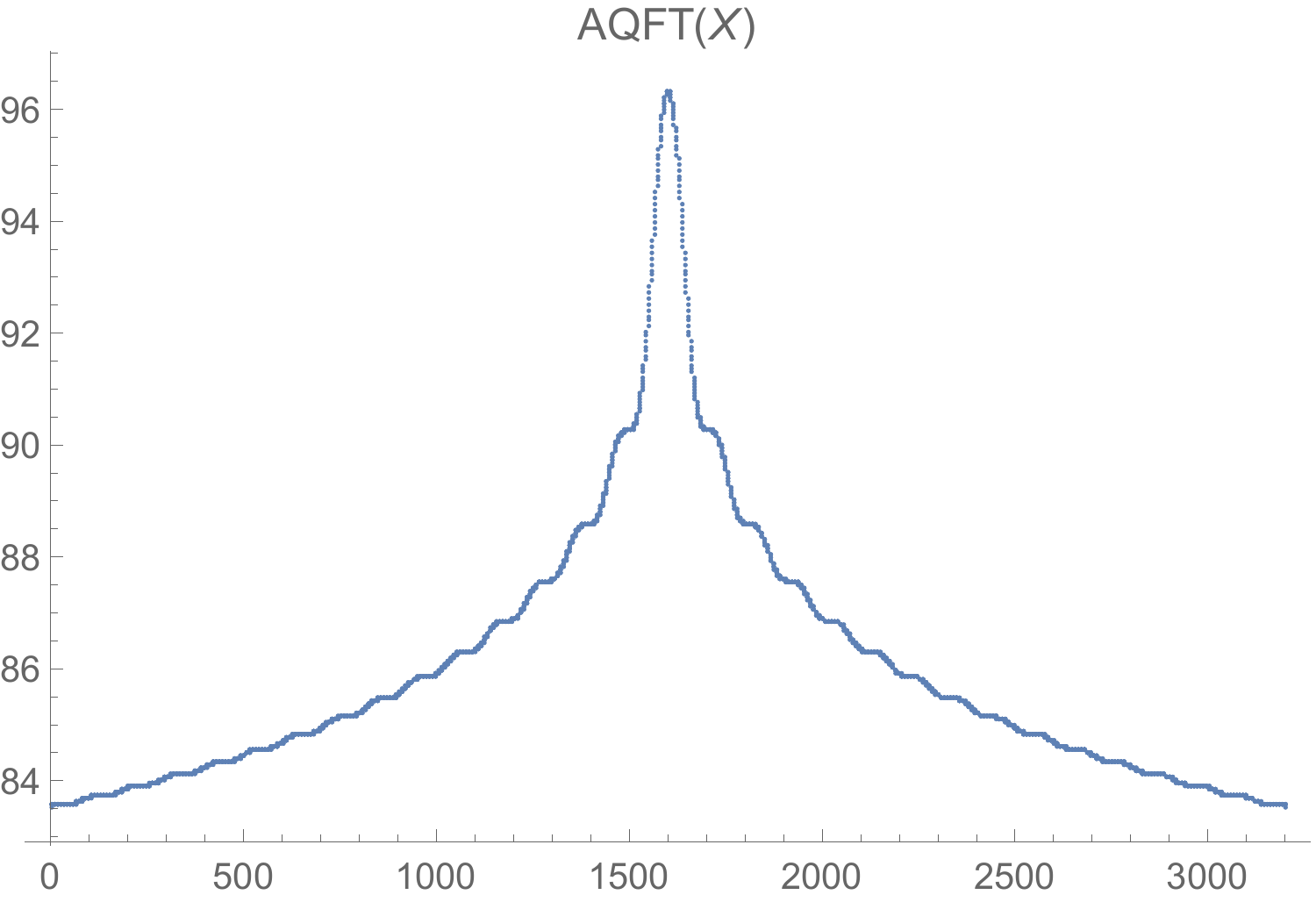}
\end{center}
\caption{For $\eta<0$, the full string theory amplitude  (summed over ${n_{D}}$) pole is spread out by order $\sim 4\pi E_{B}\alpha'$ compared to the QFT1 model in the kinematic regime described in the text.  The term by term spreading described in the text and in figure \ref{ShiftShape}\ survives the sum over the (deformed) propagators.}
\label{etanegativefull}
\end{figure}

Altogether, we find that the prediction of long range interactions via longitudinal string spreading passes a significant test.  This occurs in a regime where the amplitude decomposes into our auxiliary process, a factor that fits with $CD\to \hat{B}\hat{C}$, and an additional factor with structure in $\kps$.  The effect arises most strongly (and most unambiguously) in a kinematic regime where the amplitude does not arise as a convergent sum over propagators.   It is notable that this result is somewhat delicate, depending on kinematic regime choices (such as $\eta<0$ versus $\eta>0$ in the present examples).    As we discussed above in the $q=0$ case, this can be viewed as a parameter of the state of the off-shell detector $D$, one which evidently enters into the process in a nontrivial way.  

%We elaborate on this in a different way in the next section, exploring wider wavepackets in momentum space.  

\section{A family of more general wavepackets}\label{wide}

We have focused thus far on a small range of momentum within which the string theory amplitude contains interesting structure generating advanced interactions, using relatively narrow wavefunctions in momentum space.  Let us finally analyze the problem with more generic wavepackets.  To be specific, we will work at $q=0$, with Gaussian wavepackets for A and C of width $\sigma$ in momentum space.

%\subsection{Timelike $\kps$}

We can rewrite the Beta function containing the leading $\kps$ dependence in the amplitude as
\be\label{Betaredef}
B(\kps, \eta)=B(\kps, K) \frac{\Gamma(\eta)\Gamma(\kps+K)}{\Gamma(\kps+\eta)\Gamma(K)}=\sum_{n_{D}} \frac{(1-K)_{n_{D}}}{{n_{D}} !}\frac{1}{\kps+{n_{D}}-i\epsilon}\frac{\sin\pi(\kps+\eta)\Gamma(-\kps-\eta)\Gamma(\kps+K)}{\sin\pi\eta\Gamma(-\eta)\Gamma(K)}
\ee
in terms of a parameter $K>0$ that we can choose.  We will convolve this with a wavepacket term by term in the sum over $n_D$, and then perform the sum.

Each term is similar to the analysis in \S\ref{qzero}\ above.  We center the wavepacket at 
\be\label{kpszeroKtimelike}
\kps_0 = -\frac{K+\eta}{2},
\ee
which is the minimum of the factors $\Gamma(-\kps-\eta)\Gamma(\kps+K)$ in the amplitude.   This relation enables us to trade the arbitrary parameter $K$ for the central momentum $\kps_0$.  
Since the width of this minimum is of order $\sqrt{K-\eta}$, we further specify $\sigma \equiv\sqrt{c_\sigma (K-\eta)}/4 E_{\hat B}\ll \sqrt{(K-\eta)}/4 E_{\hat B}$ so that these factors are nearly constant within the range of support of the wavepacket.  With these specifications, the convolution of the amplitude with the wavepacket is approximately given by
\be\label{nptermqzero}
 \frac{B(C\hat{C}, \hat{B}\hat{C})}{A\hat{A}}\frac{\Gamma(\eta)\Gamma(-\alpha'\kps_0-\eta)\Gamma(\kps_0+K)}{\Gamma(K)}\sum_{n_{D}}\frac{(1-K)_{n_{D}}}{{n_{D}} !}\int \frac{d\delta \tilde p_C}{\sigma_0} \frac{e^{-\delta\tilde p_C^2/2\sigma_{\text{eff}}^2}e^{i\delta\tilde p_C(X+X_*)}e^{i\pi \eta}}{\tilde\kps+{{n_D/\alpha'}}-i\epsilon}  
\ee 
plus a similar delay term.   Here $\sigma_0$ is as defined above (\ref{sigmaeff}) and $\sigma_{\text{eff}}$ includes the width of the minimum of the factors $\Gamma(-\kps-\eta)\Gamma(\kps+K)$ in the amplitude; we will specify it explicitly below. A saddle point estimate for the integral would give a result proportional to $\exp(-(X+X_*)^2\sigma_{\text{eff}}^2/2)$, with support at $X=-X_*=-4\pi E_{\hat B}$.  We would like to determine if this early support survives the alternating sum over $n_D$.

Before getting to the sum, let us simplify the prefactor in (\ref{nptermqzero}): using (\ref{kpszeroKtimelike}) we find
\be\label{prefactor}
\frac{\Gamma(\eta)\Gamma(-\alpha'\kps_0-\eta)\Gamma(\kps_0+K)}{\Gamma(K)}\simeq \frac{\Gamma(\eta)}{(K/2)^\eta}2^{-K}\simeq \Gamma(\eta)(-\kps_0)^{-\eta} 2^{-K}
\ee
We will find a compensating factor of $2^K$ in the sum, along with a residual suppression. 

Let us work at $X=-X_*$ for simplicity.  
If we write the propagator as the Fourier Transform of a step function,
\be\label{stepFT}
\frac{-i}{z+n-i\epsilon}=\int_0^\infty dw \,e^{-i w(z+n/\alpha'-i\epsilon)}
\ee
and use
\be\label{sumnD}
\sum_n e^{-i w n/\alpha'}\frac{(1-K)_n}{n!}=\left(1-e^{-i w/\alpha'}\right)^{K-1}
\ee
then the sum in (\ref{nptermqzero}) can be written as 
\bea\label{sumint}
& & \int_0^\infty dw \int \frac{dz}{4 E_{\hat B}\alpha'\sigma_0} e^{-(z-z_0)^2/2{\hat\sigma^2}_{\text{eff}}} e^{-i w z}\left(1-e^{-i w/\alpha'}\right)^{K-1}\nonumber \\&=&\frac{1}{4 E_{\hat B}\alpha'\sigma_0}\int_0^\infty dw \, e^{-w^2\hat\sigma_{\text{eff}}^2/2} e^{i w\frac{K+\eta}{2}}\left(1-e^{-iw/{\alpha'}}\right)^{K-1}\nonumber\\ &=&\frac{2^{K-1}}{4 E_{\hat B}\alpha'\sigma_0}\int_0^\infty dw \, e^{-w^2\hat\sigma_{\text{eff}}^2/2} e^{i w\frac{\eta+1}{2}}\left(\sin\frac{w}{2}\right)^{K-1}\nonumber\\
\eea
where we 
%have introduced notation $z$ as shorthand for $\kps$, and 
have defined
\be\label{sighat}
\hat\sigma\equiv 4 E_{\hat B}\sigma_0 =\sqrt{c_\sigma (K-\eta)}, ~~~~ c_\sigma<\frac{1}{4}
\ee
and
\be\label{sigeffnew}
\frac{1}{{\hat\sigma^2}_{\text{eff}}}=\frac{1}{\hat\sigma^2}-\frac{4}{K-\eta}=\frac{1}{\hat\sigma^2}(1-4c_\sigma)
\ee
with the second term arising from the width of the minimum of the factors $\Gamma(-\kps-\eta)\Gamma(\kps+K)$ in the amplitude.
We have also used the fact that the range of $\delta\tilde p_C$ integration in (\ref{nptermqzero}) can be extended to $\pm\infty$ to good approximation since the Gaussian strongly suppresses contributions from the endpoints at $\delta\tilde p_C\sim \pm E$.  

Next, let us study the conditions for a saddle point $w_s$ in the $w$ integral, starting from the second line of (\ref{sumint}).  Neglecting the Gaussian factor, which generically varies more weakly with $w$ than the other two factors, we find 
\be\label{saddlew}
-i w_s = \log\frac{-(K+\eta)}{K-\eta}
\ee 
If $K+\eta<0$, meaning $\kps_0$ spacelike (\ref{kpszeroKtimelike}), then $w_s$ is imaginary and we find no suppression from the Gaussian wavepacket factor at the saddle, as in our previous analysis with $\kps_0=-\eta/2$.  There is a crossver at $\kps_0=0$:  for $K+\eta>0$ then $w_s$ develops a real part, and for large $K$ it is well approximated by $\pi$.

This latter case is relevant for exploring much wider momentum-space wavepackets, for which we can push to the regime $K\gg |\eta|> 1$.  The integral over $w$ has a saddle point near $w\sim\pi$, which can also be seen as  the first maximum of the strongly peaked $\sin^K\frac{w}{2}$ factor in the last line of (\ref{sumint}).  This gives a contribution to the amplitude (including all factors) of order
\be\label{sizewide}
{\cal A}\sim  \frac{B(C\hat{C}, \hat{B}\hat{C})}{A\hat{A}}\frac{1}{4 E_{\hat B}\alpha' \sigma_0} e^{-\pi^2\frac{{\hat\sigma^2}_{\text{eff}}}{2}(1-\frac{4{\hat\sigma^2}_{\text{eff}}}{K})} \left(\frac{\kps_0}{\eta}\right)^{-\eta}
\ee
The penultimate suppression factor here is approximately given by (taking $c_\sigma\ll 1$)
\be\label{suppression}
\exp\left(-\frac{\pi^2\hat\sigma^2}{2}\left[\frac{1-4\hat\sigma^2/K}{1-4\hat\sigma^2/(K-\eta)}\right] \right)=\exp\left(-\frac{X_*^2\sigma_0^2}{2}\left[\frac{1-4\hat\sigma^2/K}{1-4\hat\sigma^2/(K-\eta)} \right]\right)
\ee
where in the second expression we used $X_*=4\pi E_{\hat B}$. 

To interpret this, recall that we have centered the wavepacket at   $X=-X_*=-4\pi E_{\hat B}$.  If the expression in square brackets in (\ref{suppression}) were 1, the factor (\ref{suppression}) would coincide with the suppression factor from the position space tail of the Gaussian wavefunction which would apply if the scattering occurs at the origin (i.e. with no advance between the center of $C$ and the center of $A$).  For positive $\eta$, the factor (\ref{suppression}) suppresses the amplitude more than this, suggesting delayed scattering.  But for $\eta<0$, this factor is less suppressed than the tail factor for scattering at the origin or later (and hence the result is enhanced compared to tree-level QFT).   This is the case which exhibited spreading in the previous examples, and which has no convergent expansion in terms of QFT propagators.  Here we see the effect persists in this family of wavepackets with wider momentum-space width (\ref{sighat}) centered at (\ref{kpszeroKtimelike}).

\section{Conclusions, implications, and future directions}

In this work we performed a new S-matrix test of the prediction \cite{lennyspreading}\cite{BHpaper}\cite{Smatrixpaper}\ that strings with large center of mass energy $\sqrt{s}\sim \sqrt{p^+_{\text{detector}}p^-_{\text{string}}}$ can interact at a large separation $\Delta X^+_{\text{spreading}}\sim 1/p^-_{\text{detector}}$.  In our six point scattering process, such an interaction is predicted between strings $C$ and $D$.  Tree-level quantum field theory has no such interaction.  With appropriate wavepackets peaked at a separation $X$ between $A$ and $C$,
%combined with  wavepackets for string $B$ that are power-law localized, 
we calculated the amplitude $A(X)$ defined in the main text.  In a particular kinematic regime, one in which the string amplitude is not a convergent sum of QFT propagators, this exhibits the predicted difference between string theory and the analogous tree-level QFT models.   String theory has peaked support for interactions at the scale $\alpha' E$, whereas the relevant QFT models exhibit narrow support around $X=0$ dictated by the wavepacket, consistently with scattering on its tail.  In the string theory case, the analysis here combined with \cite{LDpaper}\ provides a nontrivial test of the possibility that the large longitudinal range in $A(X)$ arises from scattering at that scale as suggested by the longitudinal spreading in the light cone wavefunction.

%We derived these results using a wide wavepacket in the $B$ direction in order to keep the momentum space kinematics (related to $\kps$) within a regime where the momentum-space amplitude contains the structure that drives the difference between QFT and string theory.  It will be interesting to consider the implications of this for black hole infallers and thought experiments.  There, the role of the auxiliary process including $B$ is played by the geometry itself, perhaps simplifying the wavepacket analysis.  But the need for the experiment to avoid cancellations among different detector momenta $\kps$ will be interesting to incorporate into black hole dynamics.  On the other hand, the range of the effect, $\sim 4\pi E_{\hat B}$, is longer than the minimal prediction in \cite{lennyspreading, BHpaper}, so in that way our results suggest a stronger effect in black hole physics than previously estimated.  

We have emphasized that tree-level quantum field theory at the same order would not exhibit the same effect, but it is a very interesting question whether rich enough QFT interactions could do so.  This question seems particularly sharp in QFTs with string-theoretic holographic duals, into which we can embed the present calculation \cite{Byungwoo}\ in an appropriate kinematic regime.  The warp factor in the geometry makes a significant difference as compare to the flat spacetime version of the calculation, reproducing essential features of QFT correlators. But one must check carefully whether any long-range interaction survives.  It will be very interesting to see how the tension between the locality of the QFT and these new tests of the non-locality of fundamental strings plays out.        

This sharpens other questions about the role of this effect in the AdS/CFT correspondence, in systems with a perturbative string-theoretic gravity side.  The dual field theory OPE relations constrain the late-time behavior of correlators.  
%In our flat space amplitude, with incoming and outgoing free string states,  vertex operators which are close in the worldsheet OPE expansion can correspond to strings whose center of masses are separated in spacetime.   For example, on the worldsheet $C$ and $\hat C$ approach each other in the Regge limit, while in spacetime the putative spreading interaction between these strings occurs at a long range.    
In curved spacetime examples such as AdS, the details we encountered here will also likely prove important.  For example, we have seen that the strength of the amplitude $A(X)$ depends on the kinematic regime supported by the wavepackets, and we also find that background gradients can affect the spreading \cite{LDpaper}.  In AdS/CFT, the wavepackets injected by insertions of local operators on the boundary may or may not support spreading interactions.        
It will be interesting to understand how this plays out in AdS, in real-time processes describing thermalization as well as the out of time order calculables characterizing chaos \cite{chaos}.  The latter probe near the horizon of the gravity-side black hole, where the most interesting EFT-violating effects can arise.  More generally,  in the context of bulk reconstruction in AdS/CFT, this effect limits the level of locality to be recovered from the CFT; this may be particularly accessible in approaches based on scattering such as \cite{scatteringreconstruction}.   

It will be interesting to apply these results to black hole infallers.  There, the role of the auxiliary process is played by the geometry itself.  Our results provide a test of the prediction \cite{BHpaper}\ that early and late infallers interact parametrically more strongly at a separation $\Delta X^+\sim E\alpha'$ in the near horizon region than is the case for tree-level effective field theory.    
%Longitudinal spreading in perturbative string theory leads to an interesting breakdown of effective field theory in black hole backgrounds \cite{BHpaper}, where a large non-local invariant $s$ arises automatically for systems of early and late infallers, even those sent in with much smaller energies.  
Determining the level of EFT breakdown is important for resolving the thought experimental puzzle articulated in \cite{firewalls}.\footnote{See \cite{BHreviews}\ for recent reviews, and \cite{otherstringy}\ for other recent work on potential string theoretic effects.}  
If the present results correspond to longitudinal spreading induced interactions which translate to the black hole problem as in \S\ref{sizesec}, it would indicate a stronger interaction between early and late infallers than in tree-level quantum field theory.  
Since this effect sets in on a relatively short timescale, it will be interesting to see if it could ever lead to observable signatures. The regime exhibiting an EFT breakdown developed in \cite{BHpaper}, however, requires a strong condition on the local energy of the detector.  Having further tested the basic effect in the present work, exhibiting directly the predicted long-range longitudinal spreading scale, it will be interesting to return to horizon physics to analyze more systematically the scope of the deviations from general relativity that it could produce.

\section*{Acknowledgements}

We dedicate this work to John Schwarz on the occasion of his 75th birthday.  
It is a pleasure to thank Byungwoo Kang for very helpful discussions as well as progress on D-brane and holographic gauge theory applications.
We thank Don Marolf for very useful comments, in particular for a discussion in which we developed this type of setup as a mockup of the configuration relevant in horizon physics \cite{BHpaper}.  We thank Nima Arkani-Hamed, Steve Giddings, David Gross, Joe Polchinski,  Andrea Puhm, Francisco Rojas,  Steve Shenker, Milind Shyani, Douglas Stanford, Tomonori Ugajin, and Ying Zhao for interesting discussions of related issues.  We are also very grateful to Simon Caron-Huot for useful comments and criticisms at an earlier stage of this project.   We would also like to thank the KITP and the Aspen Center for Physics for hospitality during parts of this project.   The work of E.S.~was supported  in part by the National Science Foundation
under grant PHY-0756174 and NSF PHY11-25915 and by the Department of Energy under
contract DE-AC03-76SF00515. The work of M.D. was supported in part by a Stanford Graduate Fellowship.    

\newpage

\appendix

\section{Sums and subtractions in light cone gauge string theory}

In this section we clarify a somewhat subtle aspect of the light cone gauge spreading prediction \cite{lennyspreading}.  This is is somewhat out of the main thread of our analysis of the six point S-matrix amplitude, but is an interesting consistency check of the physical prediction we are testing.  In light cone gauge $X^-=(X^0-X)/\sqrt{2}\equiv x^-+p^-\tau$, the Virasoro constraint identifies the modes of $X^+=(X^0+X)/\sqrt{2}$ with the Virasoro generators of the $D-2$ transverse dimensions.    In the superstring, these Virasoro generators include contributions from worldsheet fermions, reviewed in \cite{GSW}.  This leads to  
\be\label{VarXp}
\langle (X^+(\sigma, 0)-x^+)^2\rangle \propto \frac{c_\perp}{(p^{-})^2}\sum_n n \sim \frac{n_{\text{max}}^2}{(p^-)^2}
\ee   
This is divergent, but can be cut off at a mode number $n_{\text{max}}$ corresponding to the light cone time resolution of a putative detector of the spreading, as reviewed in \cite{BHpaper}.  It leads to the prediction noted in (\ref{refinedprediction}) in the main text.

A similar sum $\propto \sum_{n>0} n$ arises in the mass shell condition determining the string spectrum, which coming from the zero point energy in the worldsheet Hamiltonian.  This quadratic divergence is the contribution of the transverse matter fields to the two dimensional cosmological constant.  A contribution $\sim n_{\text{max}}^2$ in this sum would not reproduce the mass spectrum of the string, which is obtained in the bosonic theory by subtracting this divergence using a local counterterm.   Given this, one might wonder if (\ref{VarXp}) should be cancelled by this subtraction,\footnote{We thank Ying Zhao and Milind Shyani for reminding us of this question.} If there were such a prescription, it would leave intact the transverse spreading prediction (which is logarithmic, and not removable by such a tuning).  

However, it is straightforward to separate issues using the superstring, which explicitly treats the two sums differently.   The calculation of the worldsheet mass spectrum, the leading divergence cancels between worldsheet bosons and fermions.  At the same time, as just noted, no such cancellation arises in (\ref{VarXp}).  This explains why the two infinite sums are not on equal footing, clarifying the prediction (\ref{refinedprediction}) from (\ref{VarXp}) \cite{lennyspreading}.       

\section{Kinematic details}

As described in the main text, we work in a regime with $K_{A\hat{A}}\ll 1$, and focus mainly on transverse momentum eigenstates with $q_A=-q_C\equiv q\ne 0$.  We analyze various kinematic regimes for  $\eta=K_{B\hat{B}}-K_{C\hat{C}}$, $B\hat{B}\hat{C}=K_{\hat{B}\hat{C}}+K_{B\hat{C}}+K_{B\hat{B}}$ and the range of $k_D^2$.  
%ranging from a minimal value near its leading pole to a maximal value satisfying $\alpha' k_D^2\gg K_{B\hat{B}}-K_{C\hat{C}}$.  
In this appendix, we verify that these regimes are consistent by verifying that these quantities are independently tunable to the extent required.

Let us restrict ourselves to a regime for which $E_B-E_{\hat B}, E_C-E_{\hat C},$ and $E_A-E_{\hat A}$ are all $\le { O}(q^2)\ll E^2$. Then (\ref{someKIJs}) simplifies further, and we find
\begin{align}\label{simplerKs}
K_{A\hat{A}}&=(q_B+q_C+q_{\hat B}+q_{\hat C})^2\alpha'=(q_C+q_{\hat B}+q_{\hat C})^2\alpha'\\
K_{C\hat{C}}&=(q_C+q_{\hat C})^2\alpha'\\
K_{B\hat{B}}&=(q_B+q_{\hat B})^2=q_{\hat B}^2\alpha'\\
k_D^2&=K_{C\hat{C}}/\alpha'+4E_{\hat B}(E_C-E_{\hat C})+2q_{\hat B}(q_C+q_{\hat C})+\frac{E_{\hat B}}{E_{\hat C}}(q_C^2-q_{\hat C}^2)
\end{align}
where in the last step we specialized to $q_B=0$ as in the main text, where we also set $Q=q_A+q_B+q_C=0$.

To ensure $K_{A\hat{A}}\ll 1$, let us write $q_C=-(q_{\hat B}+q_{\hat C})+\delta q$ with $\delta q\ll 1$.  Then 
\be\label{B1C2}
\eta=K_{B\hat{B}}-K_{C\hat{C}}=2 q_{\hat B}\delta q \alpha'
\ee
%For this to be $>1$ (so as to access the softness of the string amplitude), we take $q_{\hat B} > 1/(2\delta q) \gg 1$.      
Similarly, we find that 
\be\label{B12kin}
B\hat{B}\hat{C} = 2 k_C\cdot(k_A+k_{\hat A})\alpha'\simeq 2(q_A+q_{\hat A})\left[q_C-\frac{E_C}{2 E_A}(q_A-q_{\hat A})\right]\alpha'\simeq 4 q \delta q \alpha'
\ee
where again $q=q_A=-q_C$.  Thus $\eta$ and $B\hat{B}\hat C$ are independently variable.

Meanwhile the dependence of $\kps$ (\ref{simplerKs}) on $E_C-E_{\hat C}$ enables its range to be separately chosen as in the examples in the main text.
%\be\label{k1psagain}
%k_D^2=4E_{\hat B}(E_C-E_{\hat C})-q_{\hat B}^2\left(1-\frac{E_{\hat B}}{E_{\hat C}}\right)+\frac{E_{\hat B}}{E_{\hat C}}(2q_{\hat B}q_{\hat C}-2\delta q(q_{\hat B}+q_{\hat C}))
%\ee
%There remains sufficient freedom to enable us to choose the support of $\tilde p_C\sim -\tilde E_C$ to lie in the ranges described above.   
A small variation in $\tilde p_C\sim -\tilde E_C$ generates a large change in $k_D^2$ via the $E_{\hat B}$-dependence.

Finally, as described in the main text, a regime of interest will be $-\eta\gg 1$, so we are not working right on the $A\hat A$ pole, but can stay near it in order to work with a relatively simple form for the momentum space amplitude.  

\section{The role of the poles}\label{poleappendix}

In this appendix, we use a toy integral to elaborate on why the $i\epsilon$ prescription in the S-matrix does not by itself exclude time advances (measured with respect to the center of a scattering object), despite the fact that the Fourier transform of a pole is a step function.  Consider the function
\be\label{fA}
f_A(\tilde p)=\left( \frac{A}{\tilde p+a+i\epsilon}+\frac{B}{\tilde p-b-i\epsilon}\right)
\ee  
with $a, b>0$.
This is somewhat analogous to a momentum-space scattering amplitude, but with the energy conserving delta function stripped off; it does not have the square root branch cuts arising in the full propagator of an intermediate state from the on-shell frequencies $\tilde\omega=\sqrt{\tilde p^2+q^2}$ of external states.  The Fourier transform of this function is
\be\label{fTfA}
\tilde f_A(\tilde x)\sim A e^{-i(a+i\epsilon)\tilde x}\theta(-\tilde x)+B\theta(\tilde x)e^{i(b+i\epsilon)\tilde x}
\ee
That is, we get a step function contribution for positive $\tilde x$ from a pole at positive momentum $\tilde p=b$, and a step function contribution for negative $\tilde x$ from a pole at negative momentum $\tilde p=-a$.  

Even before reinstating the missing delta function and frequency dependence, it is worth noting that (\ref{fTfA}) does not imply that scattering at positive momentum only gets support from positive $\tilde x$ and vice versa, because of the uncertainty principle.  To spell that out, we introduce a toy model wavefunction $g_\Psi(\tilde p)=\theta(\tilde p)$ (with Fourier transform $\tilde g_\Psi(\tilde x)\sim 1/(\tilde x+i\epsilon)$), supported only at $\tilde p>0$.   This is supported for the sign of momentum in the pole at $\tilde p=b$ that generated the $B\theta(\tilde x)$ term in (\ref{fTfA}), while the $A\theta(-\tilde x)$ term came from the pole at $\tilde p=a<0$.  Nonetheless, the scattering will depend nontrivially on $A$.  We convolve $g_\Psi$ against the toy amplitude (\ref{fA}), computing
\be\label{fgp}
\int d\tilde p\,  f_A(\tilde p)g_\Psi(\tilde p)=\int d\tilde x\, \tilde f_A(\tilde x) \tilde g_\Psi(-\tilde x)
\ee
where we have expressed this convolution in both position and momentum space.  
Plugging in the above functions, this is
\be\label{fgpexplicit}
\int_{0}^\infty d\tilde p\,\left(\frac{A}{\tilde p+a-i\epsilon}+\frac{B}{\tilde p-b-i\epsilon}\right)=\int_{-\infty}^0 d\tilde x \,\frac{B e^{i(b+i\epsilon)\tilde x}}{\tilde x+i\epsilon}+\int_0^\infty d\tilde{x}\, \frac{Ae^{-i(a+i\epsilon)\tilde x}}{\tilde x+i\epsilon}
\ee
The contribution proportional to $A$ here, the $\tilde x<0$ term in (\ref{fTfA}) does not vanish even though the toy wavefunction only has support at $\tilde p>0$.

Next let us consider a toy problem which incorporates more of the elements of the scattering problem analyzed in the bulk of this paper.  In the extreme case of a $\tilde p_B$ eigenstate there, the toy model amplitude would be of the form 
\be\label{FA}
F_A(\tilde p)=\left( \frac{A}{\tilde p+a+i\epsilon}+\frac{B}{\tilde p-b-i\epsilon}\right)\delta(\sqrt{\tilde p^2+q^2}+\sqrt{(\tilde p-2p_0)^2+q^2}-\omega_0)
\ee  
for constant $p_0, \omega_0$.  
Denoting the two solutions to the $\delta$ function $p_\pm$, the Fourier transform of this amplitude is 
\be\label{FTFA}
\tilde F_A(\tilde x)\sim \frac{e^{ip_+\tilde x}}{|\d \hat f/d\tilde p|_{p_+}|}\left(\frac{A}{p_++a}+\frac{B}{p_+-b}\right)+(+\leftrightarrow -)
\ee
where $\hat f$ is the argument of the delta function.   In this expression, the $A$ and $B$ terms are no longer step functions even at the level of this Fourier transform.  A similar statement holds for the situation worked out in the main text, where the wavefunction for B is given by (\ref{wavepacketPsharp}); this is not a momentum eigenstate but restricts its momentum integral to lie between two closeby values.  As discussed in the main text, this gives us sufficient resolution to distinguish the behavior of our tree-level QFT and string theory amplitudes.  In this appendix, we have simply reviewed in more detail why the $i\epsilon$ prescription in the S-matrix, which is common to string theory and QFT, does not by itself sharply restrict the scattering to a step function contribution.  
\indent 

\begingroup\raggedright\begin{thebibliography}{10}
\baselineskip=14.5pt
\bibitem{lennyspreading}
L. Susskind,
``Strings, black holes and Lorentz contraction,"
Phys. Rev. D {\bf 49}, 6606-6611 (1994). \\ 
M. Karliner, I. R. Klebanov, and L. Susskind,
``Size and Shape of Strings," 
Int. J. Mod. Phys. {\bf A3} 1981 (1988). 
[hep-th/9308139].

\bibitem{grossmende}
 D.~J.~Gross and P.~F.~Mende,
  ``String Theory Beyond the Planck Scale,''
  Nucl.\ Phys.\ B {\bf 303}, 407 (1988).
  %%CITATION = NUPHA,B303,407;%%
  %776 citations counted in INSPIRE as of 11 Aug 2015

\bibitem{BHpaper}
%\bibitem{Dodelson:2015toa} 
  M.~Dodelson and E.~Silverstein,
  ``String-theoretic breakdown of effective field theory near black hole horizons,''
  arXiv:1504.05536 [hep-th].
  %%CITATION = ARXIV:1504.05536;%%
  %1 citations counted in INSPIRE as of 20 May 2015

\bibitem{Smatrixpaper}
M.~Dodelson and E.~Silverstein,
  ``Longitudinal nonlocality in the string S-matrix,''
  arXiv:1504.05537 [hep-th].
  %%CITATION = ARXIV:1504.05537;%%
  %1 citations counted in INSPIRE as of 20 May 2015
%\cite{Dodelson:2015toa}
\bibitem{SCH}

Simon Caron-Huot, unpublished notes.  

\bibitem{bpst}
R. C. Brower, J. Polchinski, M. J. Strassler, C. Tan,
``The Pomeron and gauge/string duality,"
JHEP {\bf{0712}} 005 (2007) [hep-th/0603115].

\bibitem{ACV}

D.~Amati, M.~Ciafaloni and G.~Veneziano,
  ``Superstring Collisions at Planckian Energies,''
  Phys.\ Lett.\ B {\bf 197}, 81 (1987).
  %%CITATION = PHLTA,B\hat{B}97,81;%%
  %450 citations counted in INSPIRE as of 31 Aug 2015

D.~Amati, M.~Ciafaloni and G.~Veneziano,
  ``Can Space-Time Be Probed Below the String Size?,''
  Phys.\ Lett.\ B {\bf 216}, 41 (1989).
  %%CITATION = PHLTA,B216,41;%%
  %723 citations counted in INSPIRE as of 31 Aug 2015
\bibitem{locality}
D. Lowe, J. Polchinski, L. Susskind, L. Thorlacius, J. Uglum,
``Black hole complementarity versus locality,"
Phys. Rev. D {\bf{52}} 6997-7010 (1995) [hep-th/9506138].
%\cite{Polchinski:1995ta}

\bibitem{JoeBHcomp} 
  J.~Polchinski,
  ``String theory and black hole complementarity,''
  In *Los Angeles 1995, Future perspectives in string theory* 417-426
  [hep-th/9507094].
  %%CITATION = HEP-TH/9507094;%%
  %28 citations counted in INSPIRE as of 04 Feb 2015
  
  \bibitem{ggm}
S.~B.~Giddings, D.~J.~Gross and A.~Maharana,
  ``Gravitational effects in ultrahigh-energy string scattering,''
  Phys.\ Rev.\ D {\bf 77}, 046001 (2008)
  [arXiv:0705.1816 [hep-th]].
  %%CITATION = ARXIV:0705.1816;%%
  %55 citations counted in INSPIRE as of 31 Aug 2015

\bibitem{LDpaper}

M. Dodelson, E. Silverstein, G. Torroba, ``Varying dilaton as a tracer of
classical string interactions'', to appear.    
\bibitem{backdraft}
E.~Silverstein,
  ``Backdraft: String Creation in an Old Schwarzschild Black Hole,''
  arXiv:1402.1486 [hep-th].
  %%CITATION = ARXIV:1402.1486;%%
  %13 citations counted in INSPIRE as of 17 Nov 2015

 A.~Puhm, F.~Rojas and T.~Ugajin,
  ``(Non-adiabatic) string creation on nice slices in Schwarzschild black holes,''
  arXiv:1609.09510 [hep-th].
  %%CITATION = ARXIV:1609.09510;%%

\bibitem{PeskinSchroeder}

M.~E.~Peskin and D.~V.~Schroeder,
  ``An Introduction to quantum field theory,''
  Reading, USA: Addison-Wesley (1995) 842 p
  %1028 citations counted in INSPIRE as of 09 May 2017

\bibitem{JoeBook}

J.~Polchinski,
  ``String theory. Vol. 1: An introduction to the bosonic string,''
  Cambridge, UK: Univ. Pr. (1998) 402 p

%\bibitem{strassler5pt}
%C. P. Herzog, S. Paik, M. J. Strassler, E. G. Thompson, 
%``Holographic Double Diffractive Scattering," JHEP 0808 (2008) 010 [hep-th/08060181].
%S. Giddings, private discussions

%\bibitem{sister}
%C. Barratt, ``Multi-Regge Limit of the Virasoro-Shapiro Model: A Sister for the Pomeron," Nucl. Phys. B\hat{B}26 (1977) 133;\\
%P. Hoyer, N. A. Tornqvist, B.R. Webber, ``A new Regge trajectory in the dual resonance model," Phys. Lett. B61 (1976) 191;

%\cite{Witten:2013pra}
\bibitem{Edepsilon} 
  E.~Witten,
  ``The Feynman $i \epsilon$ in String Theory,''
  JHEP {\bf 1504}, 055 (2015)
  [arXiv:1307.5124 [hep-th]].
  %%CITATION = ARXIV:1307.5124;%%
  %15 citations counted in INSPIRE as of 14 Oct 2015

\bibitem{Smatrixbook}
R. J. Eden,
P. V. Landshoff,
D. I. Olive, and
J. C. Polkinghorne, {\it The Analytic S-matrix}, Cambridge University Press, April 2002.  

\bibitem{AllanNima}

A.~Adams, N.~Arkani-Hamed, S.~Dubovsky, A.~Nicolis and R.~Rattazzi,
  ``Causality, analyticity and an IR obstruction to UV completion,''
  JHEP {\bf 0610}, 014 (2006) [hep-th/0602178].
  %%CITATION = doi:10.1088/1126-6708/2006/10/014;%%
  %381 citations counted in INSPIRE as of 02 Jan 2017
  
\bibitem{BP}

A.~Bialas and S.~Pokorski,
  ``High-energy behaviour of the bardakci-ruegg amplitude,''
  Nucl.\ Phys.\ B {\bf 10}, 399 (1969).
  doi:10.1016/0550-3213(69)90128-X
  %%CITATION = doi:10.1016/0550-3213(69)90128-X;%%
  %20 citations counted in INSPIRE as of 28 Sep 2016
  \bibitem{superstringamp}
  
  R. Medina, F. T. Brandt, and F. R. Machado, ``The open superstring 5-point amplitude revisited," arxiv:0208121 [hep-th].
  
K.~Becker, M.~Becker, I.~V.~Melnikov, D.~Robbins and A.~B.~Royston,
  ``Some tree-level string amplitudes in the NSR formalism,''
  JHEP {\bf 1512}, 010 (2015)
  doi:10.1007/JHEP12(2015)010
  [arXiv:1507.02172 [hep-th]].
  %%CITATION = doi:10.1007/JHEP12(2015)010;%%
  %1 citations counted in INSPIRE as of 28 Dec 2016  

\bibitem{Byungwoo}

M. Dodelson, B. Kang, R. Nally, E. Silverstein, et al, work in progress

%\bibitem{giddings}

%S. Giddings, private communication.  

\bibitem{chaos}
J.~Maldacena, S.~H.~Shenker and D.~Stanford,
  ``A bound on chaos,''
  arXiv:1503.01409 [hep-th].
  %%CITATION = ARXIV:1503.01409;%%
  %17 citations counted in INSPIRE as of 31 Aug 2015
%\cite{Shenker:2014cwa}

%\bibitem{Shenker:2014cwa} 
  S.~H.~Shenker and D.~Stanford,
  ``Stringy effects in scrambling,''
  JHEP {\bf 1505}, 132 (2015)
  [arXiv:1412.6087 [hep-th]].
  %%CITATION = ARXIV:1412.6087;%%
  %11 citations counted in INSPIRE as of 31 Aug 2015
%\cite{Shenker:2013yza}

%\bibitem{Shenker:2013yza} 
  S.~H.~Shenker and D.~Stanford,
  ``Multiple Shocks,''
  JHEP {\bf 1412}, 046 (2014)
  [arXiv:1312.3296 [hep-th]].
  %%CITATION = ARXIV:1312.3296;%%
  %35 citations counted in INSPIRE as of 31 Aug 2015
%\cite{Shenker:2013pqa}

%\bibitem{Shenker:2013pqa} 
  S.~H.~Shenker and D.~Stanford,
  ``Black holes and the butterfly effect,''
  JHEP {\bf 1403}, 067 (2014)
  [arXiv:1306.0622 [hep-th]].
  %%CITATION = ARXIV:1306.0622;%%
  %58 citations counted in INSPIRE as of 31 Aug 2015

%\cite{Polchinski:2015cea}

%\bibitem{Polchinski:2015cea} 
  
J.~Polchinski,
  ``Chaos in the black hole S-matrix,''
  arXiv:1505.08108 [hep-th].
  %%CITATION = ARXIV:1505.08108;%%
  %3 citations counted in INSPIRE as of 31 Aug 2015

\bibitem{scatteringreconstruction}
 
 J.~Maldacena, D.~Simmons-Duffin and A.~Zhiboedov,
  ``Looking for a bulk point,''
  arXiv:1509.03612 [hep-th].
  %%CITATION = ARXIV:1509.03612;%%
  %21 citations counted in INSPIRE as of 11 Sep 2016

N.~Engelhardt and G.~T.~Horowitz,
  ``Towards a Reconstruction of General Bulk Metrics,''
  arXiv:1605.01070 [hep-th].
  %%CITATION = ARXIV:1605.01070;%%

\bibitem{firewalls}
A.~Almheiri, D.~Marolf, J.~Polchinski and J.~Sully,
  ``Black Holes: Complementarity or Firewalls?,''
  JHEP {\bf 1302}, 062 (2013)
  doi:10.1007/JHEP02(2013)062
  [arXiv:1207.3123 [hep-th]].
  %%CITATION = doi:10.1007/JHEP02(2013)062;%%
  %386 citations counted in INSPIRE as of 17 Nov 2015

\bibitem{BHreviews}

J.~Polchinski,
  ``The Black Hole Information Problem,''
  arXiv:1609.04036 [hep-th].
  %%CITATION = ARXIV:1609.04036;%%
  %2 citations counted in INSPIRE as of 04 Nov 2016

 D.~Harlow,
  ``Jerusalem Lectures on Black Holes and Quantum Information,''
  arXiv:1409.1231 [hep-th].
  %%CITATION = ARXIV:1409.1231;%%
  %23 citations counted in INSPIRE as of 17 Nov 2015

\bibitem{otherstringy}

R.~Ben-Israel, A.~Giveon, N.~Itzhaki and L.~Liram,
  ``Stringy Horizons and UV/IR Mixing,''
  arXiv:1506.07323 [hep-th].
  %%CITATION = ARXIV:1506.07323;%%
  %3 citations counted in INSPIRE as of 17 Nov 2015

A.~Giveon and N.~Itzhaki,
  ``String Theory Versus Black Hole Complementarity,''
  JHEP {\bf 1212}, 094 (2012)
  doi:10.1007/JHEP12(2012)094
  [arXiv:1208.3930 [hep-th]].
  %%CITATION = doi:10.1007/JHEP12(2012)094;%%
  %39 citations counted in INSPIRE as of 17 Nov 2015

 I.~Bena, G.~Bossard, S.~Katmadas and D.~Turton,
  ``Non-BPS multi-bubble microstate geometries,''
  arXiv:1511.03669 [hep-th].
  %%CITATION = ARXIV:1511.03669;%%
%\cite{Bena:2015lkx}

%\bibitem{Mertens:2015adr} 
  T.~G.~Mertens, H.~Verschelde and V.~I.~Zakharov,
  ``String partition functions in Rindler space and maximal acceleration,''
  arXiv:1511.00560 [hep-th].
  %%CITATION = ARXIV:1511.00560;%%

  %%%%%%
%%%%%%

\bibitem{GSW}

 M.~B.~Green, J.~H.~Schwarz and E.~Witten,
  ``Superstring Theory. Vol. 1: Introduction,''
  Cambridge Monographs in  Mathematical Physics (1987)
  %184 citations counted in INSPIRE as of 20 Oct 2016

%%\bibitem{SpinsStrings} 
%%  S.~Caron-Huot, Z.~Komargodski, A.~Sever and A.~Zhiboedov,
%%  ``Strings from Massive Higher Spins: The Asymptotic Uniqueness of the Veneziano Amplitude,''
%%  arXiv:1607.04253 [hep-th].
  %%CITATION = ARXIV:1607.04253;%%
  %2 citations counted in INSPIRE as of 12 Oct 2016

%\bibitem{juancausality}
 %X.~O.~Camanho, J.~D.~Edelstein, J.~Maldacena and A.~Zhiboedov,
%  ``Causality Constraints on Corrections to the Graviton Three-Point Coupling,''
 % arXiv:1407.5597 [hep-th].
  %%CITATION = ARXIV:1407.5597;%%
  %28 citations counted in INSPIRE as of 15 Apr 2015

%\bibitem{sst}
%N. Seiberg, L. Susskind, and N. Toumbas,
%``Space/Time Non-Commutativity and Causality,"
%JHEP {\bf{0006}} 044 (2000)  [hep-th/0005015].

%\bibitem{HMW}
% D.~Harlow, J.~Maltz and E.~Witten,
  %``Analytic Continuation of Liouville Theory,''
%  JHEP {\bf 1112}, 071 (2011)
%  doi:10.1007/JHEP12(2011)071
%  [arXiv:1108.4417 [hep-th]].
  %%CITATION = doi:10.1007/JHEP12(2011)071;%%
  %89 citations counted in INSPIRE as of 11 Jan 2017
    %62 citations counted in INSPIRE as of 19 Dec 2014
\endgroup
\end{document}